\documentclass{jfm}

\usepackage{graphicx}
\usepackage{float} 
\usepackage{newtxtext}
\usepackage{newtxmath}
\usepackage{natbib}
\usepackage{hyperref}
\usepackage{xcolor}
\usepackage{longtable}
\hypersetup{
    colorlinks = true,
    urlcolor   = blue,
    citecolor  = green!50!black,
}

\newcommand{\RomanNumeralCaps}[1]

\title{ Droplet-triggered Faraday waves in a vertically forced bath
}

\author{F. Tanfeez Mahmood \and C. R. Constante-Amores%
\thanks{\texttt{crconsta@illinois.edu}}}

\affiliation{Department of Mechanical Science and Engineering,
University of Illinois Urbana-Champaign, USA}

\begin{document}
\maketitle

\begin{abstract}
Repeated impacts of a droplet on a vertically forced bath can produce a growing Faraday-wave response under forcing conditions for which the corresponding initially undisturbed bath remains linearly stable. We investigate this mechanism using axisymmetric two-phase direct numerical simulations 
over a broad range of  Reynolds 
and Weber numbers. 
The resulting phenomenological regime map identifies 
quasi-periodic, decaying, faint, and unstable-bath responses.
Along the parameter path considered here, the transition to the unstable-bath regime occurs over a comparatively narrow range of Weber numbers and varies only weakly with Reynolds number.
In the unstable regime, the bath develops a coherent subharmonic response at half the forcing frequency, whereas disturbances in the decaying regime remain broadband and weaken between successive rebounds.
Bath-only simulations show that a finite-amplitude perturbation representative of droplet impact can seed the confined radial mode susceptible to parametric forcing, initiating a growing subharmonic response below the linear instability threshold of the initially flat bath.
A reduced capillary-dominated model provides a scaling estimate for the impact-equivalent perturbation amplitude, and bath-only simulations initialized with these perturbations reproduce the bounded-to-growing transition observed in the coupled simulations.
Floquet analysis of the finite cylindrical bath identifies the first axisymmetric Neumann mode as having the lowest subharmonic instability threshold.
A time-resolved Bessel decomposition independently shows that the growing bath response is dominated by the same radial mode, while the droplet-induced subharmonic response projects predominantly onto this structure. These results identify a finite-amplitude route by which localized droplet impacts can access Faraday-wave dynamics below the linear onset threshold of an initially undisturbed bath.

\end{abstract}


\section{Introduction}
\label{sec:Introduction}

Liquid-drop impact on solid and liquid surfaces occurs everywhere in nature and forms the foundation of a wide spectrum of technological applications, such as inkjet printing and spray coating
\citep{Yarin_Roisman_Tropea_2017,smith2018influence,lohse2022fundamental}. The rich interfacial dynamics of drop impact have been the subject of extensive experimental, theoretical, and numerical attention since the early observations of \cite{worthington1876forms}. Depending upon the impact conditions, and the nature of the substrate underneath it, an impacting drop may spread, splash, coalesce or rebound. If the velocity is sufficiently low, the thin layer of air trapped between the incoming interfaces can 
delay or prevent contact altogether. This allows the drop to bounce or remain afloat \citep{yiantsios1990buoyancy, li2015probing}. This form of air-mediated rebound has been observed in a broad array of configurations, including solid surfaces \citep{de2015wettability}, liquid films \citep{tang2018bouncing, sanjay2023drop}, quiescent liquid pools \citep{klyuzhin2010persisting} and vertically vibrated baths \citep{molavcek2013drops, tambasco2018bouncing}.

When the underlying surface is a deformable liquid, the impact dynamics also involve the coupled deformation of the droplet and the bath. The coalescence of a droplet with a bath of the same liquid has long received experimental attention \citep{schotland1960experimental}, where the final stage is governed by the drainage of the intervening gas film. When this layer becomes sufficiently thin, intermolecular forces trigger rupture and the two interfaces merge \citep{yarin2006drop, de2012dynamics,deka2019coalescence}. However, in an identical low-impact regime, the gas layer can survive long enough to prevent coalescence, allowing the kinetic energy of the droplet to deform the bath and droplet interfaces before capillary relaxation reverses the motion and causes a rebound \citep{zou2011experimental, gopinath2001dynamics}. The dynamics of this interaction are primarily determined by the competing roles of inertia, capillarity, and viscous dissipation which are generally quantified in terms of the Weber number ($We$), Reynolds number ($Re$), and Ohnesorge number ($Oh$).
Consequently, experiments on quiescent baths have shown different coalescing and bouncing regimes where the transition depends on impact conditions and liquid properties \citep{zhao2011transition}. Increasing $We$ in the bouncing regime leads to deeper bath penetration and greater rebound heights. Meanwhile, a large fraction of the incident kinetic energy is converted to deformation and bath motion and subsequently generates capillary waves \citep{wu2020small, alventosa2023inertio, aguero2026droplet}.

Beyond single impacts, the repeated rebound on a quiescent bath offers a natural environment for investigating the coupled evolution of the droplet and the free surface. In non-wetting impact models it has been shown that the deformation of the bath generated during the rebound generates capillary waves and facilitates the exchange of momentum between the impactor and the surface of the liquid \citep{galeano2017non, galeano2019quasi}. When the bath is instead excited with a vertical oscillation, this interaction is sustained by an external forcing. The dynamics then also depend on the peak bath acceleration ($\gamma$) and driving angular frequency ($\omega=2 \pi f$), where $f$ is the forcing frequency. These are typically expressed using dimensionless acceleration ($\Gamma=\gamma/g$) and vibration number ($\Omega=\omega/\omega_D$), where $g$ is the gravitational acceleration and $\omega_D=\sqrt{\sigma/(\rho R^3)}$   is the characteristic capillary oscillation frequency scale  of a droplet of radius $R$, surface tension $\sigma$, and density $\rho$ \citep{molavcek2013drops}. At weak forcing, the bath cannot replenish the energy lost by viscous dissipation and wave generation, so the droplet eventually coalesces. However, beyond a bouncing threshold, momentum gained from the vibrating bath during impact overcomes these losses, allowing for persistent periodic non-coalescing rebounds where the intervening air layer is refilled after each collision \citep{walker1978drops, molavcek2013drops}. When the forcing is further increased, simple bouncing may transition to higher period modes, walking states controlled by interaction with the impact-generated wave field \citep{couder2005walking, primkulov2025diffraction}, and more complicated or chaotic motion \citep{cristea2018walking, saenz2018statistical}. Relevant experimental and numerical studies of droplet rebound, along with the key parameter ranges and number of explored rebounds, are outlined in table \ref{tab: intro}. 

The upper forcing limit of the vertically bouncing regime is determined by the Faraday instability of the bath itself. If a liquid layer is excited by a sufficiently strong periodic vertical forcing ($\omega$), its initially flat interface becomes parametrically unstable to sustained standing capillary-gravity waves, referred to as Faraday waves \citep{faraday1831peculiar, benjamin1954stability}.
Above the critical forcing threshold,  $\Gamma_c$, instability occurs, with the primary response generally being subharmonic, oscillating at half the forcing frequency, $f/2$.
Meanwhile, the associated wavelength is selected using the capillary-gravity dispersion relation \citep{benjamin1954stability}. Each admissible surface mode is reduced to a Mathieu equation by classical inviscid theory \citep{lamb1930hydrodynamics} and alternating harmonic and subharmonic instability tongues are predicted in the forcing frequency-amplitude space.  Classical inviscid theory neglects
viscosity; later analyses treated
viscous damping explicitly.
An extended model was proposed
by \cite{Kumar_Tuckerman_1994} to account for the finite viscous damping, making the threshold finite, while changing the selected wavelengths and the instability boundaries. They performed a complete Floquet analysis of the fully viscous two-fluid problem in a horizontally unbounded domain. However, in laterally confined containers, the sidewalls discretize the admissible spatial modes, and thereby the contact-line orientation of the free surface becomes crucial to the linear stability problem. Finite cylindrical configurations have thus attracted special attention, including 
studies of axisymmetric viscous interfacial oscillations represented by Bessel modes \citep{farsoiya2017axisymmetric}, and
brimful systems with a pinned contact line \citep{bongarzone2022subharmonic}, while three-dimensional simulations have revealed the strong influence of the geometry of the container on the resulting Faraday-wave patterns \citep{perinet2009numerical}. In this study we  consider axisymmetric vertical bouncing, and choose forcing conditions below the classical Faraday threshold of the corresponding undisturbed bath. For the finite cylindrical geometry considered here, the relevant linear modes are thereby axisymmetric and subject to a freely sliding $90^\circ$ contact line. This motivates a corresponding finite-domain extension of the viscous Floquet model.

The aforementioned stability theories indicate the amplification or decay of infinitesimal disturbances to an otherwise undisturbed bath. However, a bouncing droplet continuously perturbs this base state and each impact generates a localized disturbance, resulting in a wave field that can persist into subsequent impacts. Experiments and models of bouncing droplets have analyzed these impact-generated waves and their feedback on the vertical dynamics of the droplet \citep{molavcek2013drops, milewski2015faraday, damiano2016surface}. Experimental work conducted at forcing levels slightly below the Faraday threshold showed that individual impacts can generate slowly damped Faraday waves that persist between successive impacts \citep{tadrist2018faraday}. Studies of walking droplets have shown that the superposition of waves produced by previous impacts can guide the subsequent droplet motion and give rise to wave-mediated horizontal dynamics and path memory \citep{protiere2006particle, harris2014droplets}. Therefore, the effect of the impact-generated wave field on the droplet dynamics is well known.

Despite  these advances, the role of a localized droplet impact as a finite-amplitude perturbation to the parametrically 
forced bath remains less understood. Prior work on subthreshold Faraday waves has focused primarily on the generation and subsequent damping of impact-induced waves below the linear instability threshold, particularly in the context of walking droplets \citep{tadrist2018faraday}. In contrast, the present study remains within the vertically bouncing regime and investigates whether the impact-generated disturbance can trigger a developing subharmonic response under forcing for which the corresponding initially undisturbed bath remains linearly stable, and how this disturbance projects onto the discrete radial modes of a confined bath.
Here, we analyze this coupling using fully resolved axisymmetric two-phase direct numerical simulations across a wide range of $We$ and $Re$.
By comparing  simulations with and without the droplet, we show that a sufficiently large impact-equivalent perturbation can initiate a growing subharmonic bath response at forcing below the linear instability threshold of the initially flat bath. We then develop a reduced scaling estimate for the characteristic deformation generated by impact and test the resulting perturbations independently in bath-only simulations. Finally, an axisymmetric extension of the viscous Floquet formulation of \cite{Kumar_Tuckerman_1994},  together with a Bessel decomposition of the simulated bath response, is used to identify the discrete radial mode governing the onset of the subharmonic response.
 The paper is organized as follows.
Section \ref{sec:Numerical Methods} presents the governing equations, numerical framework, computational setup and validations.
Section \ref{sec: results} first establishes the $We$--$Re$ regime map, and then studies the impact-driven bath response, the mechanism of Faraday-wave onset and its axisymmetric modal dynamics,
and compares it to the decaying regime. Conclusions and remarks are outlined in section \ref{sec: conclusion}.

\begin{table}
\centering
\caption{Summary of prior studies on droplet rebound dynamics and governing parameters 
}
\begin{tabular}{p{3.0cm} p{4.0cm} p{3.5cm} p{2.0cm}}
\hline
Reference & Type of study  & Parameters & Rebounds\\
\hline

\citet{zhao2011transition} & Experimental & 
$1 \le We \le 60$ \newline
$71 \le Re \le 1940$ \vspace{0.5em}
& Single \\

\citet{zou2011experimental} & Experimental & 
$3.5 \le We \le 7.6$ \newline
$740 \le Re \le 1600$ \vspace{0.5em}
& Single \\

\citet{molavcek2013drops} & Experimental &  $0.003 \le We \le 10$ \newline
  $0.156 \le Re \le 1.56$  \vspace{0.5em}
& Multiple \\

\citet{galeano2017non} & Numerical & 
$We = 9.04$, $Re = 61.54$ \vspace{0.5em}
& Multiple \\

\citet{tadrist2018faraday} & Experimental and theoretical & 
$We = 0.124$, $Re = 1.635$ \vspace{2em}
& Multiple 
\\

\citet{wu2020small} & Experimental & 
$1 \le We \le 12$ \newline
$We > 12$ (No bouncing) \vspace{0.5em}
& Single \\

\citet{galeano2021capillary} & Experimental and numerical &  
$1 \le We \le 14$ \newline
  $250 \le Re \le 1225$ \vspace{0.5em}
& 2--3 \\

\citet{alventosa2023inertio} & Experimental and numerical & $0.5 \le We \le 8.0$ \newline
  $15 \le Re \le 280$  \vspace{0.5em}
& Single \\

\citet{aguero2026droplet} & Experimental and numerical & 
$0.0088 \le We \le 7.76$ \newline
$0.6 \le Re \le 310$ \vspace{0.5em}
& Single \\


\end{tabular}
\label{tab: intro}
\end{table}



\section{Problem formulation and numerical method}\label{sec:Numerical Methods}

\subsection{Governing Equations}
\label{sec:Gov. Eqns}

The incompressible two-phase Navier-Stokes equations were solved employing the one-fluid formulation implemented in the open-source software Basilisk \citep{ popinet2018numerical}.
A  liquid droplet of radius $R$ with  density $\rho_{l}$, viscosity $\mu_{l}$, and surface tension $\sigma$  bounces onto a liquid pool with the same material properties, as shown in figure \ref{fig:1}a. The surrounding gas with $\rho_{g}$ and $\mu_{g}$, and the liquid are assumed to be immiscible, incompressible Newtonian fluids.
All variables 
are rendered non-dimensional by using the following scaling, with tildes denoting dimensionless quantities

\begin{equation}
\tilde{r} = \frac{r}{R}, \quad
\tilde{t} = \frac{t}{R/U_0}, \quad
\tilde{\boldsymbol{u}} = \frac{\boldsymbol{u}}{U_0}, \quad
\tilde{p} = \frac{p}{\rho_{l} U_0^{2}}, \quad
\tilde{\rho} = \frac{\rho}{\rho_{l}}, \quad
\tilde{\mu} = \frac{\mu}{\mu_{l}}.
\label{eq:2.1}
\end{equation}
where $t$, $\boldsymbol{u}$, and $p$ stand for dimensional time, velocity, and pressure, respectively.  The length and time
scales are normalized by the droplet radius $R$ and the convection time scale $R/U_0$, respectively.
We acknowledge that the other time scales used in our study include the forcing period ($2\pi/\omega_d$), the gravitational time scale ($\sqrt{R/g}$), the capillary time scale ($\sqrt{\rho_{l}R^{3}/\sigma}$), and the viscous timescale ($\rho_{l}R^{2}/\mu_{l}$).

As a result of this scaling, the continuity and momentum equations become 
\begin{equation}
    \nabla \cdot \boldsymbol{\tilde{u}} = 0,
    \label{eq:2.2}
\end{equation}
\begin{equation}
\tilde{\rho} \left( \frac{\partial \tilde{\boldsymbol{u}}}{\partial \tilde{t}} + \tilde{\boldsymbol{u}} \cdot \nabla \tilde{\boldsymbol{u}} \right) + \nabla \tilde{p} 
= \frac{1}{Re} \, \nabla \cdot \left[ \tilde{\mu} \left( \nabla \tilde{\boldsymbol{u}} + \nabla \tilde{\boldsymbol{u}}^{T} \right) \right] 
+ \frac{1}{We} \, \boldsymbol{\tilde{f}}_{\sigma} 
- \frac{\tilde{\rho}}{Fr^2} \left( 1 + \tilde{F} \sin( \tilde{\omega} t ) \right) \hat{\boldsymbol{z}}.
\label{eq:2.3}
\end{equation}
where the dimensionless parameters present in equation (\ref{eq:2.3}) are defined as follows:
\begin{gather}
\begin{aligned}
We = \frac{\rho_{l} U_0^{2} R}{\sigma}, \quad
Re = \frac{\rho_{l} U_0 R}{\mu_{l}}, \quad
Fr = \frac{U_0}{\sqrt{g R}}, \quad
\tilde{F} = \frac{\gamma}{g} = \frac{A_{0} \omega_d^{2}}{g}, \quad
\tilde{\omega} = \frac{\omega_d R}{U_0},
\end{aligned}
\label{eq:2.4}
\end{gather}
here $We$ signifies the Weber number, representing the ratio of inertial to capillary forces; $Re$ denotes the Reynolds number, which quantifies the ratio of inertial forces to viscous forces; and $Fr$ refers to the Froude number, which measures inertia relative to gravity.
Further the $We$ and $Re$ are related through the Ohnesorge number, $Oh=\sqrt{We}/Re=\mu/\sqrt{\rho_l\sigma R}$, which characterizes the interplay between capillary and viscous forces in the liquid. $\tilde{F}$ and $\tilde{\omega}$
correspond to the dimensionless forcing acceleration amplitude and the dimensionless angular forcing
frequency, respectively. They account for the vibrations of the bath, whose dimensional terms are normalized by the gravitational acceleration $g$, peak forcing amplitude $A_{0}$, and time scale $U_0/R$.
Finally, 
density and viscosity ratios are represented by $\rho^{*} ={\rho_{g}}/{\rho_{l}}$ and $\mu^{*}={\mu_{g}}/{\mu_{l}}$, respectively, where the former is fixed to air-water and the latter is varied through $\mu_l$ in this work.


\subsection{Numerical framework}
The axisymmetric direct numerical simulations (DNS) are performed utilizing the open-source software Basilisk C \citep{popinet_basilisk}, which solves the incompressible Navier-Stokes equations (see \ref{eq:2.2} and \ref{eq:2.3}) employing a second-order accurate finite-volume method \citep{popinet2018numerical}.
A volume-of-fluid (VoF) formulation is applied to describe the multiphase flow system. 
In order to guarantee non-coalescence between the droplet and the bath, we adopt two independent VoF tracer regions, $f_{d}$ and $f_{b}$, which denote the droplet and bath liquids, respectively 
\begin{equation}
\left( \frac{\partial}{\partial t} + \boldsymbol{u} \cdot \nabla \right) 
\left\{ f_{d}, \, f_{b} \right\} = 0,
\label{eq:2.6}
\end{equation}
while the surrounding gas is represented implicitly through $f_{a}= 1 - f_{d}-f_{b}$ \citep{ramirez2020lifting,sanjay2023drop}.
This two-colour formulation prevents numerical merging of the droplet and bath interfaces and therefore imposes non-coalescence in the absence of an explicit rupture model.

The material properties (including dimensionless effective density $\tilde{\rho}$ and viscosity $\tilde{\mu}$, respectively) are described under the one-fluid formulation assumption  by
\begin{equation}
\tilde{\rho} = \left(f_{d}+ f_{b} \right)+\rho^{*}\left(1-f_{d}- f_{b} \right),
\end{equation}
\begin{equation}
\tilde{\mu} = 
\left(f_{d}+ f_{b} \right)+
\mu^{*}\left(1-f_{d}- f_{b} \right),
\end{equation}

The surface tension force $\boldsymbol{\tilde{f}}_{\sigma}$ is introduced at the interfaces to address the dynamic boundary condition present across them. The force is modeled approximately in accordance with the methodologies established by \cite{brackbill1992continuum} and \cite{tryggvason2006direct}, using the height function approach to calculate the interfacial curvature. 
To circumvent spurious currents near the interface,
the surface tension force is expressed as
\begin{equation}
\boldsymbol{\tilde{f}}_{\sigma} \approx \tilde{\mathcal{K}}_{d} \, \nabla f_{d} + \tilde{\mathcal{K}}_{b} \, \nabla f_{b},
\end{equation}
where  $\tilde{\mathcal{K}}_{d}$ and $\tilde{\mathcal{K}}_{b}$ are the curvatures corresponding to $f_{d}$ and $f_{b}$, respectively. 

\begin{figure}
\centering
\begin{tabular}{cc}
\includegraphics[width=0.5\textwidth]{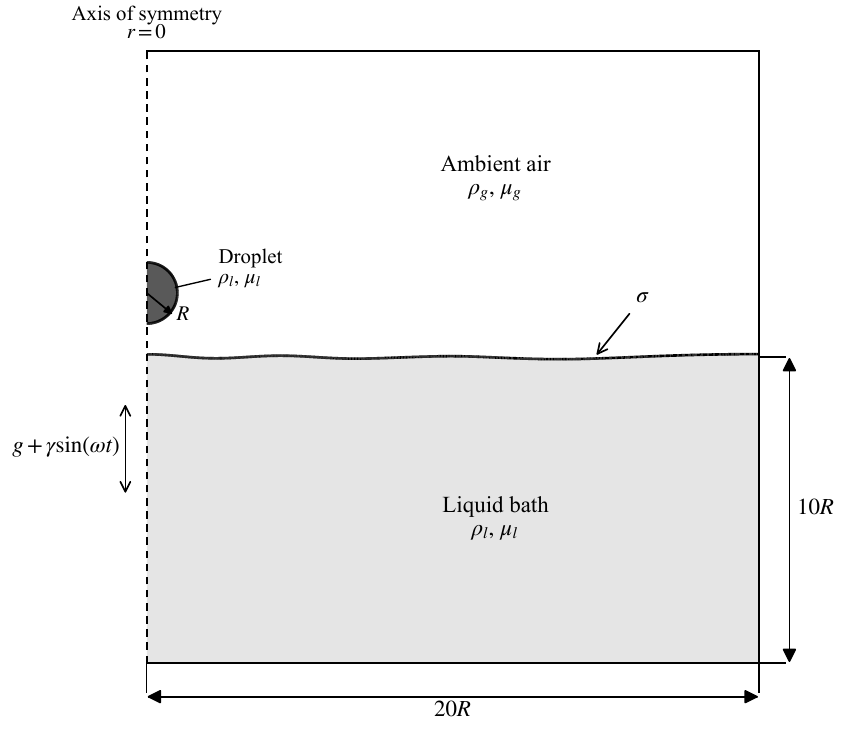}&
\includegraphics[width=0.37\textwidth]{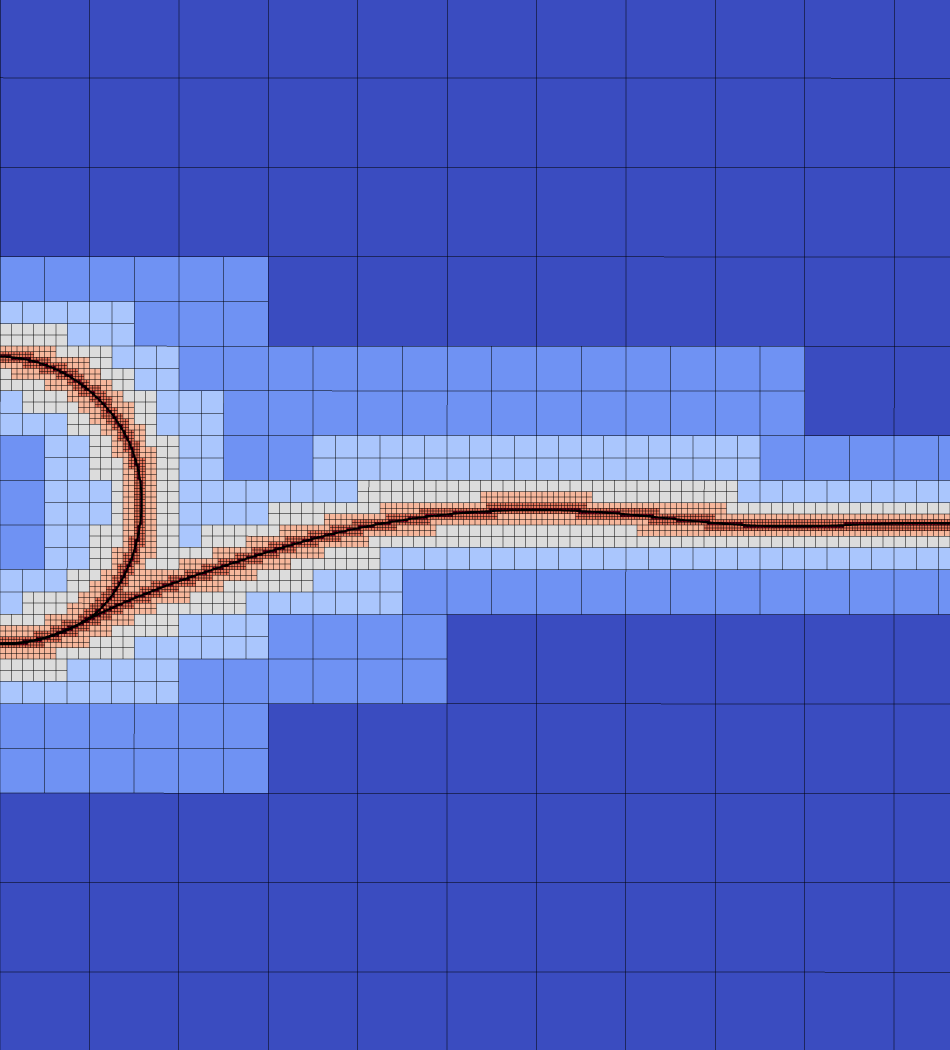}\\
(a) & (b)\\
\end{tabular}
\caption{Axisymmetric computational domain and adaptive mesh. (a) Schematic of the axisymmetric domain, with the dashed line showing the axis of symmetry, $r=0$. (b) Representative mesh refinement 
around the droplet and bath interfaces. }
\label{fig:1}
\end{figure}

The present axisymmetric 2D computational domain is discretized using the quadtree-based finite volume grid \citep{popinet2015quadtree}, which allows for non-uniform resolution. As illustrated in figure \ref{fig:1}b, an adaptive mesh refinement (AMR) strategy has been utilized to dynamically resolve the critical interfacial regions (e.g., the droplet interface and the bath surface) and capture the bouncing mechanism accurately. The AMR approach follows the wavelet-based error minimization method \citep{popinet2009accurate}, with refinement tolerances of $10^{-3}$ and $10^{-4}$, for the VoF tracers and velocity field, respectively. To further guarantee the stability of the solution in the presence of rapidly evolving interfaces, the time step is dynamically controlled to maintain a maximum Courant number of CFL = 0.25. 


\subsection{Numerical setup, validation, and parameters}

A spherical droplet is initiated at a vertical distance $h=R$ above the bath surface with a velocity $U_0$. The bath has a depth of $H=10R$ (pool height), which is sufficiently deep to avoid any artificial boundary effects on wave generation during impact and rebound. The computational domain is square and spans a size of $20R\times20R$, consistent with previous studies
\citep{galeano2021capillary,alventosa2023inertio}. A wide range of surface tension $\sigma$ and dynamic viscosity $\mu_l$ values have been considered to effectively capture the competition of viscous and surface tension forces.
Complete details of the fluid properties are outlined in table \ref{tab:1}.

The cylindrical bath is subjected to a sinusoidal vertical acceleration,
\begin{equation}
\xi(t)=g+\gamma\sin(\omega_dt),
\end{equation}
where $\gamma = A_{0} \omega_d^{2}$ is the peak forcing acceleration and $\omega_d$ is the imposed angular forcing frequency.
No explicit perturbation is imposed on the bath interface in the coupled droplet simulations so that the initial disturbance is generated self-consistently by droplet impact. Since the forcing acceleration is comparable to gravity  
($\tilde{F}=1$), we restrict attention to  the bouncing droplet regime.

For the fluid properties, geometry and forcing frequency considered here, bath-only simulations and the Floquet analysis, described  in section \ref{Floquet_analysis_section},
give a linear onset at $\tilde{F}_c = 1.17$.
The coupled droplet simulations are performed at $\tilde F=1$, below this linear instability threshold of the initially undisturbed bath.
We acknowledge that \cite{molavcek2013drops} similarly report a subcritical bouncing regime below their Faraday threshold, although their threshold is not directly comparable due to differences in fluid properties and forcing conditions.

An impermeable no-slip condition is imposed at the pool bottom, while the artificial vertical wall is treated as free-slip and impermeable. The upper computational
boundary is assigned an outflow condition.
To accurately encompass the extended-time dynamics of the bouncing droplet (e.g., spanning over 10 bounces), 
the bath interface is prescribed to meet the vertical wall at a contact angle of $90^\circ$.
Finally, an axisymmetric condition imposed at the domain centerline ($r = 0$), allows the use of a partial computational domain. We acknowledge that our setup  suppresses azimuthal symmetry breaking,  and therefore cannot capture lateral walking states \citep{protiere2006particle, harris2014droplets}. Consequently, the current simulations are restricted to vertically bouncing states.

\begin{table}
  \begin{center}
\def~{\hphantom{0}}
  \setlength{\tabcolsep}{15pt}
  \begin{tabular}{lcc}
      \textbf{Parameter}  & \textbf{Symbol}  & \textbf{Value} \\[3pt]
       Impact speed       & $U_0$              & ~~50 cm/s~ \\
       Drop radius        & $R$              & ~~0.035 cm \\
       Initial drop height & $h$             & ~~0.035 cm \\
       Container radius        & $R_c$              & ~~0.7 cm \\
       Bath height        & $H$              & ~~0.35 cm \\
       Density (liquid)   & $\rho_{l}$       & ~~998 kg/m$^{3}$ \\
       Density (gas)      & $\rho_{g}$       & ~~1.21 kg/m$^{3}$ \\
       Dynamic Viscosity (liquid)   & $\mu_{l}$       & ~~$0.873 \times 10^{-3}$ -- $3.493 \times 10^{-3}$ Pa.s \\
       Dynamic Viscosity (gas)   & $\mu_{g}$       & ~~$1.81 \times 10^{-5}$ Pa.s \\
       Surface Tension (liquid)   & $\sigma$   & ~0.0087 -- 0.87 N/m \\
       Gravitational acceleration   & $g$          & 9.81 m/s$^{2}$  \\
       Forcing amplitude   & $A_{0}$          & $0.297$mm \\
       Forcing frequency    & $\omega_d$          & 181.9 rad/s  \\
       \hline
       Weber Number   & $We$          & 0.1 -- 10 \\
       Reynolds Number   & $Re$          & 50 -- 200  \\
       Ohnesorge Number   & $Oh$          & 0.0016 -- 0.0632  \\
       Froude Number   & $Fr$          & 8.53  \\
       Density ratio   & $\rho^{*}$          & $1.2 \times 10^{-3}$  \\
       Viscosity ratio   & $\mu^{*}$          & $0.52 \times 10^{-2}$ -- $2.07 \times 10^{-2}$  \\
       Forcing frequency    & $\tilde{\omega}$          & $0.127$  \\
       Forcing acceleration    & $\tilde{F}$          & 1.0  \\
  \end{tabular}
  \caption{Parameters and their characteristic range of values relevant to this study.}
  \label{tab:1}
  \end{center}
\end{table}

In the present work,  simulations use a maximum refinement level of $LVL=11$,  maintaining a minimum grid size of
$\Delta=R/102.4$, 
resulting  in approximately 102 grid cells per droplet radius, as used previously by 
\citet{galeano2021capillary,alventosa2023inertio}. In selected cases, a higher resolution of $\Delta = R/204.8$ was used to accurately resolve the pressure distribution within the thin gas layer. 
We have carried out extensive grid independence studies  to ensure that changing the grid size does not influence the results;   see Appendix \ref{App D} for more details.

We have employed different strategies for the validation of our numerical framework. 
First, in the absence of external forcing ($\tilde{F}=0$), the droplet rebound dynamics were compared against the experimental results of \citet{alventosa2023inertio}. This comparison shows excellent agreement in droplet trajectory, confirming that the solver accurately captures the experimental dynamics (see section \ref{App A1} from Appendix \ref{App A} for more  details). Second, the bath response without a droplet is validated independently against the linear stability theory of \citet{Kumar_Tuckerman_1994}.
The DNS growth rates agree well  with the theoretical predictions over the forcing range considered (see section \ref{App A2} from Appendix \ref{App A}).
Third, 
the long-time  dynamics of a droplet on a vibrating bath
are  compared with the experimental regime structure reported by  \citet{molavcek2013drops}. The simulations reproduce  the qualitative signatures associated with the experimental coalescence, sustained bouncing, and Faraday-wave-dominated responses.
For the coalescence case, we recover the loss of sustained rebound associated experimentally with the coalescence region.  For the sustained bouncing we  capture the $(1,1)^1$ bouncing mode, and the corresponding bouncing period (see section \ref{App A3} from Appendix \ref{App A} for more details).

Hereafter, all quantities are dimensionless unless stated otherwise, and tildes are dropped for simplicity.

\section{Results}
\label{sec: results}

\subsection{Regime structure and energetic signatures}

\begin{figure}
\centering
\includegraphics[width=\linewidth]{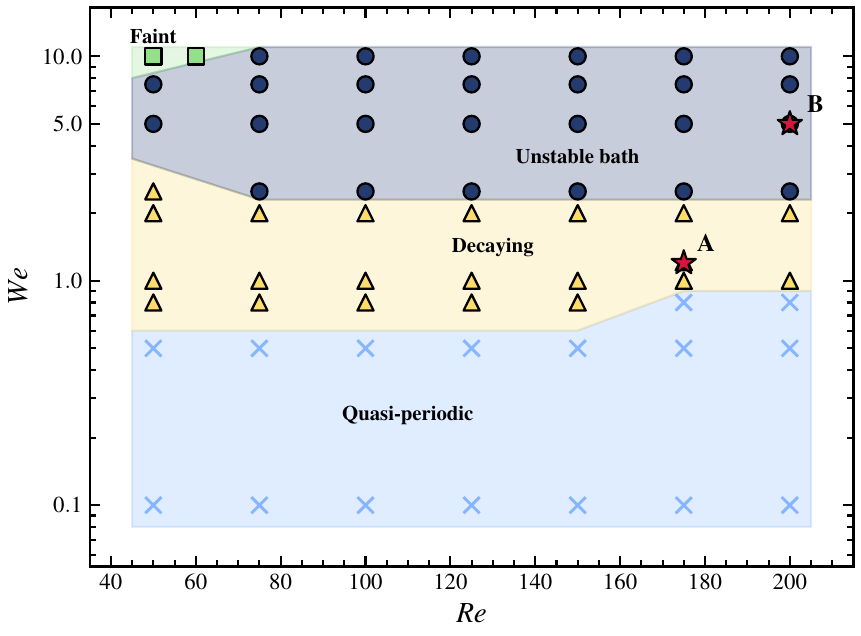}
\caption{Phenomenological regime map of the droplet--bath response for $50\leq Re\leq200$ and $0.1\leq We\leq10$. 
Markers A 
and B
highlight the representative decaying and unstable-bath cases discussed below in the text, respectively.}
\label{fig:regime}
\end{figure}

We first characterize how the long-time droplet-bath dynamics change across the parameter space before examining the origin of the unstable state in detail in the next section. Figure~\ref{fig:regime} summarizes the response over $0.1\leq We\leq10$ and $50\leq Re\leq200$,
spanning conditions motivated by the experiments of \citet{alventosa2023inertio}. Four qualitatively distinct responses are observed: quasi-periodic bouncing, decaying bouncing, an unstable-bath state, and a faint bouncing state.

\begin{figure}
\centering
\begin{tabular}{ccc}
\hline
&$We=0.80$ and $Re=175$ &\\
\includegraphics[width=0.33\linewidth]{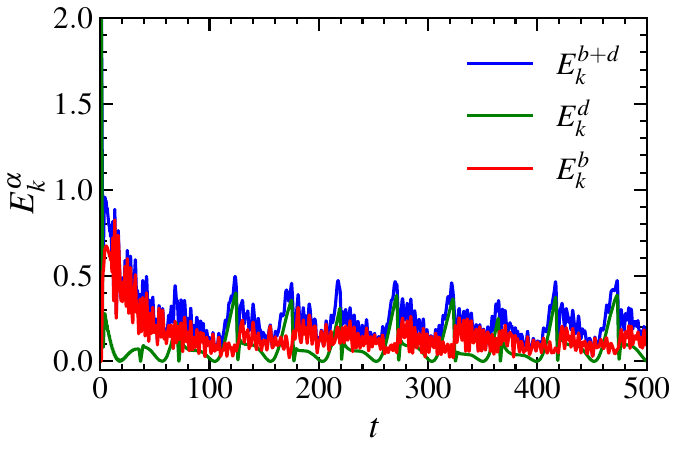}&
\includegraphics[width=0.33\linewidth]{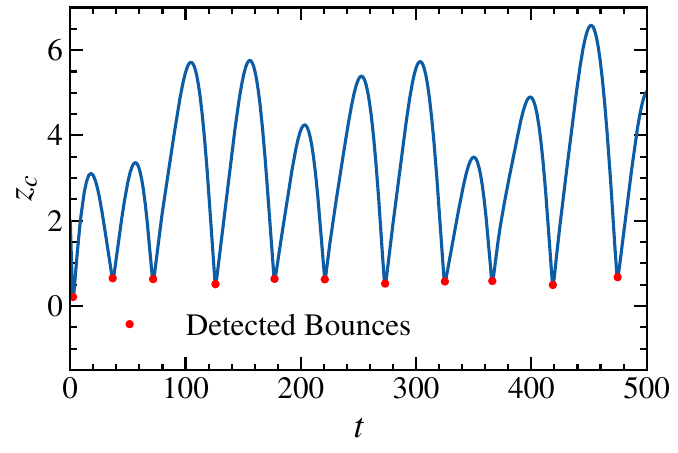}&
\includegraphics[width=0.37\linewidth]{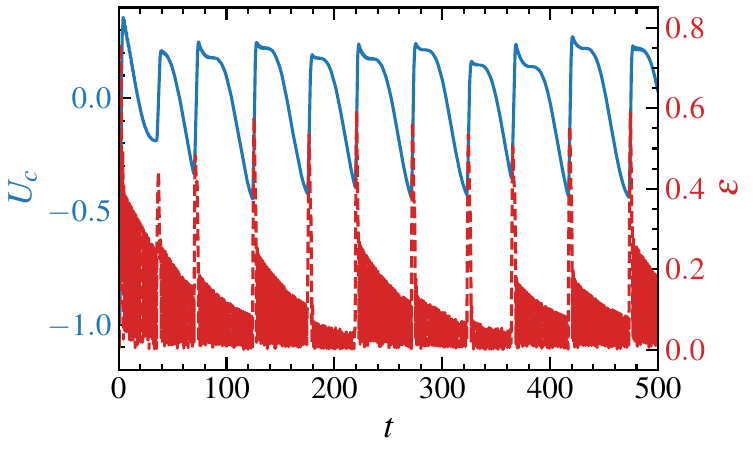}\\
&(a)& \vspace{5pt}\\

&$We=1.21$ and $Re=175$ &\\
\includegraphics[width=0.33\linewidth]{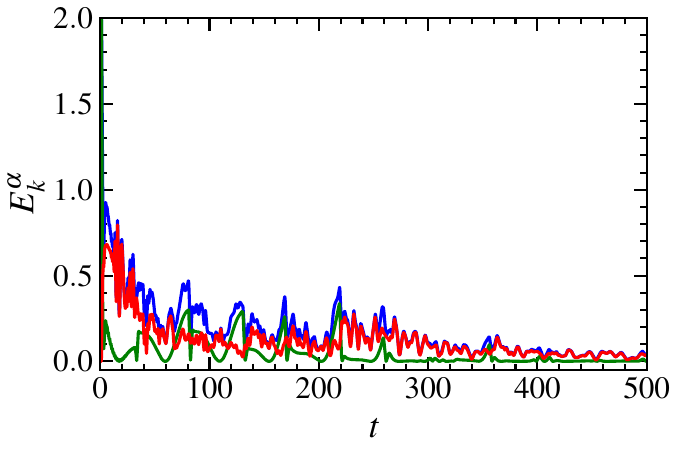}&
\includegraphics[width=0.33\linewidth]{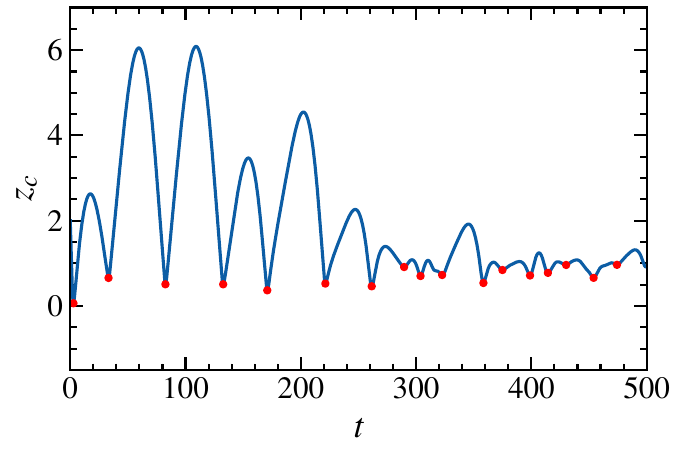}&
\includegraphics[width=0.37\linewidth]{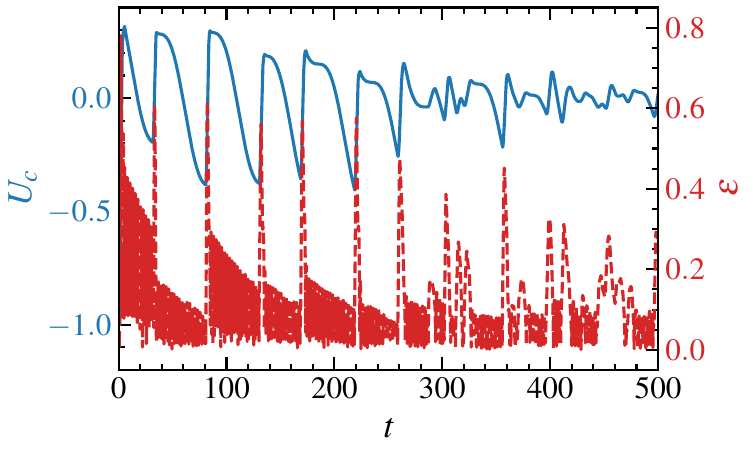}
\\
&(b)& \vspace{5pt}\\
&$We=5.00$ and $Re=200$ &\\
\includegraphics[width=0.33\linewidth]{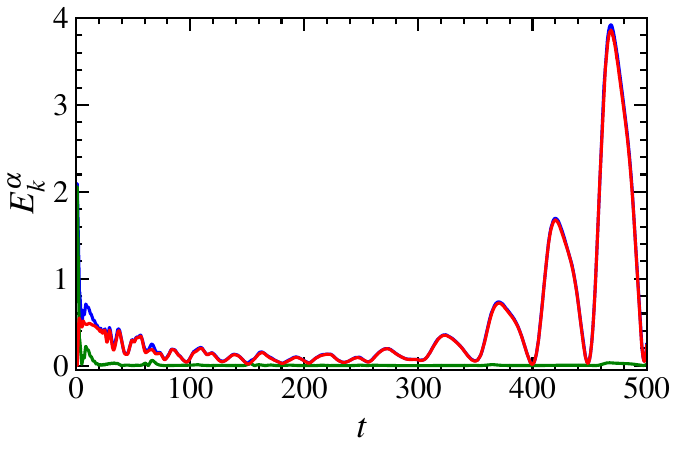}&
\includegraphics[width=0.33\linewidth]{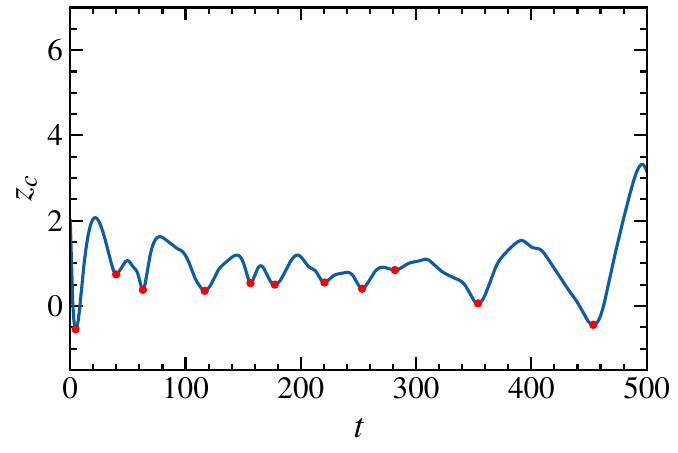}&
\includegraphics[width=0.37\linewidth]{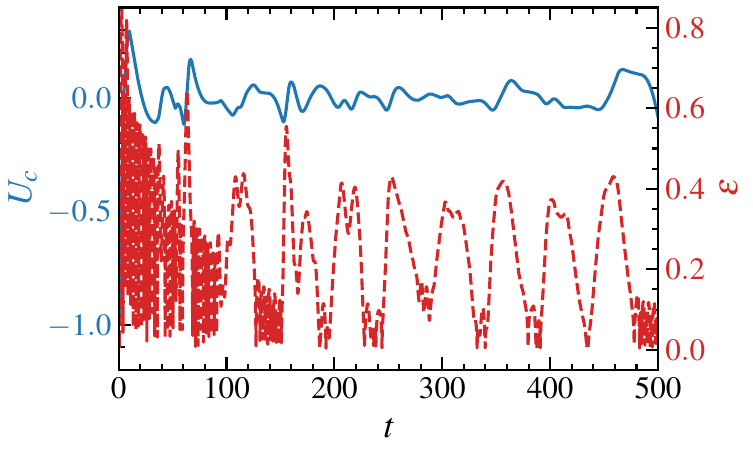}
\\
&(c)& \vspace{5pt}\\
&$We=10$ and $Re=50$ &\\
\includegraphics[width=0.33\linewidth]{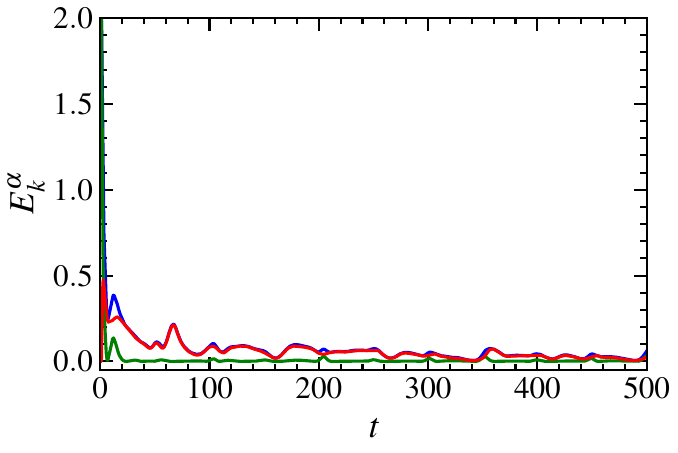}&
\includegraphics[width=0.33\linewidth]{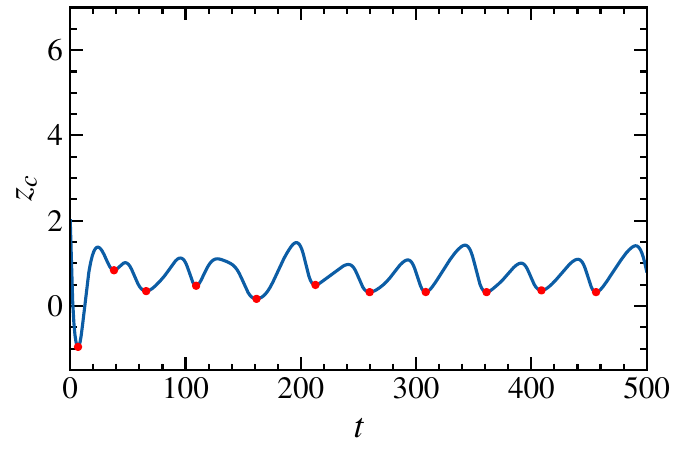}&
\includegraphics[width=0.37\linewidth]{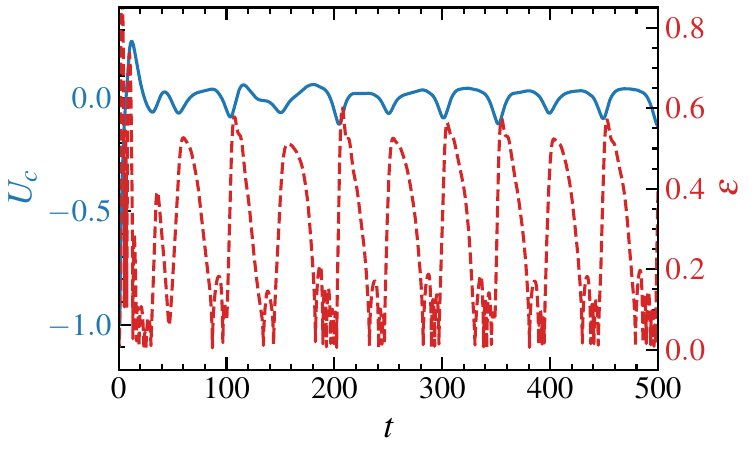}\\
&(d)&\\
\end{tabular}
\caption{
Time-resolved diagnostics used to distinguish the dynamical regimes in figure \ref{fig:regime}. 
Rows correspond to
(a) quasi-periodic bouncing, $(We,Re)=(0.80,175)$;
(b) decaying bouncing, $(1.21,175)$;
(c) unstable-bath dynamics, $(5.00,200)$; and
(d) faint decaying bouncing, $(10,50)$.
The left column shows the kinetic-energy evolution of the liquid phase $E_k^{b+d}$ and its droplet $E_k^d$ and bath $E_k^b$ contributions. 
The middle column shows the vertical center-of-mass position of the droplet, $z_c(t)$, with detected bounce events marked by red symbols. 
The right column compares the center-of-mass vertical velocity $U_c(t)$ with the droplet ellipticity $\varepsilon(t)$. 
}
\label{fig:map2}
\end{figure}

The classification in figure~\ref{fig:regime} was obtained from the time-resolved diagnostics shown in figure~\ref{fig:map2}. For every simulation, we examined the kinetic-energy partition between the droplet
and bath together with the droplet centre-of-mass trajectory. The centroid velocity and ellipticity, defined in Appendix~\ref{App_droplet_measures}, were used as complementary measures of rebound coherence and deformation. All simulations were continued to at least
$t=750$. Quasi-periodic states are characterized by sustained rebound events, bounded kinetic energy and a repeatable relationship between translation and deformation. Decaying states retain identifiable
rebounds but exhibit a progressive reduction in their amplitude and energetic content. The unstable-bath state is distinguished by a late-time increase of the bath kinetic energy that is not accompanied by increasingly energetic droplet rebounds. Finally, the faint state contains only small-amplitude vertical motion and weak kinetic energy, without the late-time growth characteristic of the unstable bath. Figure~\ref{fig:map2} illustrates these distinctions using four representative trajectories.

For $(We,Re)=(0.80,175)$, the system establishes a persistent bouncing state following the initial impact transient (see figure~\ref{fig:map2}a). The centre-of-mass trajectory exhibits repeated rebounds with a characteristic interval $t_p\simeq45$, while the droplet and bath kinetic energies remain bounded over the duration of the simulation. Each collision is accompanied by a rapid reversal of the centroid velocity ($U_c$) and a pronounced change in ellipticity, indicating a repeatable coupling between the vertical motion and droplet deformation. The rebound sequence persists throughout the simulation without any systematic growth or decay in amplitude.

A different response occurs for $(We,Re)=(1.21,175)$,  which is classified as decaying (figure~\ref{fig:map2}b). Several energetic rebounds are initially observed, but their amplitude progressively decreases. This decay is evident not only in the centre of-mass excursions, but also in the kinetic-energy partition, in which the bath kinetic-energy peaks weaken progressively as the rebound sequence develops. The sharp reversals of the centroid velocity likewise become weaker and less regularly spaced,
while the ellipticity loses its repeatable synchronization with the bounce events. Thus, both the droplet rebound and the bath response are progressively attenuated. The decreasing bath motion is consistent with a reduced ability of the bath to return momentum to the droplet during subsequent encounters, ultimately preventing the bouncing cycle from being sustained.

The response changes more fundamentally for $(We,Re)=(5.00,200)$, which we classified as an unstable bath regime (see figure~\ref{fig:map2}c). By looking at the  centre of-mass, the droplet  remains close to the bath and undergoes only weak and irregular vertical excursions. The ellipticity likewise displays large and irregular excursions rather than the repeatable deformation cycles observed in the bouncing state. Interestingly, the kinetic energy of the droplet is very small in comparison to the bath. 
At longer times ($t>300$), the bath contribution begins to increase and develops successively larger peaks. In  Section ~\ref{Sub: Unstable bath}, we  will show that the growth of the system energy is due to  the excitation of a subharmonic Faraday mode.

The faint state, illustrated by $(We,Re)=(10,50)$ in figure~\ref{fig:map2}d, exhibits a distinct kinetic-energy signature. Following the initial transient, the system rapidly settles into a weakly energetic response, without developing a sequence of large rebounds. The centre-of-mass motion remains confined to small-amplitude,
comparatively regular oscillations, while the kinetic energy remains small and is dominated by the bath contribution. The ellipticity, however, retains a pronounced periodic modulation, indicating that
droplet deformation persists despite the weak translational motion. This behavior differs from the decaying regime, which exhibits a sequence of pronounced kinetic-energy peaks associated with successive rebounds whose amplitudes progressively decrease in time. In contrast, the faint state shows no comparable energetic peaks and remains weakly excited throughout the simulation.

The parameter dependence of these four responses is summarized in
figure~\ref{fig:regime}. Increasing $We$ along the present parameter path first produces a band of decaying dynamics and subsequently an unstable-bath response over much of the explored Reynolds-number range. The transition occurs for $We=O(1)$ and varies comparatively weakly with $Re$, except at the lowest Reynolds numbers. We emphasize, however, that $We$ is varied here primarily through changes in surface tension. Consequently, changes in $We$ simultaneously modify both the impact dynamics and the intrinsic capillary–gravity spectrum of the bath.
The approximate collapse of the transition in Weber number
motivates the following section.

Finally, we acknowledge that because the two-colour formulation prevents coalescence once the gas film becomes under-resolved, the decaying and weak-response classifications describe the simulated non-coalescing dynamics. In a physical system, some of these states may ultimately terminate by film rupture and coalescence.

\subsection{Unstable bath regime}
\label{Sub: Unstable bath}

\begin{figure}
\centering
\begin{tabular}{cccc}
\includegraphics[width=0.23\linewidth]{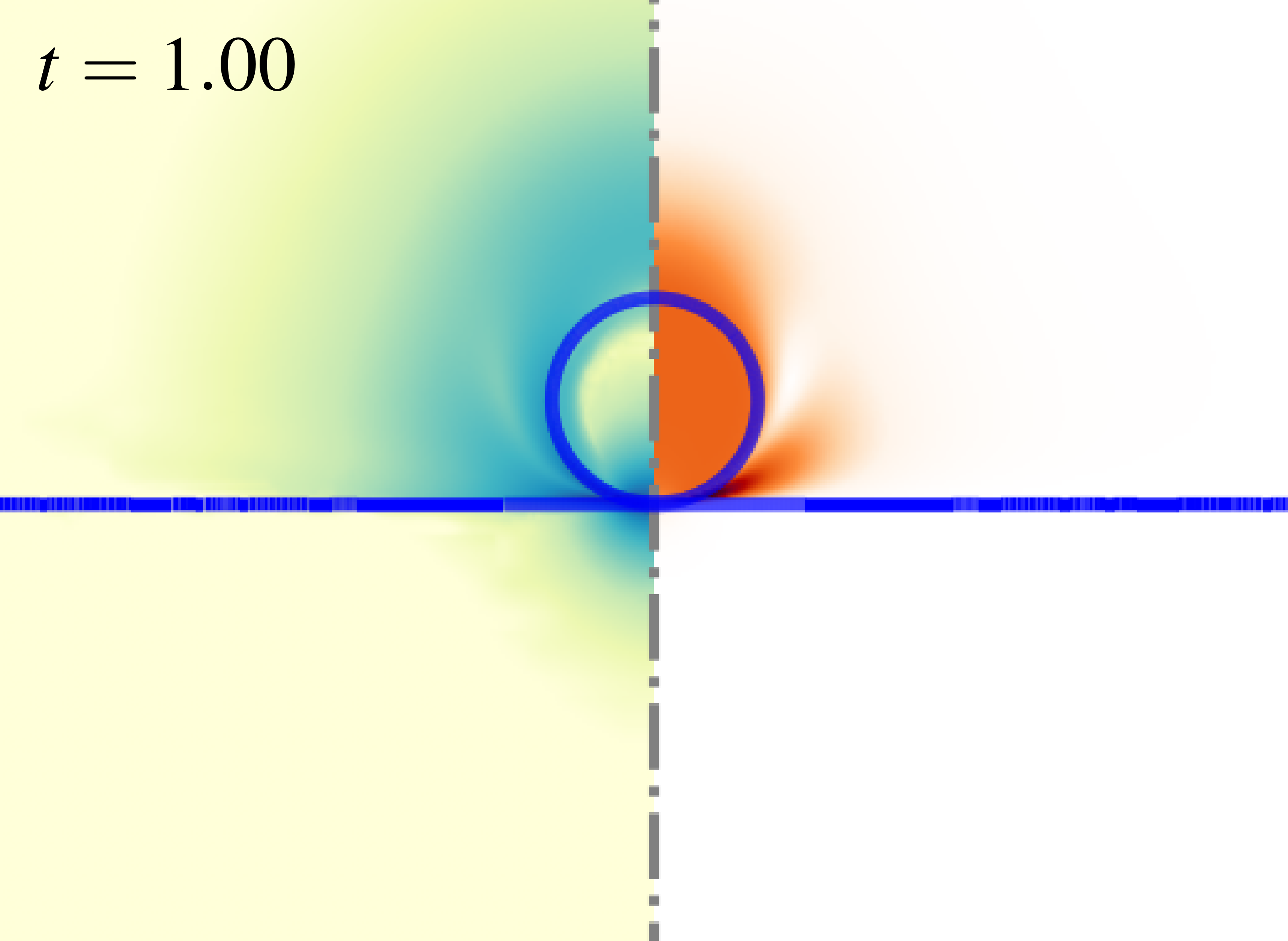}&
\includegraphics[width=0.23\linewidth]{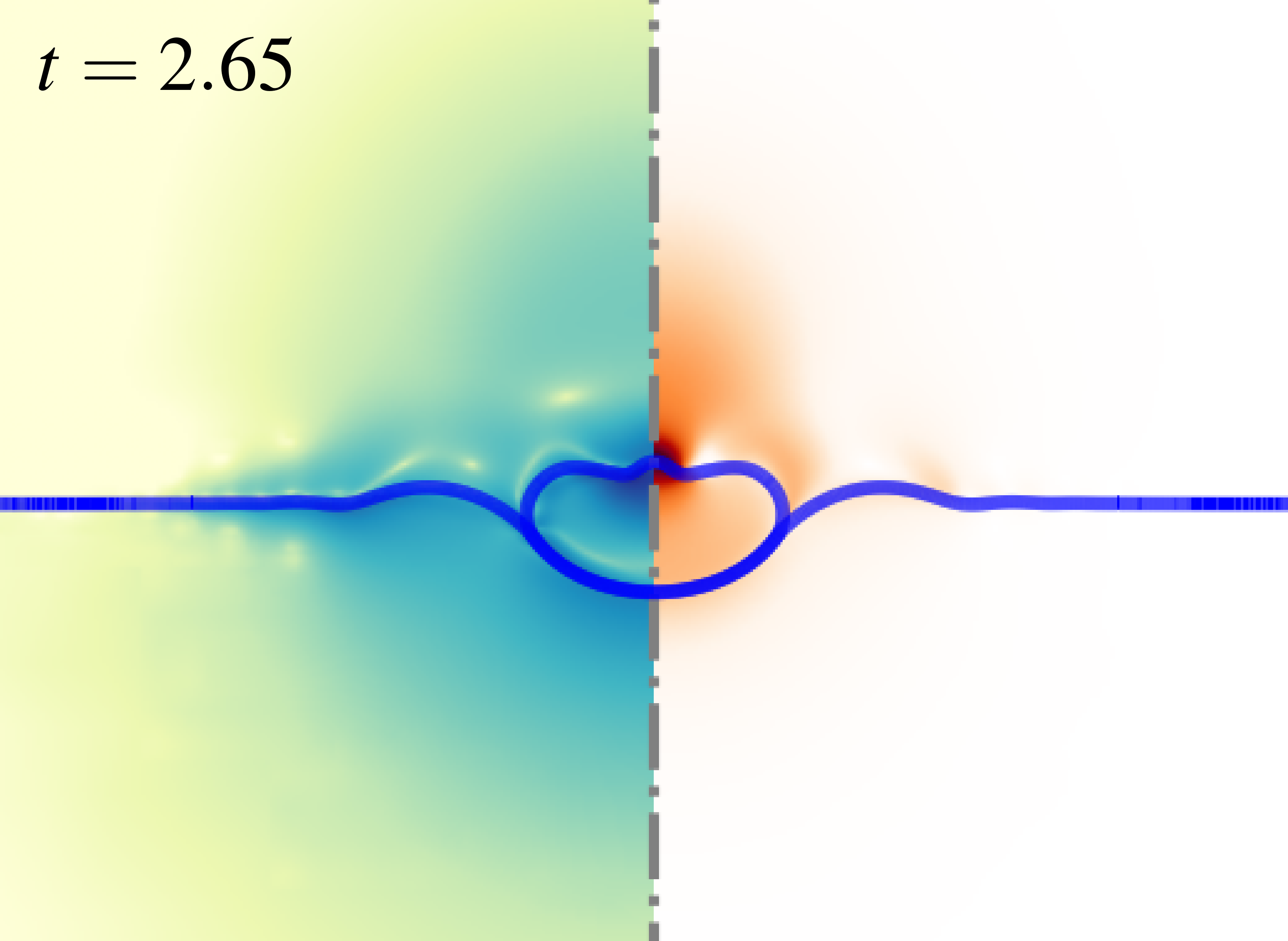}& 
\includegraphics[width=0.23\linewidth]{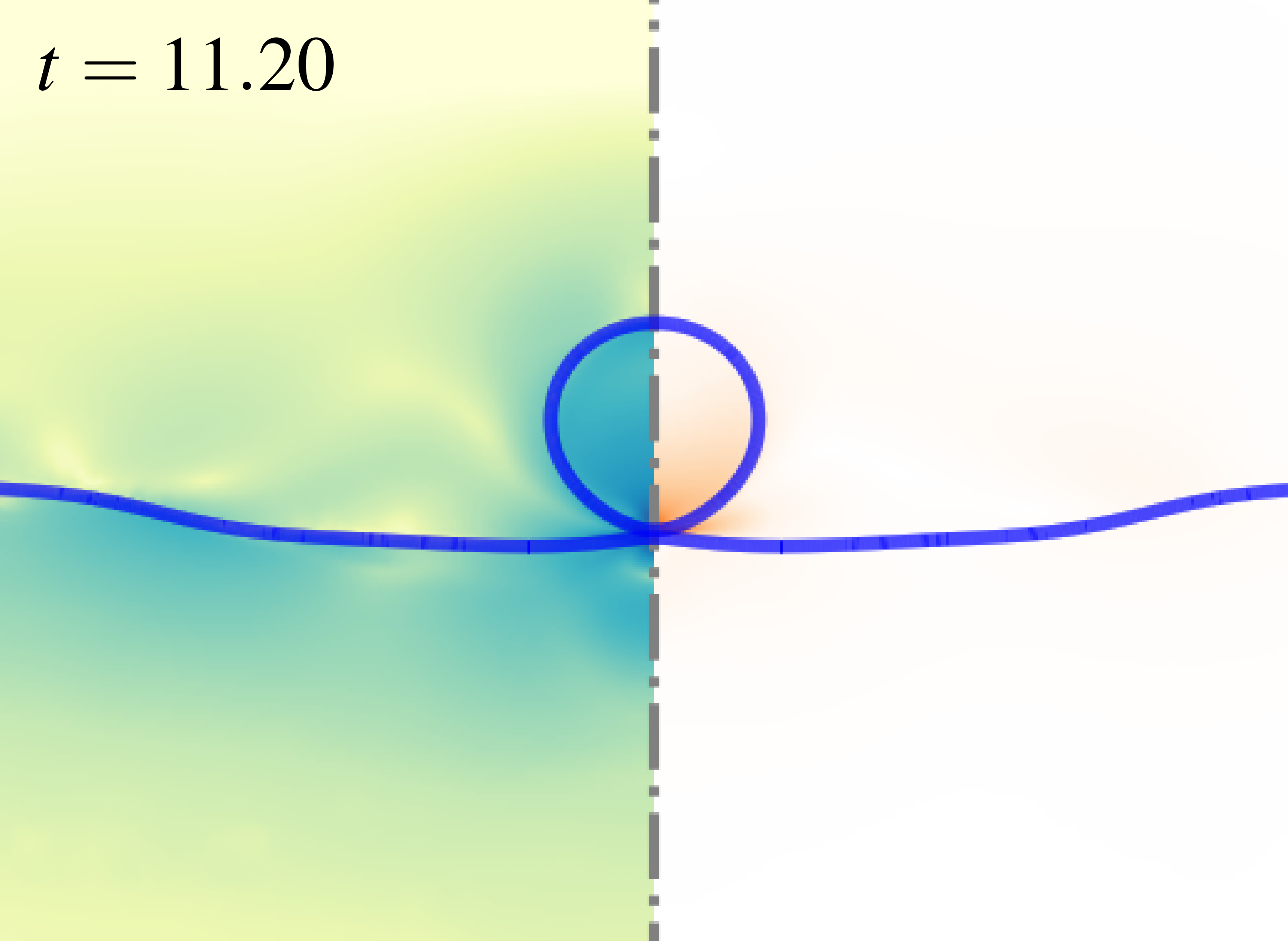}&
\includegraphics[width=0.23\linewidth]{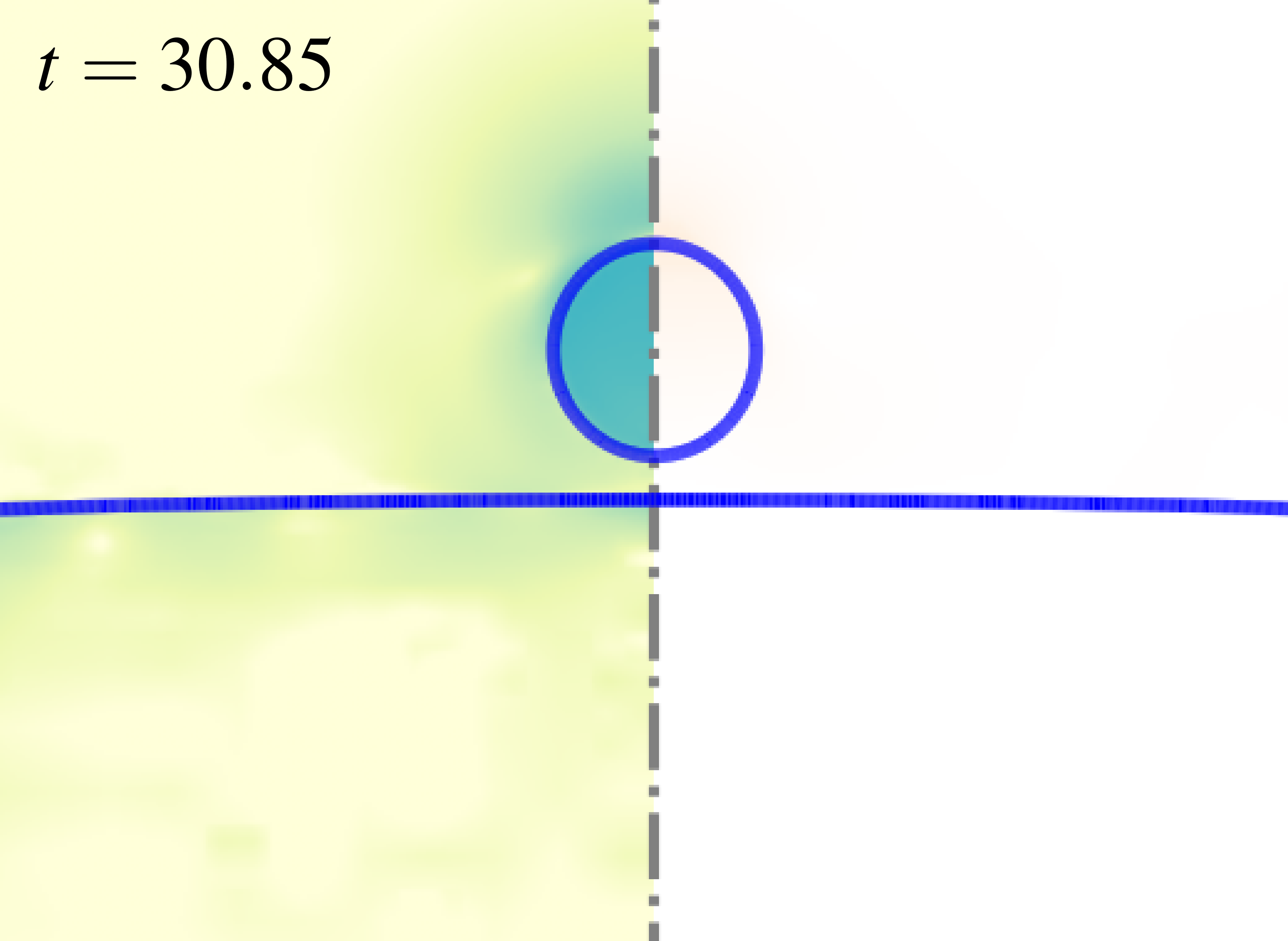}\\
\end{tabular}
\begin{tabular}{c}
(a)\\
\\
\end{tabular}
\begin{tabular}{cccc}
\includegraphics[width=0.23\linewidth]{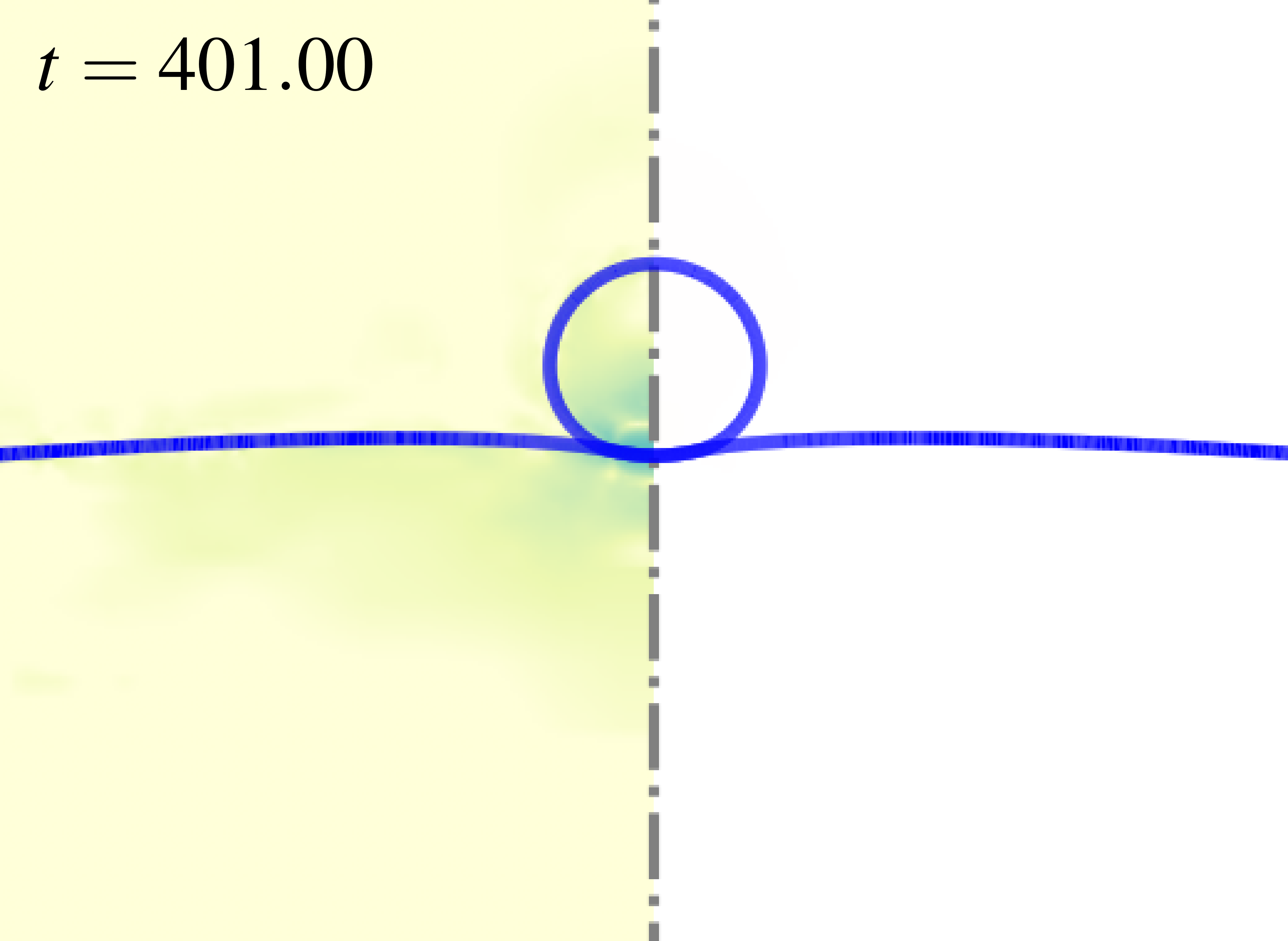}&
\includegraphics[width=0.23\linewidth]{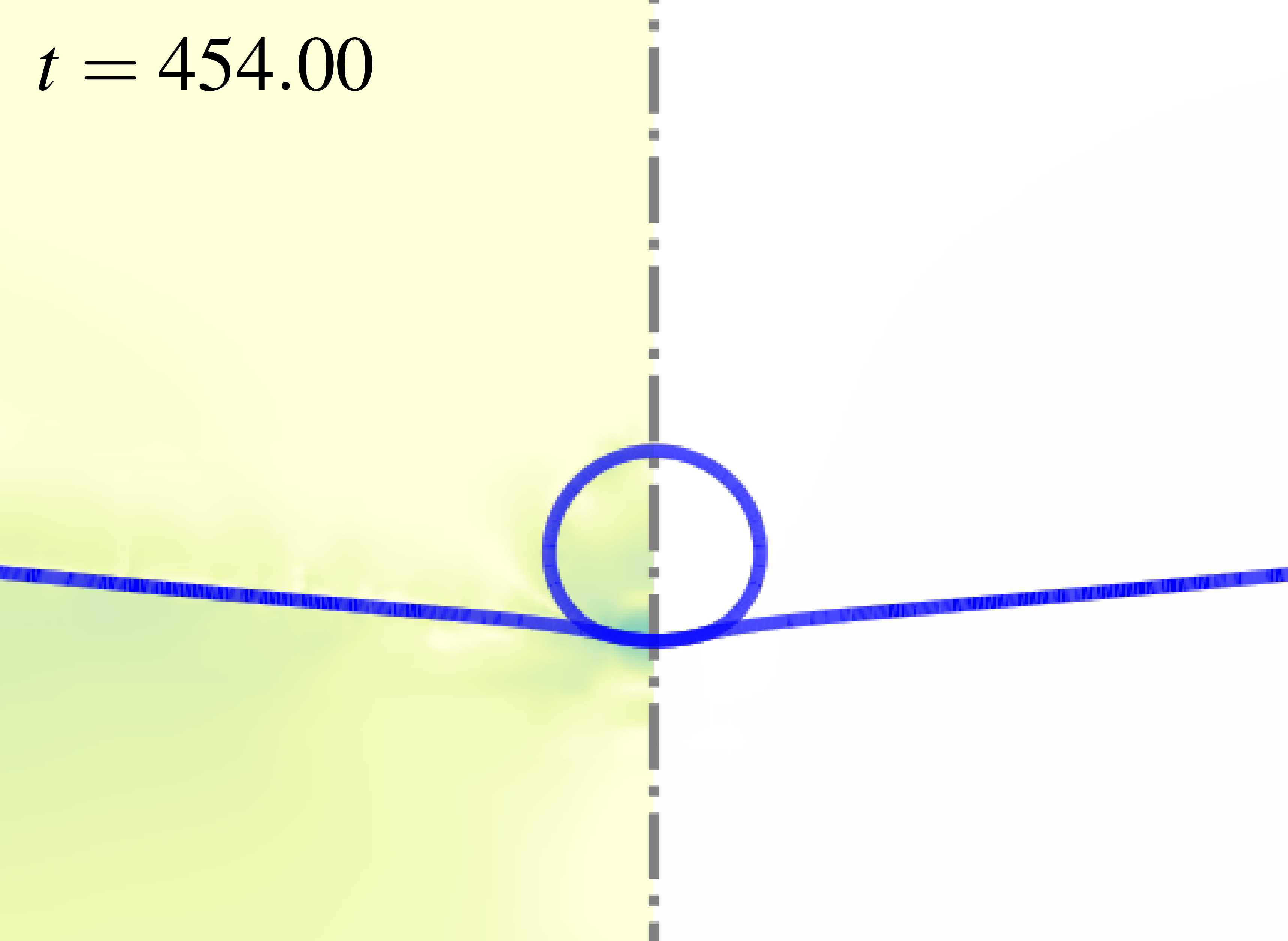}& 
\includegraphics[width=0.23\linewidth]{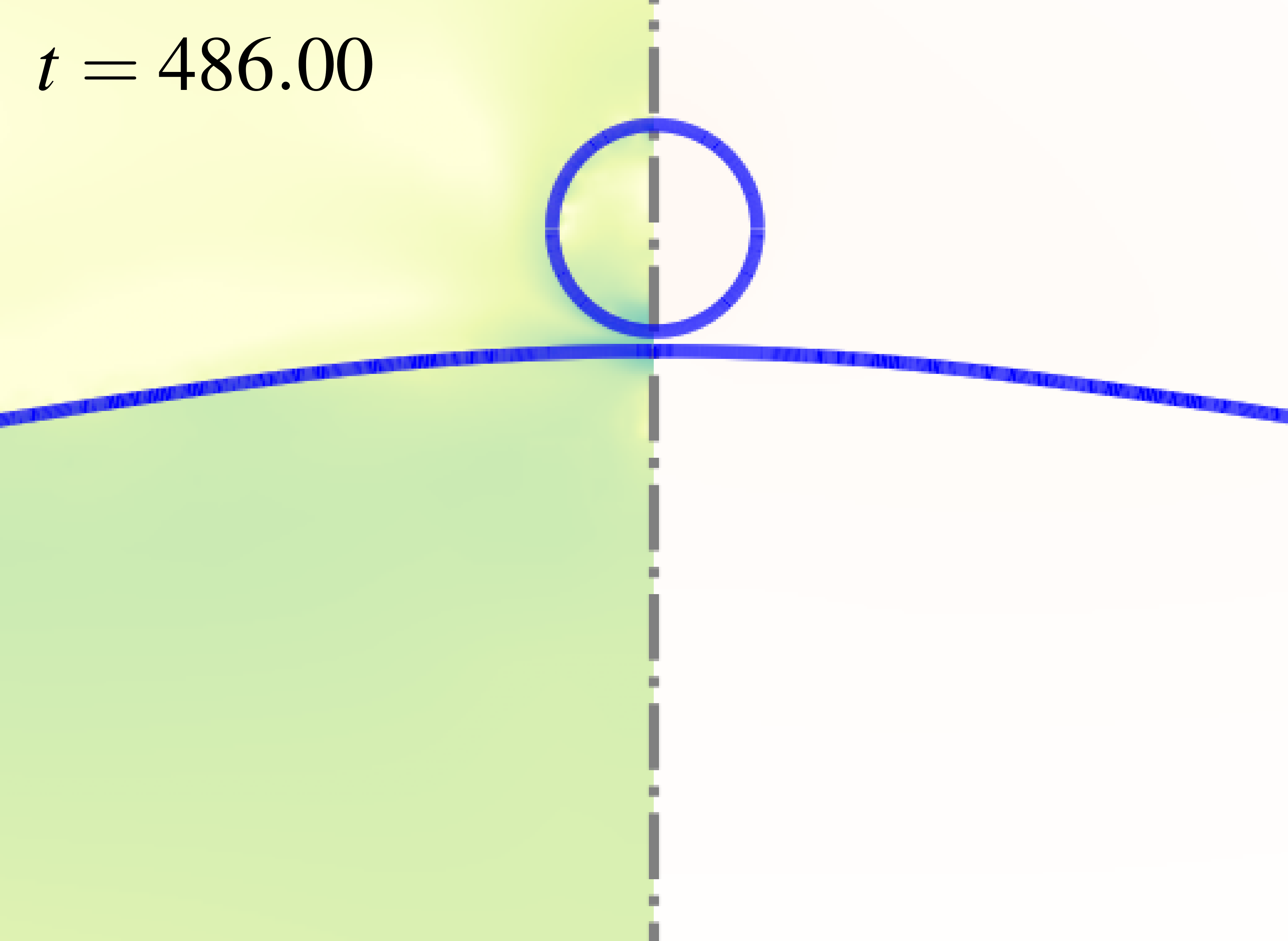}&
\includegraphics[width=0.23\linewidth]{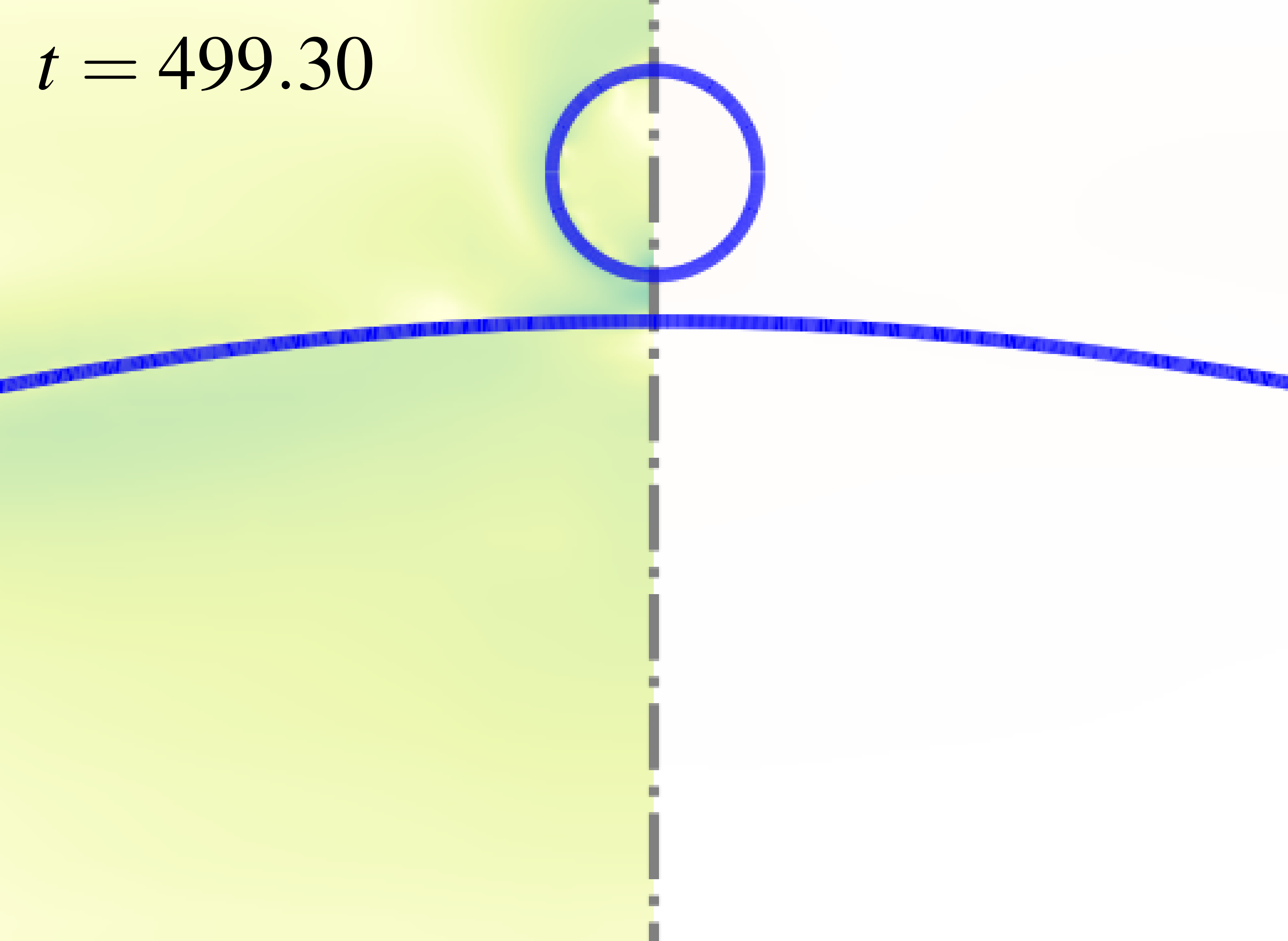}\\
\end{tabular}
\begin{tabular}{c}
(b) \\ 
\includegraphics[width=0.5\linewidth]{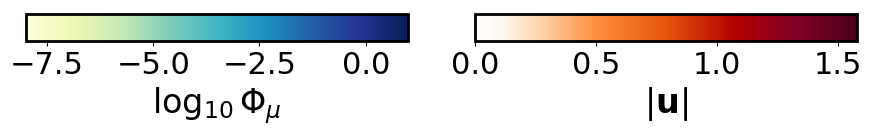}\\
\end{tabular}
\caption{Snapshots of the droplet--bath interface in the unstable bath regime with $We=5.00$ and $Re=200$. Four stages of the bounce cycle
are shown: impact, maximum spread, lift-off, and descent, corresponding to one-to-four columns, respectively. The first bounce (top row) and bounce number 11 at $t=458$ (bottom row) are shown. The left and right color bars represent the logarithm of the dimensionless
viscous-dissipation rate per unit volume, $
\Phi_\mu$, and 
the 
velocity magnitude, $|\mathbf u|$, respectively.
The blue curves denote the liquid--air interfaces and the dashed line marks the axis of symmetry.}
\label{fig: unstable_snapshots}
\end{figure}

\begin{figure}
\centering
\begin{tabular}{c}
\includegraphics[width=\textwidth]{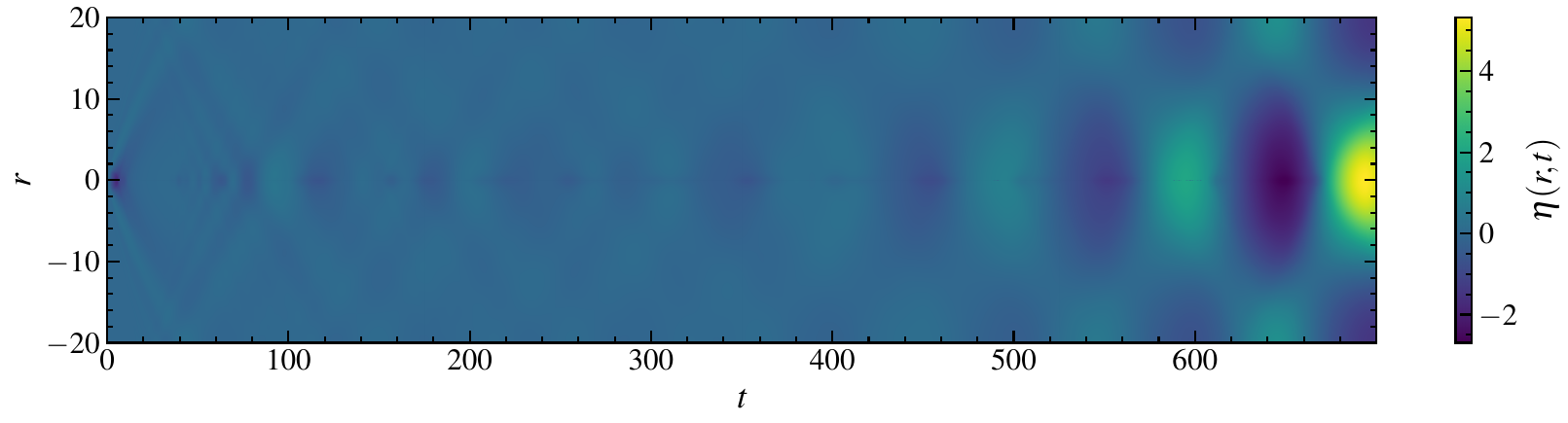} \\
(a) \vspace{6pt}\\
\end{tabular}
\begin{tabular}{cc}
\includegraphics[width=0.47\linewidth]{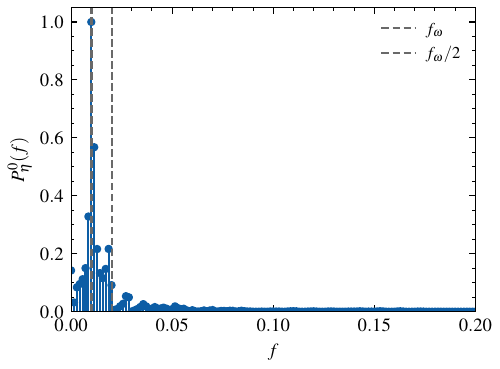} &  \includegraphics[width=0.47\linewidth]{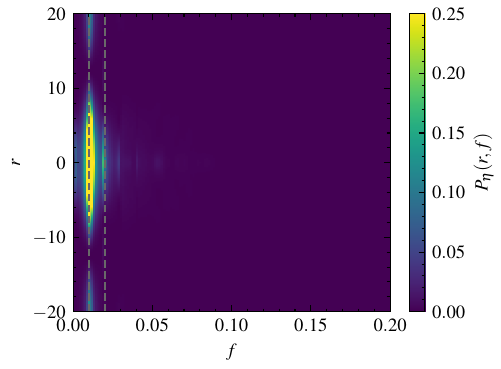}\\
(b) & (c) \vspace{6pt}\\
\end{tabular}
\begin{tabular}{c}
\includegraphics[width=\textwidth]{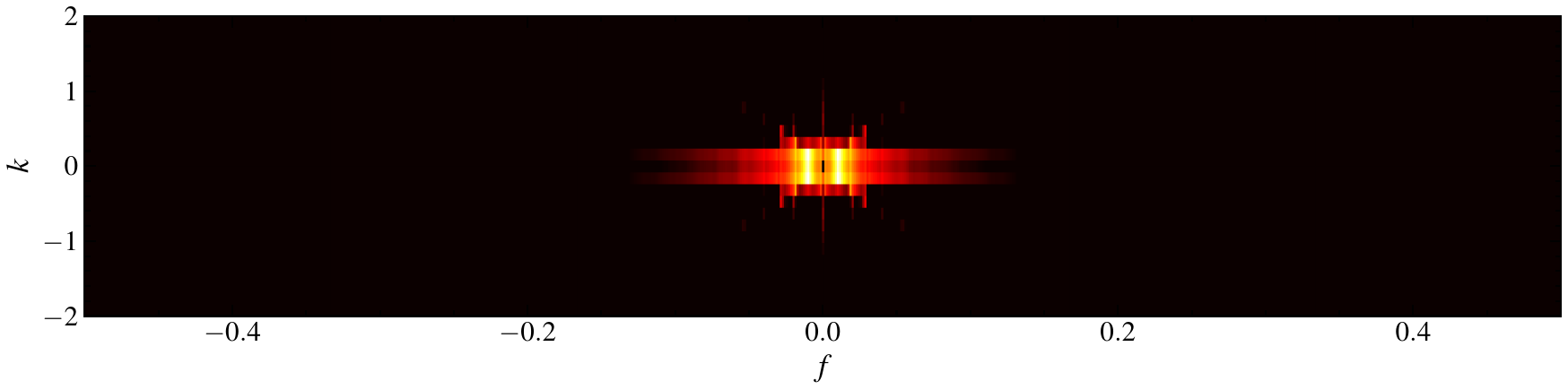}\\
(d) \\
\end{tabular}
\caption{Spectral analysis of the bath for the unstable bath at $We=5.00$ and $Re=200$.
(a) Space--time diagram of the interface elevation
$\eta(r,t)=z_{\mathrm{int}}(r,t)-z_0$.
(b) Temporal spectrum of the centerline elevation, $P_{\eta}^{0}(f)$, with the forcing frequency $f_\omega$ and its subharmonic $f_\omega/2$ represented by the vertical dashed lines.
(c) Radial distribution of the temporal spectrum, $P_{\eta}(r, f)$.
(d) Two-dimensional power spectrum, $P_{\eta}^{2D}(k, f)$, showing combined spatial and temporal frequency content of the bath response on a logarithmic scale.
}
\label{fig: fft_bath}
\end{figure}

\subsubsection{Evolution of the unstable-bath response and its Faraday-wave signature}

In this section we will consider the case of $(We,Re)=(5.00,200)$ as a representative unstable-bath case.
Figure \ref{fig: unstable_snapshots} shows snapshots of the droplet-bath interfaces for two different bounce instances.
To characterize both the fluid motion and its dissipation each snapshot displays the dimensionless velocity magnitude, $|\mathbf{u}|$, and the logarithm of the local viscous dissipation rate,
$\Phi_\mu = 2\mu (\mathcal{D}:\mathcal{D})/Re$, where $
\mathcal{D}=1/2 \,
\left(\nabla\mathbf{u}+\nabla\mathbf{u}^{T}\right)$.
%
During the first bounce, the bath response is localized near the impact point (see $t=1$),  where large $|\mathbf{u}|$ and $\Phi_\mu$ indicate intense fluid motion and viscous dissipation.
As the droplet spreads and rebounds ($t=2.65$), the impact generates an outward-propagating capillary wave, extending the velocity field along the interface while dissipation remains concentrated in regions of strong strain.
At later times ($t=11.20$ and $30.85$), both quantities decay substantially, leaving only a weak residual bath deformation.
By bounce 11, the dynamics differ markedly from the first impact. The bath undergoes a substantial time-dependent deformation over the bounce cycle, while the droplet impacts and rebounds with a weaker velocity field (see $t=486$). In contrast to the first bounce, no pronounced outward-propagating capillary wave is generated after impact; instead, the interface exhibits a persistent oscillatory motion that evolves between successive bounces. This change suggests that the droplet is interacting with an increasingly active bath wave field, whose possible connection to Faraday-type dynamics is examined below.

Figure \ref{fig: fft_bath}a shows the space--time
evolution of the bath elevation, $\eta(r,t)=z_{\mathrm{int}}(r,t)-z_0$, where $z_{\mathrm{int}}(r,t)$ is the instantaneous axial position of the bath-air interface 
and $z_0$ represents the undisturbed bath level. 
Unlike the transient capillary waves generated by individual impacts in the other regimes, here
a coherent oscillation develops over many impacts and grows after $t\gtrsim450$. Thus, rather than decaying between impacts, the bath disturbance accumulates over many forcing cycles.

To quantify this response, figure~\ref{fig: fft_bath}b shows the normalized temporal spectrum at the axis,
$$
P_{\eta}^{0}(f) =
\frac{|\mathcal{F}_{t}[\eta(0,t)]|^{2}}
{\max_{f} |\mathcal{F}_{t}[\eta(0,t)]|^{2}},
$$
where $\mathcal{F}_t[\cdot]$ denotes the temporal Fourier transform.
The imposed angular frequency is ${\omega}=0.127$, corresponding to the frequency $f_\omega={\omega}/(2\pi)\simeq 2.0\times10^{-2}$. 
 The dominant bath response 
occurs near 
$f_F \simeq f_\omega/2 \simeq 1.0\times10^{-2}$,
revealing a 
period-doubled response characteristic of Faraday-wave dynamics.

The same subharmonic component is
 spatially extended.
Figure~\ref{fig: fft_bath}c shows
$$
P_{\eta}(r, f) =
\frac{|\mathcal{F}_{t}[\eta(r, t)]|^{2}}
{\max_{r, f} |\mathcal{F}_{t}[\eta(r, t)]|^{2}},
$$
for which the spectral power is concentrated near $f_F$ over a finite radial interval.
The response is therefore not restricted to the vertical
motion beneath the droplet, but corresponds to an extended bath mode excited most strongly near the impact region.

Figure \ref{fig: fft_bath}d shows the  two-dimensional spectrum. 
Here, the mean-subtracted bath elevation is transformed in both space and time. {We define the dimensionless radial wavenumber as $k= 2 \pi f_r$, where $f_r$ denotes the spatial frequency conjugate to the radial coordinate $r$. Thus, the normalized two-dimensional power is given as}
$$
P_{\eta}^{2D}(k, f) =
\frac{|\mathcal{F}_{r,t}[\eta(r,t) - \langle \eta \rangle]|^{2}}
{\max_{k,f} |\mathcal{F}_{r,t}[\eta(r,t) - \langle \eta \rangle]|^{2}}.
$$
The spectral energy is concentrated around the same subharmonic frequency and spans a finite band of spatial wavenumbers, rather than exhibiting the broadband signature expected from isolated impacts.
Together, figures~\ref{fig: fft_bath}a-d show the emergence of a growing, spatially extended subharmonic bath response consistent with  Faraday wave dynamics.


\subsubsection{Role of droplet impacts in triggering the bath instability}
\label{sub: droplet impact bath instb}

\begin{figure}
\centering
\begin{tabular}{cc}
\includegraphics[width=0.47\linewidth]{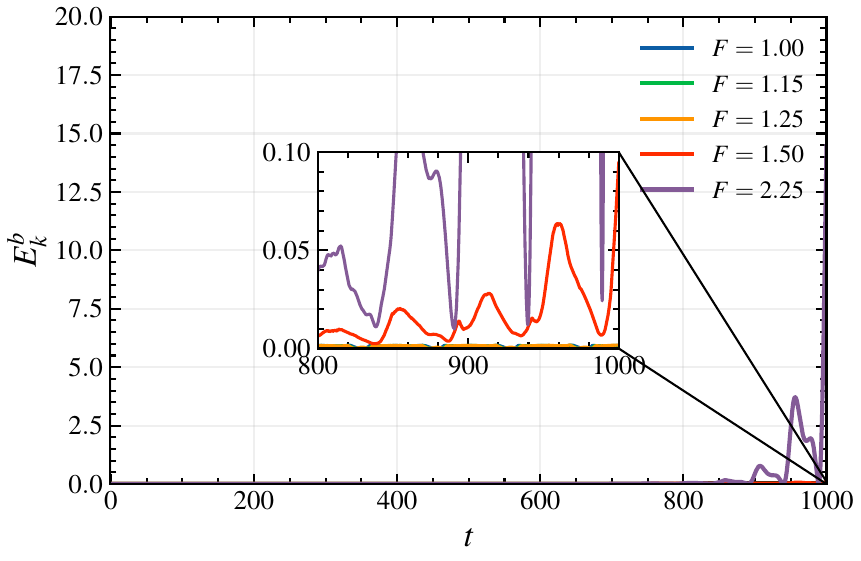} &  
\includegraphics[width=0.47\linewidth]{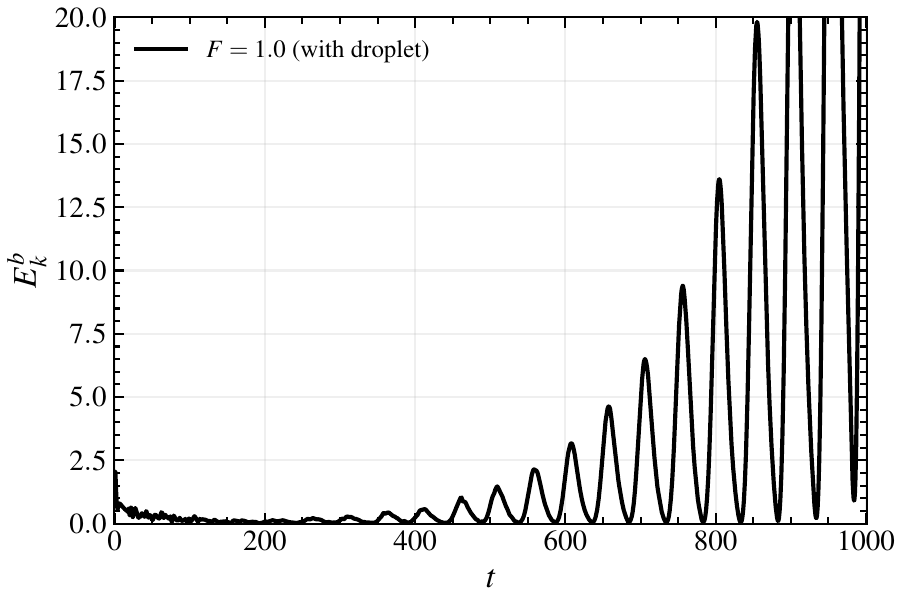}\\
(a) & (b) \\
\end{tabular}
\caption{Role of droplet impacts in the onset of the bath instability for $We=2.5$ and $Re=200$. (a) Bath kinetic energy in the absence of the droplet for different forcing
amplitude, 
the inset highlighting the onset of instability for $F \ge 1.50$. 
(b) Corresponding droplet-present simulations at  
$F=1.0$, showing earlier instability growth around $t \approx 400$.
}
\label{fig:KE_no_drop}
\end{figure}

To isolate the role of the droplet in the onset of sustained subharmonic growth in the bath, we repeat the $We=2.5$ and $Re=200$ simulation with the droplet removed, while retaining the same bath properties, geometry, and vertical forcing. In these droplet-free calculations, $We$ and $Re$ are retained as reference quantities based on the original droplet radius, impact velocity, and liquid properties.  
The case $We=2.5$ is selected because it corresponds to the transition identified in figure~\ref{fig:regime}. In the droplet-containing simulations, the coupled response remains bounded at $We=2.0$, whereas at $We=2.5$ the bath develops sustained growth.

Figure \ref{fig:KE_no_drop}a compares bath-only calculations at $(We,Re)=(2.5,200)$ over a range of forcing amplitudes. In the absence of the droplet, the bath remains bounded at $F=1.25$, whereas clear exponential growth is observed at $F=1.50$ after a long transient. 
Fits to the early-time exponential growth rates yield $F_c^{\mathrm{DNS}}\simeq1.3$, while the corresponding Floquet analysis gives $ F_c=1.17$.  The two estimates therefore differ by approximately 10\%, indicating reasonable quantitative agreement between the DNS and linear Floquet predictions (e.g., a similar interpolation was used by \citet{kahouadji2015numerical}).
The behavior changes markedly when the droplet is present.
Figure \ref{fig:KE_no_drop}b shows that, at the same $We$ and $Re$, the bath already becomes unstable at 
$F=1.0$, and the instability emerges much earlier in time (${t} \approx 400$). 
Since the bath without the droplet remains stable under identical forcing conditions, 
the instability cannot be attributed to the imposed oscillation
alone. Instead, droplet impacts provide a finite-amplitude disturbance that promotes the onset of the subharmonic bath response at a forcing level below the droplet-free threshold. This comparison therefore identifies the droplet
as a key trigger of the unstable-bath regime observed at the transition in figure~\ref{fig:regime}.

We next develop a reduced model that represents the droplet impact as an effective initial bath disturbance, and relates its amplitude to the governing impact parameters.

\subsubsection{Scaling of the impact-induced bath perturbation}
\label{Sec: scaling analysis}

We start from the same linear quasipotential bath model used by \citet{galeano2017non} and \citet{alventosa2023inertio}. The bath velocity potential satisfies
\begin{equation}
\nabla^2 \phi = 0,
\end{equation}
with no penetration at the bottom and linearized kinematic and dynamic conditions at the free surface. In dimensional form, these conditions are
\begin{equation}
\frac{\partial \eta}{\partial t}
=
\frac{\partial \phi}{\partial z}
+
2\nu \nabla^2 \eta,
\end{equation}
and
\begin{equation}
\frac{\partial \phi}{\partial t}
=
-g\eta
-2\nu \frac{\partial^2 \phi}{\partial z^2}
+\frac{\sigma}{\rho_b}{\mathcal{K}}
-\frac{p_s}{\rho_b},
\end{equation}
where $\eta$ is the bath displacement, $p_s$ is the droplet-induced pressure, and ${\mathcal{K}}$ is the linearized curvature of the bath interface. 
Following \cite{farsoiya2017axisymmetric}, we expand the interface axisymmetric Bessel modes as
\begin{equation}
\eta(r,t)=\sum_p a_p(t)J_0(\kappa_p r),
\end{equation}
where $a_p(t)$ is the time-dependent amplitude of the $p$-th axisymmetric radial mode, $J_0$ is the zeroth-order Bessel function of the first kind and $\kappa_p$ is the corresponding radial wavenumber, reduces the free-surface problem to a set of forced, weakly damped oscillators,
\begin{equation}
\ddot{a}_p
+
4\nu \kappa_p^2 \dot{a}_p
+
\omega_p^2 a_p
=
F_p(t),
\end{equation}
where $F_p(t)$ is the projection of the droplet-induced pressure onto the $p$-th bath mode.
The corresponding modal frequency is
\begin{equation}
    \omega_p^2 = \left(g + \frac{\sigma}{\rho_b}\kappa_p^2 \right) \kappa_p \mathrm{tanh}(\kappa_p H)
\end{equation}
where $\rho_b$ denotes the bath density. Each bath mode behaves as a forced capillary-gravity oscillator, with gravity and surface tension providing the restoring mechanisms.

To estimate the characteristic deformation generated locally beneath the droplet during impact, we reduce the modal description to a single representative length scale associated with the impact footprint.
In this limit the projected pressure forcing is removed, $F_p(t)=0$, and viscous damping is neglected at leading order. Retaining a single representative mode associated with the impact length scale then gives
\begin{equation}
    \ddot{\zeta}+\omega_L^2\zeta=0,
\label{Eq harmonic}
\end{equation}
where $\zeta(t)$ denotes the characteristic local bath displacement and $\omega_L$ is the capillary-gravity frequency. Under a deep bath approximation, $\mathrm{tanh}(k_LH) \approx 1$, so the dispersion relation reduces to
\begin{equation}
    \omega_L^2 = gk_L + \frac{\sigma}{\rho_b}k^3,
\end{equation}
where $k_L$ is the characteristic wavenumber corresponding to the localized deformation.

When the droplet impacts the bath, rather than exciting the bath surface uniformly its impact pressure remains restricted to a localized contact region beneath it. The characteristic length scale of this disturbed region is written as  
\begin{equation}
    L=BR,
\end{equation}
where $B$ is the dimensionless radial extent of the impact-pressure footprint measured from the DNS.
In the current work, we obtain $B$ directly from the spatio-temporal pressure maps (e.g., see figure \ref{fig: Pressure map_unstable}),
where we identify the radial extent over which droplet-bath interaction occurs during impact. Since the dominant disturbance is localized over the length scale $L$, the corresponding characteristic wavenumber scales as $k_L \sim 1/L$. Substituting this into the capillary-gravity dispersion relation gives
\begin{equation}
    \omega_L^2 \sim \frac{g}{L} + \frac{\sigma}{\rho_b L^3},
\label{eq: dispersion}
\end{equation}

\begin{table}
\centering
\setlength{\tabcolsep}{15pt}
\begin{tabular}{cccc}
$We$  & $B$ & $\chi$ & $\varepsilon_{\mathrm{eff}}$\\
\hline
5.0   & 0.52 & 0.51 & 0.43\\
2.5   & 0.48 & 0.61 & 0.32\\
2.0   & 0.57 & 0.47 & 0.28\\
\end{tabular}
\caption{ Impact-equivalent perturbation amplitudes predicted by
Eq.~\ref{Eq init. perturb} for representative Weber numbers.}
\label{tab:predicted epsilon_eff}
\end{table}

To assess the relative importance of gravity and surface tension at this scale, we define a Bond number based on the impact footprint,
\begin{equation}
    Bo_L=\frac{\rho_bgL^2}{\sigma}=B^2 Bo_R,
\end{equation}
where $Bo_R=\rho_bgR^2/\sigma$ is the Bond number based on the droplet radius. For the present system, $Bo_R \approx 0.00137 - 0.137$ and the pressure footprint gives $B \approx 0.4 - 0.5$ (see figures \ref{fig: Pressure map_unstable} and \ref{fig: Pressure map_decaying}), yielding
  $  Bo_L \simeq (0.219 - 34) \times 10^{-3}$.
Since $Bo_L\ll1$ over the cases considered, capillary restoring effects dominate gravity at the impact-footprint scale. The corresponding characteristic frequency  in \ref{eq: dispersion} therefore reduces at leading order to
\begin{equation}
    \omega_L \sim \left(\frac{\sigma}{\rho_b L^3}\right)^{1/2}.
\label{Eq cap-dominated}
\end{equation}

Next, we model the droplet impact as an impulsive transfer of velocity to the localized bath. Immediately following impact, the interface displacement is assumed to remain small while having a finite initial velocity,
\begin{equation}
    \zeta(0) = 0, \qquad \dot\zeta(0)=u_i,
\end{equation}
where $u_i$ denotes the initial velocity transferred to the bath. Solving the harmonic oscillator \ref{Eq harmonic} with these initial conditions results in
\begin{equation}
    \zeta(t) = \frac{u_i}{\omega_L} \mathrm{sin}(\omega_L t),
\end{equation}
from which the maximum displacement generated by the impulse scales as
\begin{equation}
    A_{\mathrm{eff}} \sim \frac{u_i}{\omega_L}.
\label{Eq impluse}
\end{equation}
To relate the transferred bath velocity to the droplet impact conditions, we introduce the velocity transfer coefficient $\chi$, 
\begin{equation}
    u_i=\chi U_0,
\end{equation}
where $U_0$ is the incoming droplet velocity. The coefficient $\chi$ represents the fraction of the impact velocity effectively transferred to the localized bath response, determined from the energy-transfer analysis presented in Appendix \ref{App B}. Substituting the velocity transfer relation into the prior expression \ref{Eq impluse} yields
\begin{equation}
    A_{\mathrm{eff}} \sim \frac{\chi U_0}{(g/L + \sigma/(\rho_b L^3))^{1/2}}.
\end{equation}
Implementing the capillary-dominated approximation derived in Eq. \ref{Eq cap-dominated}
and normalizing by the droplet radius gives
\begin{equation}
    \frac{A_{\mathrm{eff}}}{R} \sim \chi B^{3/2} \left(\frac{\rho_b U_0^2R}{\sigma} \right)^{1/2}.
\label{eq: A-eff}
\end{equation}
Using the Weber number definition used in section \ref{sec:Gov. Eqns} and noting that the droplet and bath are composed of the same fluid ($\rho_b = \rho_l)$, equation \ref{eq: A-eff} gives the initial perturbation amplitude $\varepsilon_{\mathrm{eff}}$
\begin{equation}
    \varepsilon_{\mathrm{eff}} \equiv \frac{A_{\mathrm{eff}}}{R} \sim {\chi B^{3/2}} \sqrt{We}.
\label{Eq init. perturb}
\end{equation}

Equation~\ref{Eq init. perturb} estimates the amplitude of the finite bath disturbance generated by impact; it does not itself constitute an instability criterion. 
For fixed $B$ and $\chi$, the estimate scales as
$\varepsilon_{\mathrm{eff}}\propto We^{1/2}$ and contains no explicit 
dependence on $Re$.
In the present simulations, both $B$ and $\chi$ vary between cases, and the resulting $\varepsilon_{\rm eff}$ increases monotonically with $We$. Whether a given disturbance subsequently grows or decays depends not only on its initial amplitude but also on the susceptibility of the periodically forced bath.
However, this scaling is consistent with the regime map in figure~\ref{fig:regime}, where the transition from the decaying to the unstable-bath regime occurs at approximately the same Weber number over the range of Reynolds numbers considered. 

We acknowledge that  equation  \ref{Eq init. perturb} is an estimate of the characteristic maximum interface displacement associated with the droplet impact, rather than  a complete specification of the post-impact state. In particular, the actual impact generates both interface displacement and velocity, with a phase determined by the forcing cycle. Here we use the predicted maximum displacement to construct a simplified bath-only initial condition and test whether a disturbance of the corresponding amplitude can access the growing subharmonic response.

\begin{figure}
\centering
\includegraphics[width=0.7\linewidth]{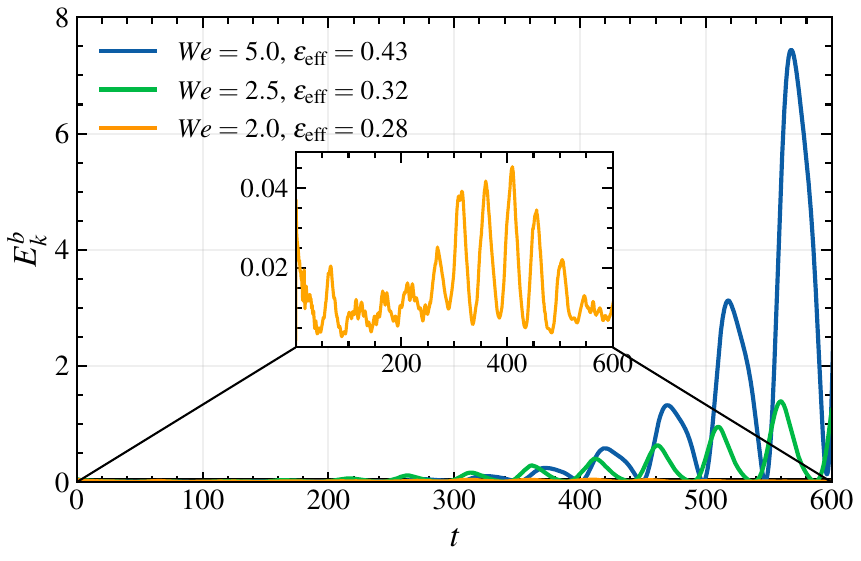}
\caption{Bath kinetic energy in droplet-free simulations initialized with the impact-equivalent perturbations predicted by equation \ref{Eq init. perturb}. The inset provides a magnified view of the $We=2.0$ case, highlighting its small and bounded response after the initial transient.
}
\label{fig:perturb}
\end{figure}

To test this scaling independently of the droplet dynamics, we initialize bath-only simulations with the predicted interface deformation
\begin{equation}
    \zeta(r,0)=-\varepsilon_{\mathrm{eff}} R \mathrm{sech}^2 \left(\frac{r}{BR}\right),
\end{equation}
where the negative sign represents the initial downward bath depression generated by the impacting droplet.  The corresponding values of $B$, $\chi$ and $\varepsilon_{\mathrm{eff}}$ are summarized in table~\ref{tab:predicted epsilon_eff}.

Figure \ref{fig:perturb} compares bath-only simulations initialized with the impact-equivalent deformations predicted by equation \ref{Eq init. perturb}. All three calculations are performed at $F=1<F_c$, for which the initially undisturbed bath is linearly stable.
For $Re=200$, the  bath remains bounded at $We=2.0$ ($\varepsilon_{\rm eff}=0.28$), whereas it exhibits a growing subharmonic response at   $We=2.5$ ($\epsilon_{\rm eff}=0.32$) and $We=5.0$ ($\varepsilon_{\rm eff}=0.43$). In other words, increasing $We$ yields larger impact perturbations, with the transition to growth occurring between $We=2.0$ and $2.5$,  consistent with figure~\ref{fig:regime}. 
Since all cases use $F=1<F_c$,  
the growth of both disturbances represents a finite-amplitude instability. In summary, the droplet impact provides a finite-amplitude seed for Faraday waves, but
whether it triggers sustained growth depends on the bath’s parametric susceptibility. The reduced model thus captures the impact’s triggering role and reproduces the observed switch between decaying and unstable bath regimes.

\subsubsection{Modal structure of the droplet-induced subharmonic response \label{Floquet_analysis_section}}

The preceding analysis shows that droplet impact provides a finite-amplitude disturbance capable of triggering the subharmonic bath response. We now
determine how this disturbance projects onto the discrete axisymmetric modes of the cylindrical bath.

In this section, we extend the Floquet analysis of \citet{Kumar_Tuckerman_1994} to the finite cylindrical model used in this work. Following the quasi-potential formulation,
the interface displacement was expanded as
$$
\eta(r,t)=\sum_p a_p(t) J_0(\kappa r),
$$
while the present analysis considers the parametric stability of each of these modes individually. We express a generic axisymmetric mode as
\begin{equation}
    \eta(r,t)=\hat{\eta}(t) J_0(\kappa r),
\end{equation}
where $\hat{\eta}(t)$ is the time-dependent modal amplitude and  $\kappa$ is the radial wavenumber. Regularity at the symmetry axis requires a finite displacement and zero radial slope at $r=0$, both of which are automatically satisfied by $J_0$. We apply a static $90^\mathrm{o}$ contact-angle condition at the wall, giving a zero interfacial slope,

\begin{equation}
    \frac{\partial \eta}{\partial r} \bigg|_{r=R_c}=0.
\end{equation}
where $R_c$ is the radius of the cylindrical bath. Since $dJ_0(\kappa r)/dr=-\kappa J_1(\kappa r)$, the admissible radial wavenumbers are discrete and thus fulfill
\begin{equation}
    J_1(\kappa_p R_c) = 0, \qquad \kappa_p=\frac{\beta_p}{R_c}, \qquad p = 1,2,....,
\end{equation}
where $\beta_p$ is the $p$-th positive zero of $J_1$. 
For each discrete radial wavenumber $\kappa_p$, we solve the finite-domain Floquet problem obtained by replacing the continuous horizontal Fourier wavenumber of \citet{Kumar_Tuckerman_1994} with the admissible axisymmetric Bessel wavenumbers of the cylindrical bath. 
The complete derivation is presented in Appendix \ref{App E}.

Before solving the full Floquet problem, we first identify which discrete radial modes are expected to be most susceptible to parametric excitation. For a forcing frequency $\omega$, the primary Faraday response is subharmonic, so resonance is expected for modes whose natural frequency lies close to $\omega/2$. For the dimensional comparison, the numerical forcing $\omega=0.127$ corresponds to $\omega_d= \omega U_0/R \simeq 181.9~\mathrm{rad~s^{-1}}$.
Since the finite cylindrical bath admits only the discrete wavenumbers $\kappa_p=\beta_p/R_c$, we evaluate the inviscid
natural frequency associated with each admissible radial mode using the two-fluid capillary--gravity dispersion relation \citep{Kumar_Tuckerman_1994}
\begin{equation}
    \omega_p^2=\frac{\left(\rho_l-\rho_g \right)g \kappa_p + \sigma \kappa_p^3}{\rho_l+\rho_g} \mathrm{tanh}(\kappa_p H).
\end{equation}
where $\rho_l$ and $\rho_g$ denote liquid and gas densities, respectively, and $H$ is the layer height.

\begin{figure}
\centering
\begin{tabular}{cc}
\includegraphics[width=0.47\linewidth]{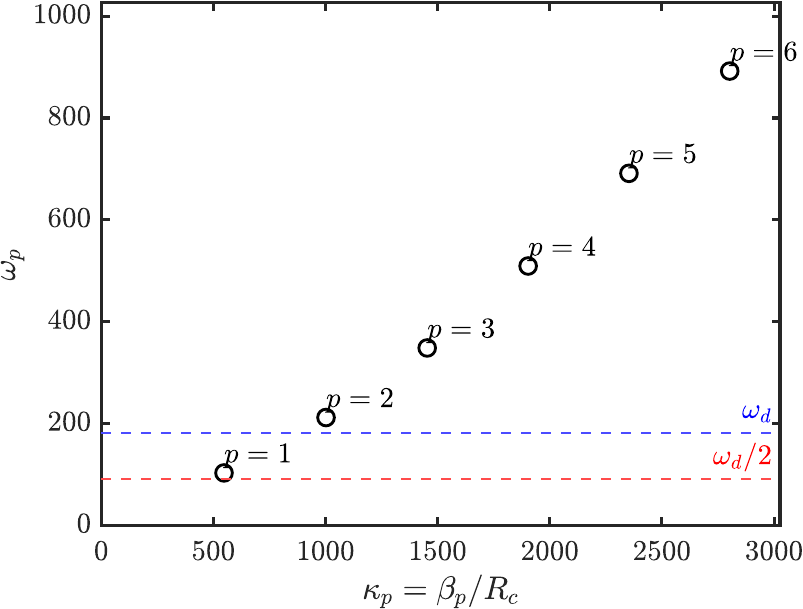}   &  
\includegraphics[width=0.45\linewidth]{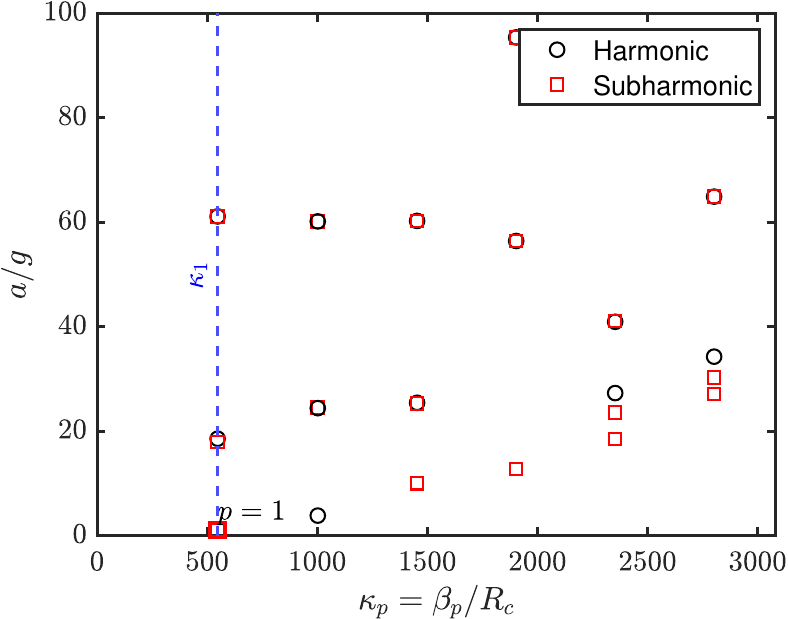}\\
(a)   &  (b)\\
\end{tabular}
\begin{tabular}{c}
\includegraphics[width=0.55\linewidth]{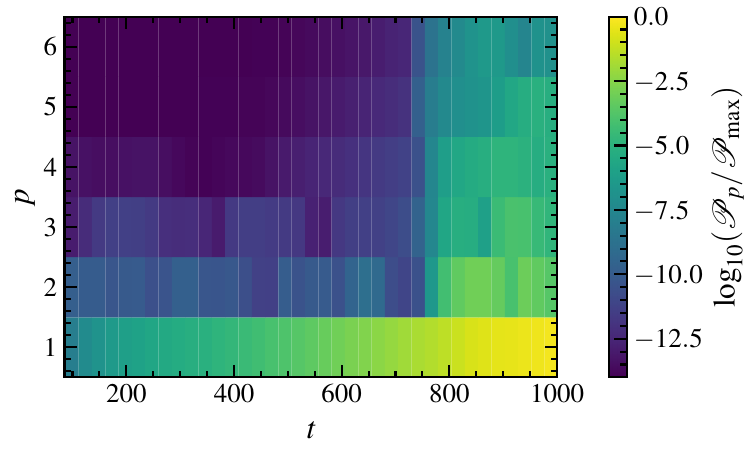} \\  
(c)\\
\end{tabular}
\caption{
Axisymmetric linear stability analysis and bath-only modal response at $We=2.5$ and $Re=200$.
(a) Inviscid dispersion relation for the first six discrete axisymmetric radial modes in a cylindrical bath, showing the dimensional natural frequencies $\omega_p$ as a function of radial wavenumbers $\kappa_p=\beta_p/R_c$, where $\beta_p$ is the $p$-th positive zero of $J_1$. The horizontal dashed lines show the dimensional forcing angular frequency $\omega_d$ and its subharmonic $\omega_d/2$. (b) Marginal stability thresholds for the corresponding discrete radial modes obtained from the axisymmetric extension of the Floquet stability analysis of \citet{Kumar_Tuckerman_1994}. Circles and squares denote the harmonic and subharmonic instability thresholds, respectively, while the dashed vertical line marks the location of the first admissible Neumann wavenumber $\kappa_1$. The highlighted square identifies the lowest subharmonic threshold, corresponding to $p=1$. (c) Temporal evolution of the normalized modal power for the first six radial modes from the bath-only simulation at $F=1.5$.
} 
\label{fig:KT94_Bessel_freq}
\end{figure}

Figure~\ref{fig:KT94_Bessel_freq}a shows the resulting discrete spectrum. For the present geometry, the first two admissible modes
$\kappa_1 =\beta_1/R_c\simeq 547~ \mathrm{m^{-1}}$ and
$\kappa_2 \simeq1002~ \mathrm{m^{-1}}$,
corresponding to 
$\omega_1 \simeq 103~\mathrm{rad~s^{-1}}$ and 
$\omega_2 \simeq 212~\mathrm{rad~s^{-1}}$, respectively.
The dimensional subharmonic frequency $\omega_d/2\simeq 91~\mathrm{rad~s^{-1}}$, therefore lies close to the first admissible radial mode $p=1$. Meanwhile, the harmonic forcing frequency $\omega_d \simeq 181.9~\mathrm{rad~s^{-1}}$ lies close to $p=2$. 
These inviscid frequencies identify where the harmonic and subharmonic resonances are expected to occur but do not determine their instability thresholds. Viscous damping and parametric forcing are therefore retained in the full Floquet calculation described below.

Figure \ref{fig:KT94_Bessel_freq}b  shows the corresponding neutral stability thresholds obtained from the full viscous Floquet analysis. Because the cylindrical geometry admits only the discrete wavenumbers $\kappa_p=\beta_p/R_c$, the instability spectrum consists of isolated modal branches.
Among the axisymmetric modes considered here, the first Neumann mode, $p=1$, has the lowest linear forcing threshold, $F_c=1.17$, on the subharmonic branch, while its harmonic threshold remains substantially larger, consistent with the near-resonant condition independently identified in figure~\ref{fig:KT94_Bessel_freq}a.

This modal selection is tested independently using the bath-only simulation at $F=1.5$, for which the imposed forcing lies above the predicted $p=1$ subharmonic threshold. The first 13 non-uniform radial modes are used to project the bath interface onto the axisymmetric Neumann basis at each instant. The modal power $\mathcal{P}_p$ is normalized by its global maximum $\mathcal{P}_\mathrm{max}$ over all retained modes and times, and the resulting modal coefficients are conditioned at the subharmonic frequency $f_\omega /2$. Figure \ref{fig:KT94_Bessel_freq}c only shows the first six radial modes for clarity. The growing response is dominated by $p=1$, while the higher modes remain significantly weaker during the onset, which independently confirms the mode chosen by the Floquet analysis.

\begin{figure}
\centering
\begin{tabular}{ccc}
\includegraphics[width=0.33\linewidth]{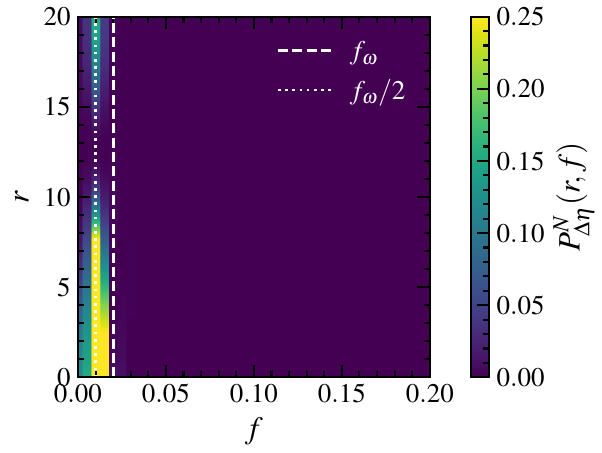}  &
\includegraphics[width=0.31\linewidth]{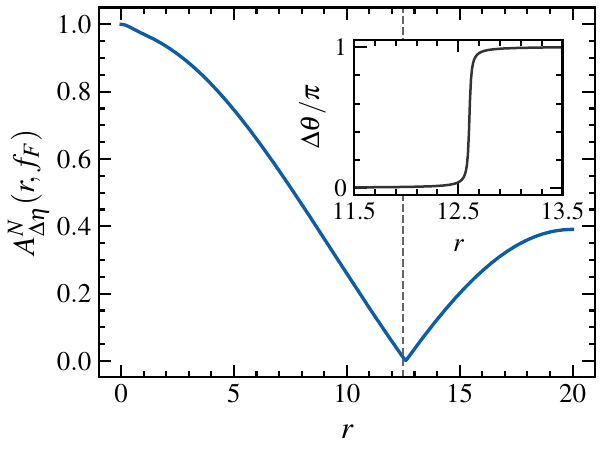} &
\includegraphics[width=0.31\linewidth]{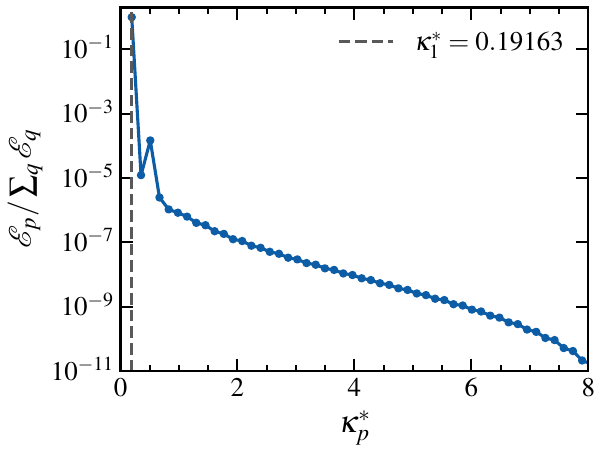}\\
(a) & (b) & (c)\\ 
\end{tabular}
\caption{Droplet-induced modal response. 
(a) Normalized temporal power spectrum, $P_{\Delta\eta}^{N}(r,f)$ over the entire radial domain, with the  dashed and dotted vertical lines denoting the harmonic frequency $f_\omega$ and its subharmonic $f_\omega/2$, respectively. (b)
Normalized radial amplitude, $A_{\Delta\eta}^{N}(r,f_F)$,
conditioned at the subharmonic frequency  $f_F=f_\omega/2$. The inset shows the corresponding relative phase, $\Delta \theta/ \pi$, revealing an approximately $\pi$-phase reversal across the radial node at $r \approx 12.5$. (c) Axisymmetric Bessel decomposition of the conditioned spatial mode, showing the modal-energy fraction $\mathcal{E}_p/\sum_q\mathcal{E}_q$ for the dimensionless Neumann wavenumber $\kappa_p^*$. The dominant wavenumber reveals the first radial mode, $\kappa_1^*\simeq0.19163$.
} 
\label{fig:KT94_Bessel_LSA}
\end{figure}

To isolate the interface response introduced by the droplet during the first impact, we define
\begin{equation}
    \Delta \eta(r,t)=\eta_{\mathrm{with~drop}}(r,t)-\eta_{\mathrm{without~drop}}(r,t),
\end{equation}
where the two systems are subjected to the same vertical forcing.
This difference removes the portion of the bath response common to the two simulations and provides a measure of the additional response associated with the presence of the droplet. Because the two systems subsequently evolve nonlinearly, $\Delta\eta$ cannot separate independent impact disturbance at long times.
The temporal spectrum, assessed over the complete radial domain in figure \ref{fig:KT94_Bessel_LSA}a, is already strongly concentrated at the subharmonic frequency $f_\omega/2$.
The additional bath response associated with the droplet is strongly concentrated on the subharmonic temporal branch.
Conditioning $\Delta \eta(r,t)$ at  $f_\omega/2$ gives the spatial amplitude depicted in figure \ref{fig:KT94_Bessel_LSA}b. The amplitude shows a decreasing trend from the axis up until $r \approx 12.5$, after which it increases again towards the wall. The inset reveals an accompanying phase change of $\pi$ across the same location, confirming that the conditioned spatial field changes sign at this radial node. This spatial structure is consistent with the first axisymmetric Neumann eigenfunction,
\begin{equation}
    \Delta \eta(r,t) \approx A_{\Delta} (t) J_0 (0.19163r)
\end{equation}
where $A_{\Delta} (t)$ is the time-dependent amplitude of the droplet-induced mode and $\kappa_1^* \approx0.19163$ is its dimensionless dominant radial wavenumber (see figure \ref{fig:KT94_Bessel_LSA}c). In dimensional form, it represents $\kappa_1=\kappa_1^*/R\approx 547~ \mathrm{m^{-1}}$, corresponding exactly to the first admissible point in the axisymmetric stability diagram (see figure \ref{fig:KT94_Bessel_freq}b), marked by $p=1$. The observed radial node follows directly from the first zero $j_{0,1}=2.4048$ of $J_0$, with $\kappa_1^* r_{\mathrm{node}}=j_{0,1}$, and hence $r_{\mathrm{node}} \approx 12.5$ is in agreement with figure \ref{fig:KT94_Bessel_LSA}b.  The Bessel decomposition provides further quantitative confirmation that approximately $99.8 \%$ of the conditioned subharmonic modal energy  is contained in the first radial mode $p=1$.

Taken together, the linear stability analysis, the bath-only simulation and the droplet-present difference field identify the same dominant radial structure. The axisymmetric Floquet calculation predicts $p=1$ as the lowest-threshold subharmonic mode, and the above-threshold bath-only simulation shows that this mode dominates during linear growth. The subharmonic component of the additional response associated with the droplet also projects predominantly onto the same first Neumann eigenfunction. These observations indicate that droplet impact couples efficiently to the radial structure to which the periodically forced bath is most susceptible. At $F=1<F_c$, this finite-amplitude excitation provides a route to a growing Faraday-wave response even though the initially flat bath remains linearly stable.

\subsubsection{Evolution of impact dynamics and energy redistribution}

We conclude the analysis of the unstable-bath regime by examining how the impact dynamics change between the first and second bounces. We first compare the gas-pressure distribution beneath the droplet, which characterizes the
air-cushioning dynamics, and then quantify the associated redistribution of energy between the droplet and the bath. Together, these diagnostics show how the deformation and residual motion generated by the first impact modify the subsequent collision. We note that the results presented in this section  were obtained
at grid refinement level $LVL = 12$, to ensure a better resolution of the air film, and subsequently the pressure distribution in the gas phase.

\begin{figure}
\centering
\begin{tabular}{ccc}
\includegraphics[width=0.32\linewidth]{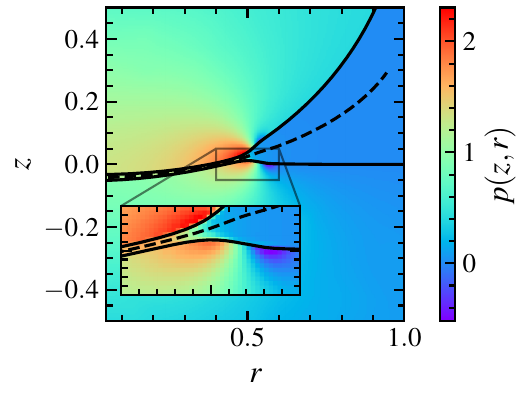} &  
\includegraphics[width=0.32\linewidth]{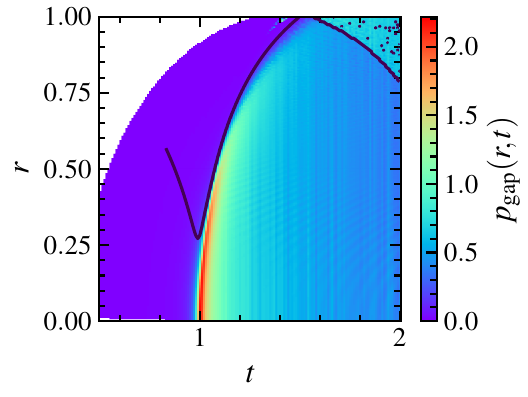} &
\includegraphics[width=0.35\linewidth]{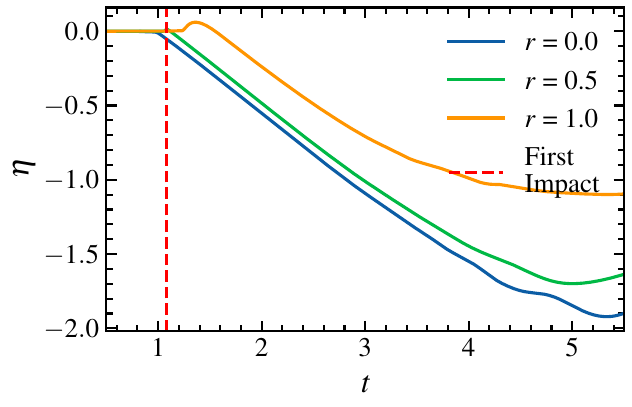} \\
  &  (a)  & \\
\includegraphics[width=0.32\linewidth]{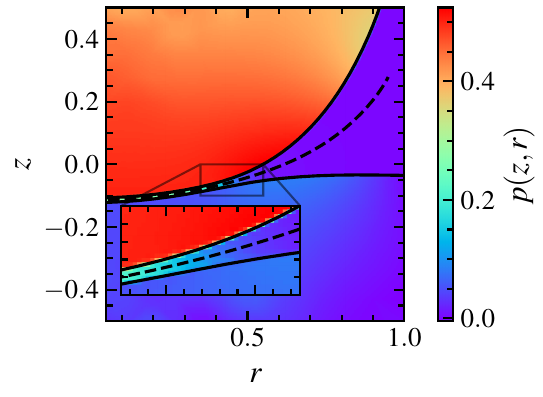} &  
\includegraphics[width=0.32\linewidth]{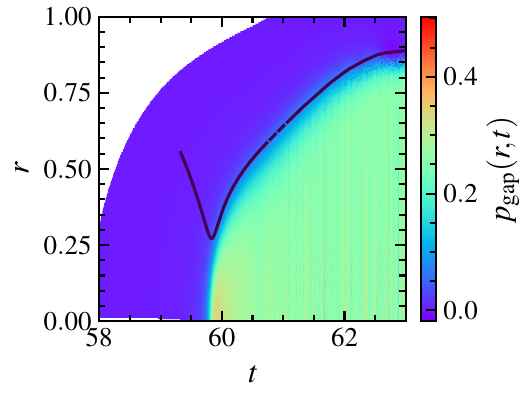} &
\includegraphics[width=0.35\linewidth]{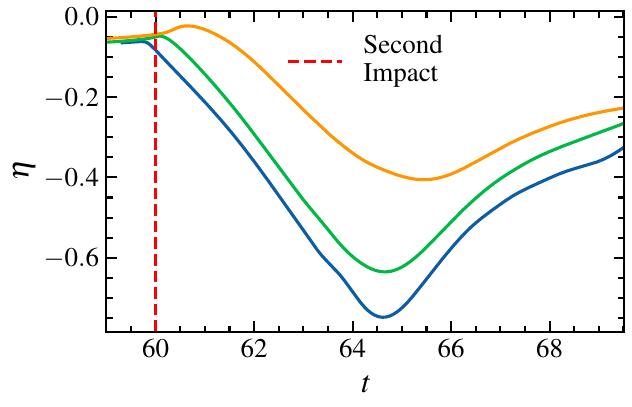} \\
  &  (b)  & \\
\end{tabular}
\caption{Impact-resolved pressure distribution, air-gap pressure evolution, and temporal bath deformation for the unstable bath at $We=5.0$ and $Re=200$. Rows (a) and (b) illustrate the first and second impact events, respectively. The left column shows the pressure field during impact, $p(z,r)$, with the inset enlarging the pressure distribution in the intervening gas layer. The middle column shows the spatio-temporal pressure evolution at the air-gap, $p_{\mathrm{gap}}(r,t)$, with the solid black line representing the radial location of the minimum droplet-bath separation, $r_{\mathrm{min}}(t)$. The right column shows the bath elevation, $\eta(r,t)$, at $r=0,~0.5,$ and 1.0, with the red dashed line indicating the corresponding impact instant.}
\label{fig: Pressure map_unstable}
\end{figure}

During the first impact, depicted in figure \ref{fig: Pressure map_unstable}a, the pressure field is dominated by an air-cushioning event in the thin gas layer separating the drop and the bath. The high-pressure region is confined to the underside of the drop and is amplified near the radial location where the gap is smallest. This is seen in the radial pressure trace, where $p_{\rm gap}$ remains large over the inner portion of the contact region and then drops rapidly once the gap opens. 
The spatio-temporal map confirms that this pressure impulse is localized in time and follows the moving region of minimum separation.
The bath response produced by this first impact is strongly non-uniform. The 
bath at $r\in[0,0.5]$ move downward almost monotonically after impact, whereas the outer radial location responds later and with a smaller initial deflection. Thus, the first collision imprints a radially structured deformation and velocity field onto the bath. 

The second impact exhibits a substantially weaker and broader gas pressure distribution (see figure \ref{fig: Pressure map_unstable}b), with the peak pressure reduced by a factor of approximately $4$-$5$. 
The pre-existing deformation and residual motion of the bath modify both the effective gap thickness and the relative approach velocity, thereby weakening the air-cushioning pressure during the subsequent impact.
%
At the centerline ($r=0$), the incremental depression produced by the second impact is approximately three times smaller than that generated by the first impact (see figure \ref{fig: Pressure map_unstable}a), with a comparable reduction observed away from the axis. This weaker deflection is consistent with the reduced gas-pressure during the second impact, considering the bath is already deformed during the collision and contains some residual motion.

\begin{figure}
\centering
\begin{tabular}{c}
\includegraphics[width=\linewidth]{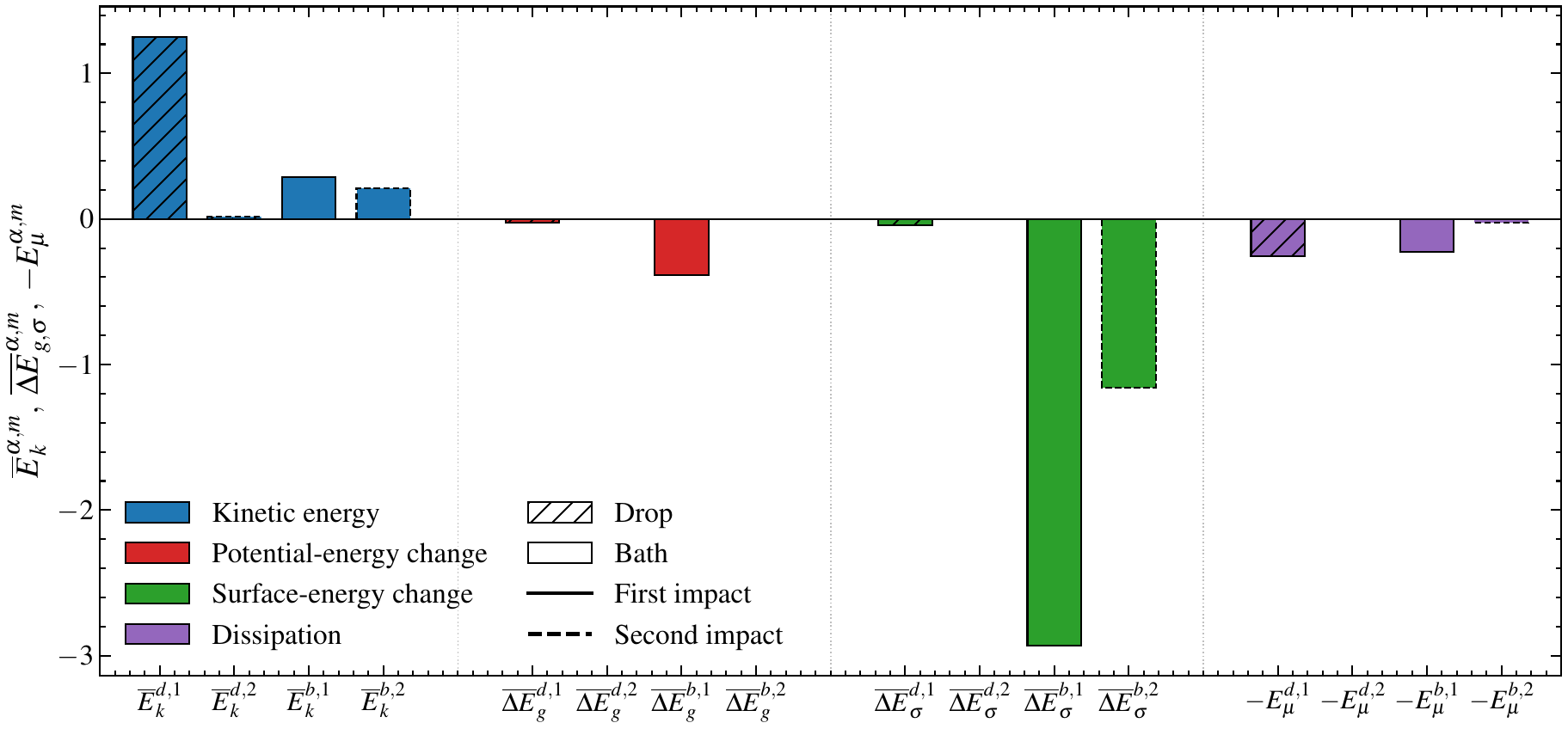}\\
(a)\\
\end{tabular}
\begin{tabular}{cc}
\includegraphics[width=0.49\linewidth]{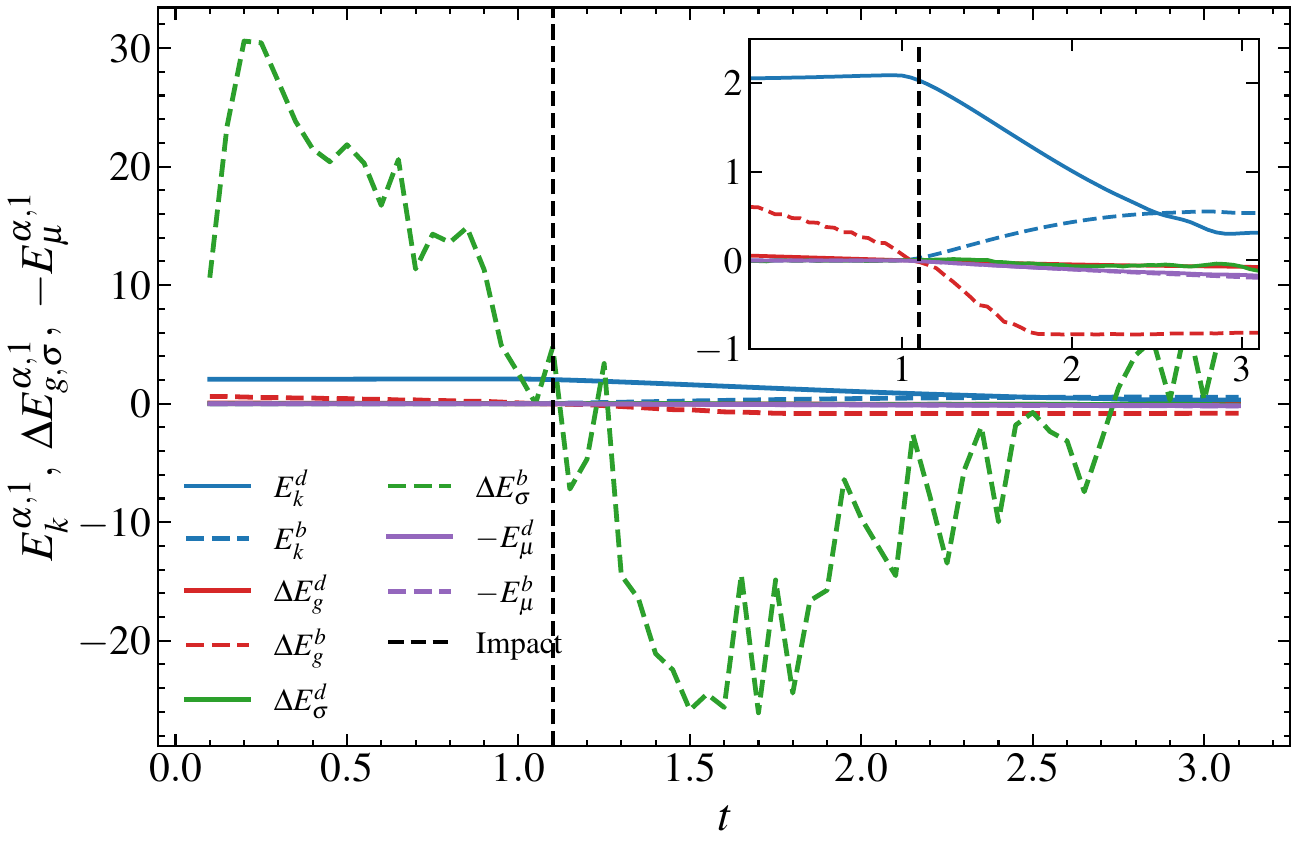}&
\includegraphics[width=0.49\linewidth]{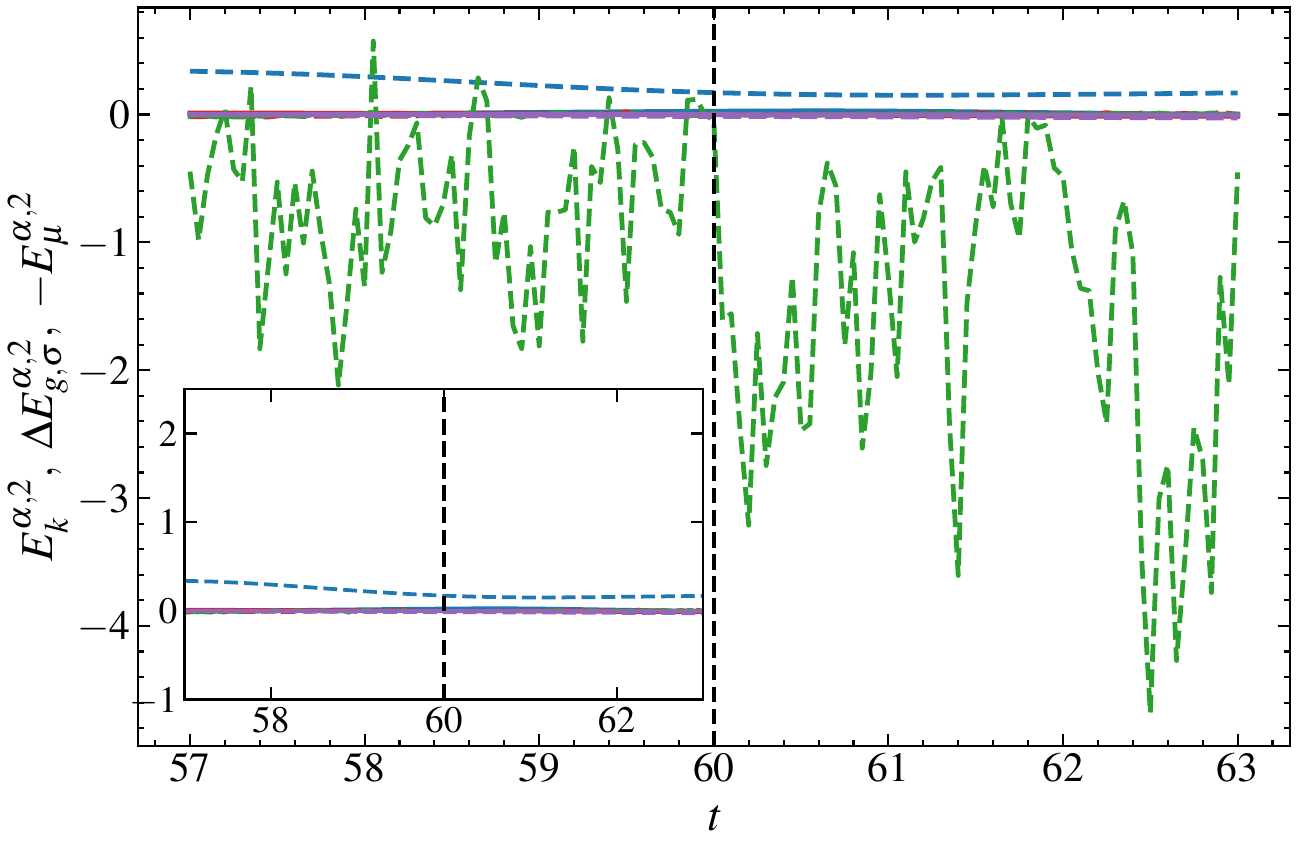}\\
(b)&(c)\\
\end{tabular}
\caption{
Energy redistribution during successive droplet impacts for $We=5.0$ and $Re=200$. (a) Phase-decomposed  contributions for the first $(m=1)$ and second $(m=2)$ impacts: event-averaged kinetic-energy $\overline{E}_k^{\alpha,m}$, gravitational-potential energy change $\Delta\overline{E}_g^{\alpha,m}$, surface-energy change $\Delta\overline{E}_\sigma^{\alpha,m}$, and cumulative viscous dissipation $-E_\mu^{\alpha,m}$, where $\alpha=d,b$ represents the droplet and bath, respectively. Hatched and unhatched bars distinguish the droplet and bath, while solid and dashed outlines show the first and second impacts. Panels (b) and (c) illustrate the corresponding temporal energy contributions during first and second impacts, respectively, with the dashed line marking the reference time $t_r^{(m)}$ immediately preceding impact.}
\label{fig: energy budget}
\end{figure}

To characterize how the energetic state of the droplet-bath system changes between successive impacts, we examine the kinetic, gravitational, interfacial and viscous contributions associated with each phase. Because the bath is continuously subjected to time-dependent vertical forcing, these quantities do not constitute a closed mechanical-energy budget unless the work performed by the forcing is included explicitly. We therefore use them here as impact-resolved energy diagnostics, primarily to compare the redistribution of stored and dissipated energy between successive collisions.
The complete derivation of the phase-decomposed energy terms 
is presented in detail in Appendix \ref{App C1}.

Let $m\in \{1,2\}$ represent first and second impact events, respectively, and $\alpha \in \{d,b\}$ denote the droplet and bath contributions, respectively. Each impact is evaluated over $t \in [t_a^{(m)} , t_b^{(m)} ]$, with duration
\begin{equation}
    \Delta t_m=t_b^{(m)} -t_a^{(m)}.
\end{equation}
 The instantaneous kinetic, gravitational-potential, and surface energies of phase $\alpha$, represented by $E_k^\alpha$, $E_g^\alpha$, and $E_\sigma^\alpha$, respectively, are defined in Appendix \ref{App C1}. The kinetic energy is retained in absolute form, with its event-averaged value given by
\begin{equation}
    \overline{E}_k^{\alpha,m} = \frac{1}{\Delta {t}_m} \int_{{t}_{a}^{(m)}}^{{t}_{b}^{(m)}} E_k^{\alpha} ({t}) d{t}.
\end{equation}
The gravitational and surface energies contain large contributions associated with the equilibrium bath volume and interfacial area. Although these contributions are physically meaningful, they remain nearly constant over the duration of an individual collision and obscure the much smaller impact-induced variations. We therefore introduce a reference time $t_r^{(m)}$ immediately preceding impact $m$, with $t_r^{(1)}=1.05$ and $t_r^{(2)}=60.00$, and define
\begin{equation}
\Delta E_j^{\alpha,m}(t)
=
E_j^{\alpha}(t)-E_j^{\alpha}\left(t_r^{(m)}\right),
\qquad
j\in\{g,\sigma\}.
\label{eq:energy_increments}
\end{equation}
Their event-averaged variations are then given by
\begin{equation}
    \Delta\overline{E}_j^{\alpha,m} = \frac{1}{\Delta {t}_m} \int_{{t}_{a}^{(m)}}^{{t}_{b}^{(m)}} \Delta E_j^{\alpha,m}(t)  d{t}, \qquad j \in ({g,\sigma}).
\end{equation}
Finally, representing the instantaneous viscous-dissipation rate of phase $\alpha$ by $\varepsilon_\mu^\alpha(t)$, the cumulative viscous dissipation energy during impact event $m$ is
\begin{equation}
    E_\mu^{\alpha,m} =  \int_{{t}_{a}^{(m)}}^{{t}_{b}^{(m)}} \varepsilon_\mu^\alpha(t) d{t}.
\end{equation}
Since viscous dissipation denotes an irreversible loss of mechanical energy, it is represented by $-E_\mu^{\alpha,m}$.

Accordingly, figure~\ref{fig: energy budget} compares
$\overline{E}_k^{\alpha,m}$,
$\Delta\overline{E}_g^{\alpha,m}$,
$\Delta\overline{E}_\sigma^{\alpha,m}$, and
$-E_\mu^{\alpha,m}$ for the two successive impact events. The phase-decomposed energy contributions are verified at $t=0$ against the corresponding analytical values obtained from the prescribed initial geometry, velocity field, and fluid properties.
During the first collision, the droplet initially carries substantially more kinetic energy than the bath (see figure~\ref{fig: energy budget}b).
Following impact, the droplet kinetic energy
decreases while the bath kinetic energy increases, accompanied by a pronounced excursion in the bath surface energy. These changes are consistent with impact-induced redistribution toward bath motion and interfacial deformation. The surface-energy contribution exhibits the largest temporal variation, consistent with the strong bath deformation and outward- propagating capillary disturbance observed in figure \ref{fig: unstable_snapshots}. The accompanying decrease in bath gravitational potential energy is consistent with the impact-induced downward displacement of the bath. Over the first impact interval, viscous dissipation remains smaller than the dominant droplet kinetic- and bath surface-energy contributions (see figure~\ref{fig: energy budget}a).
The second collision occurs with substantially less droplet kinetic energy (see figure~\ref{fig: energy budget}c), and the corresponding bath surface-energy excursion is markedly reduced. At the same time, the bath enters the second impact with non-negligible kinetic energy, indicating that the collision occurs against a pre-existing moving and deformed bath rather than the initially quiescent state. Together with the reduced pressure and interface deformation observed during the second collision, these results indicate that the first impact provides the dominant initial finite-amplitude perturbation, whereas the second produces a smaller incremental disturbance to the evolving bath state.

\subsection{Decaying  regime}

\begin{figure}
\centering
\begin{tabular}{cccc}
\includegraphics[width=0.23\linewidth]{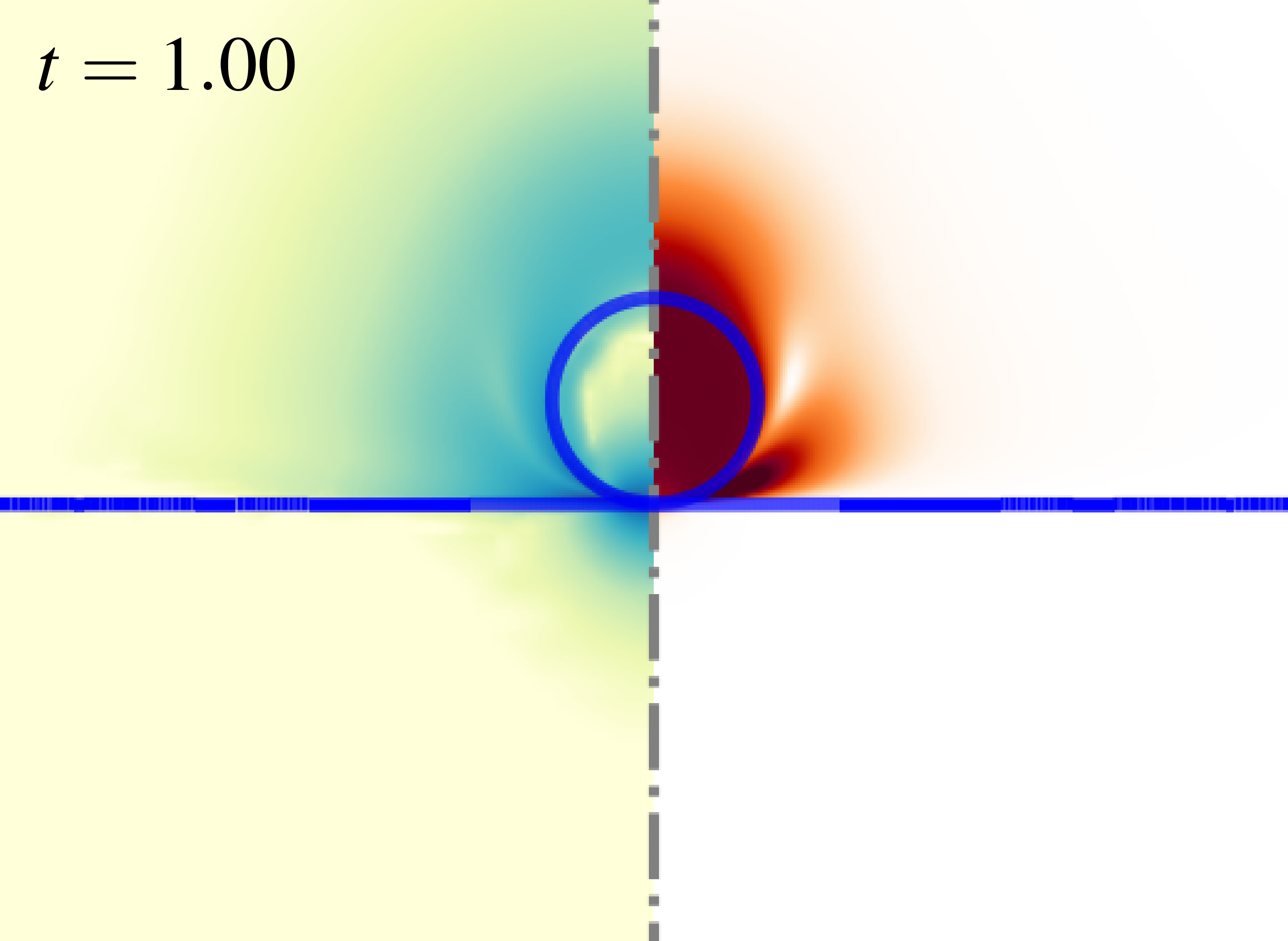}&
\includegraphics[width=0.23\linewidth]{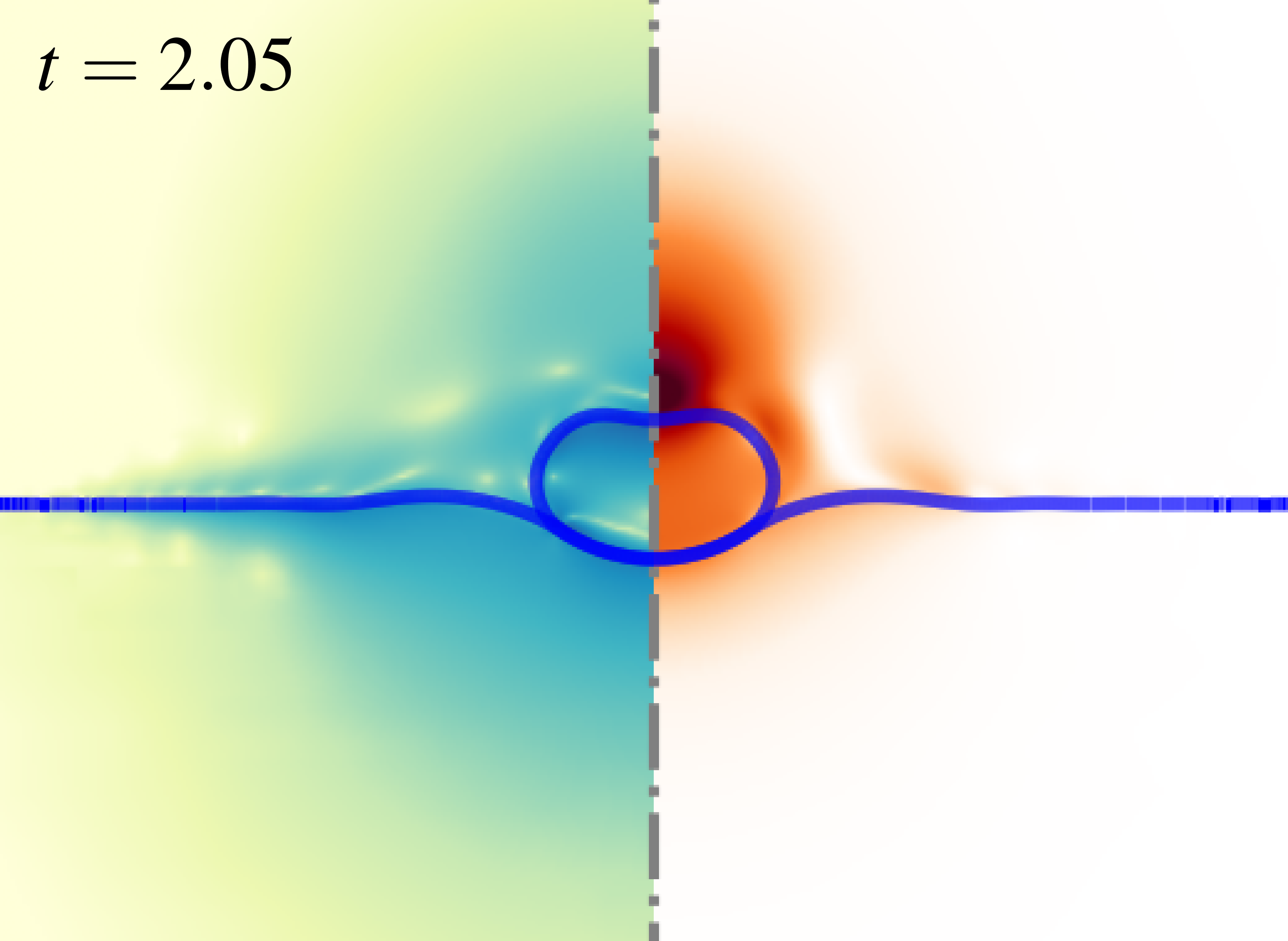}&
\includegraphics[width=0.23\linewidth]{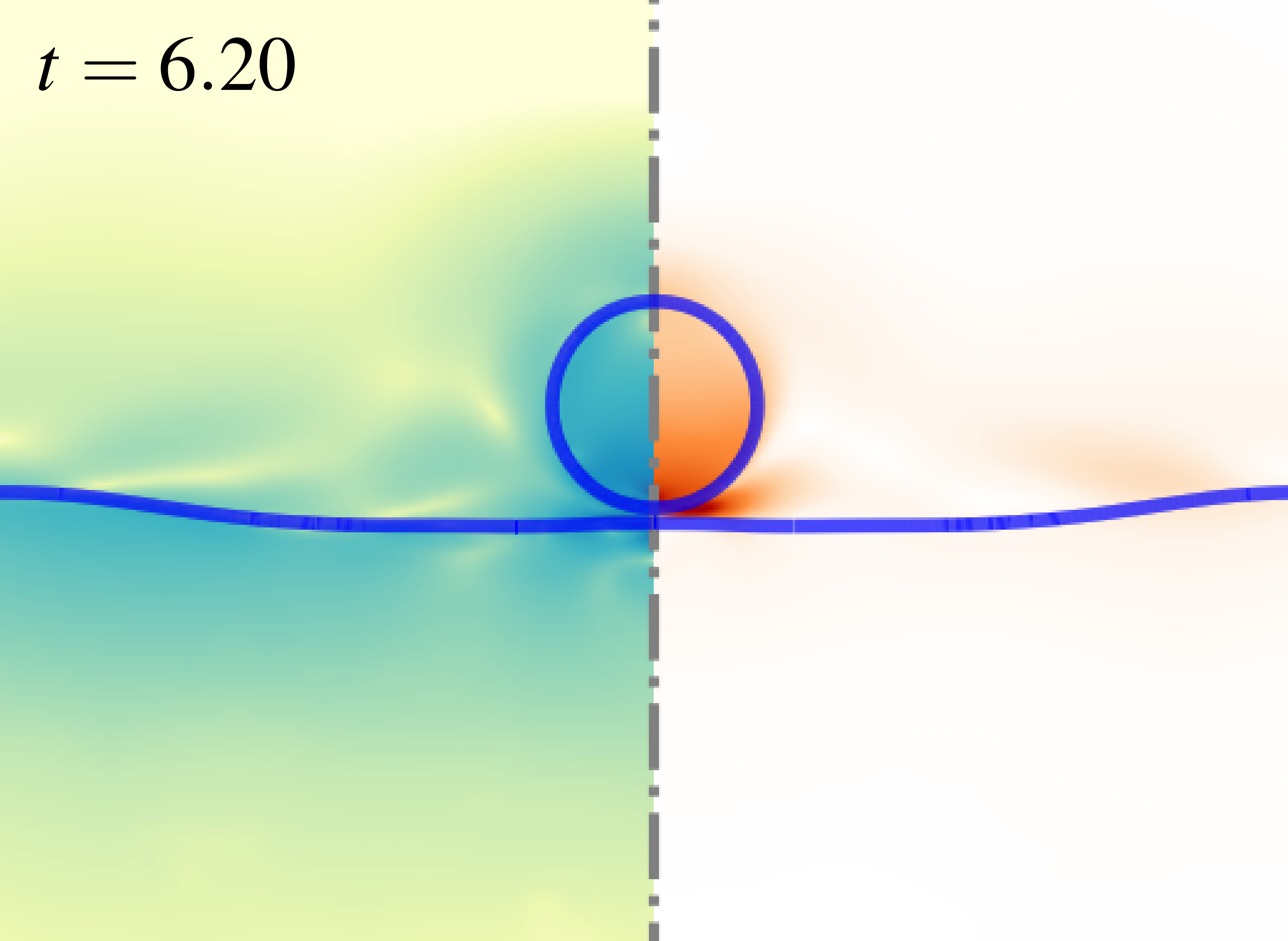}&
\includegraphics[width=0.23\linewidth]{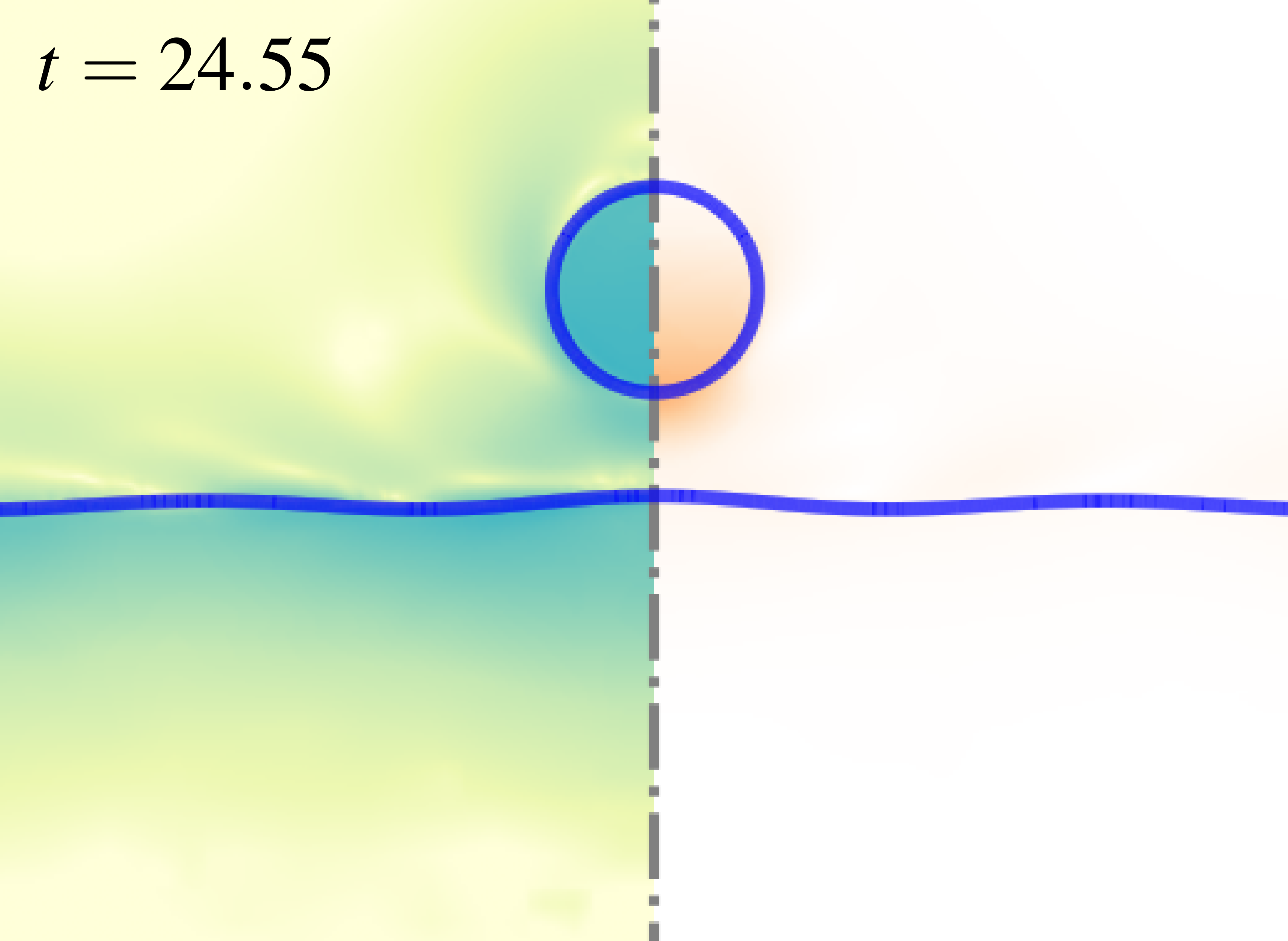}\\
\end{tabular}
\begin{tabular}{c}
(a) 
\\
\\
\end{tabular}
\begin{tabular}{cccc}
\includegraphics[width=0.23\linewidth]{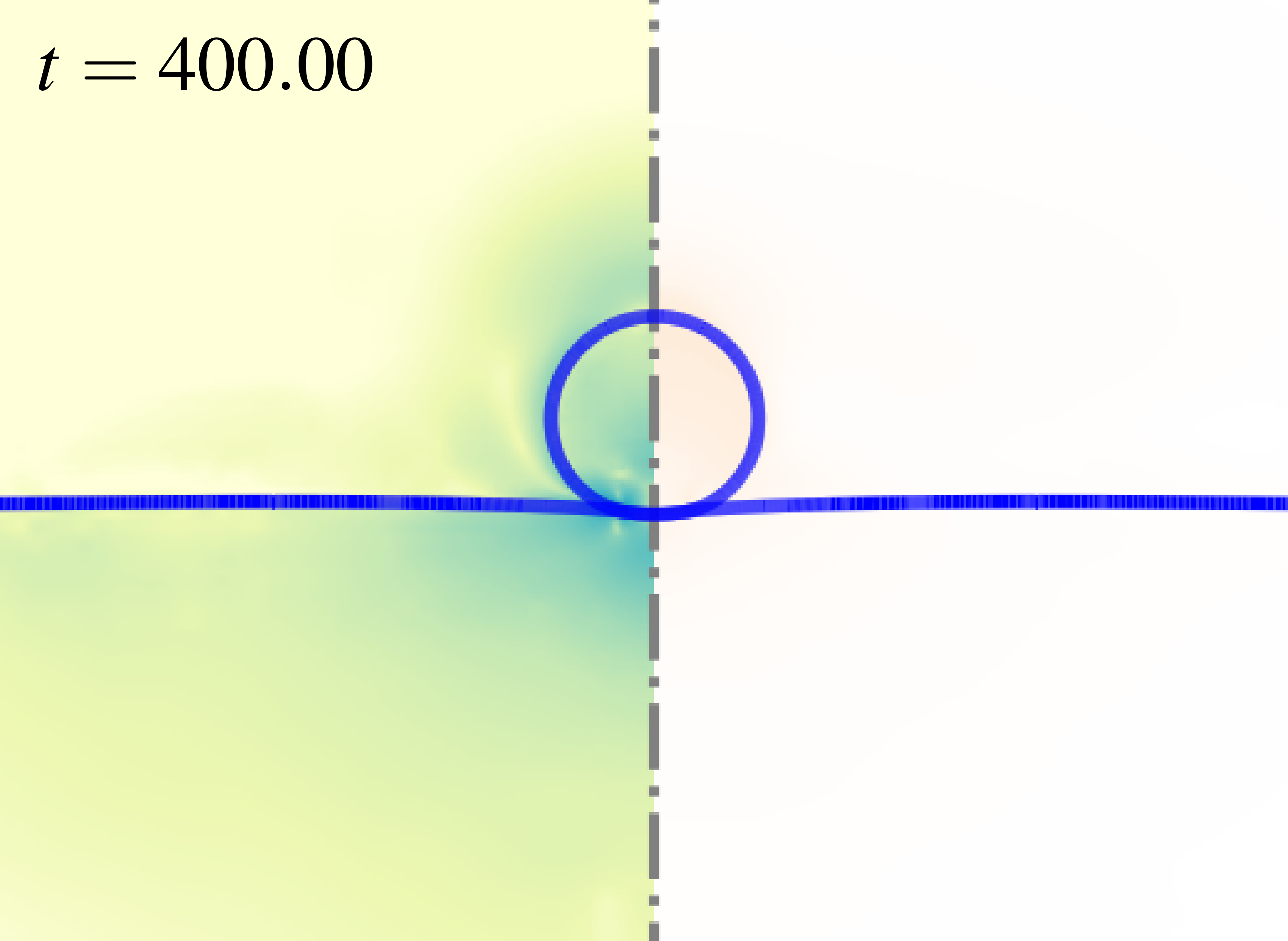}&
\includegraphics[width=0.23\linewidth]{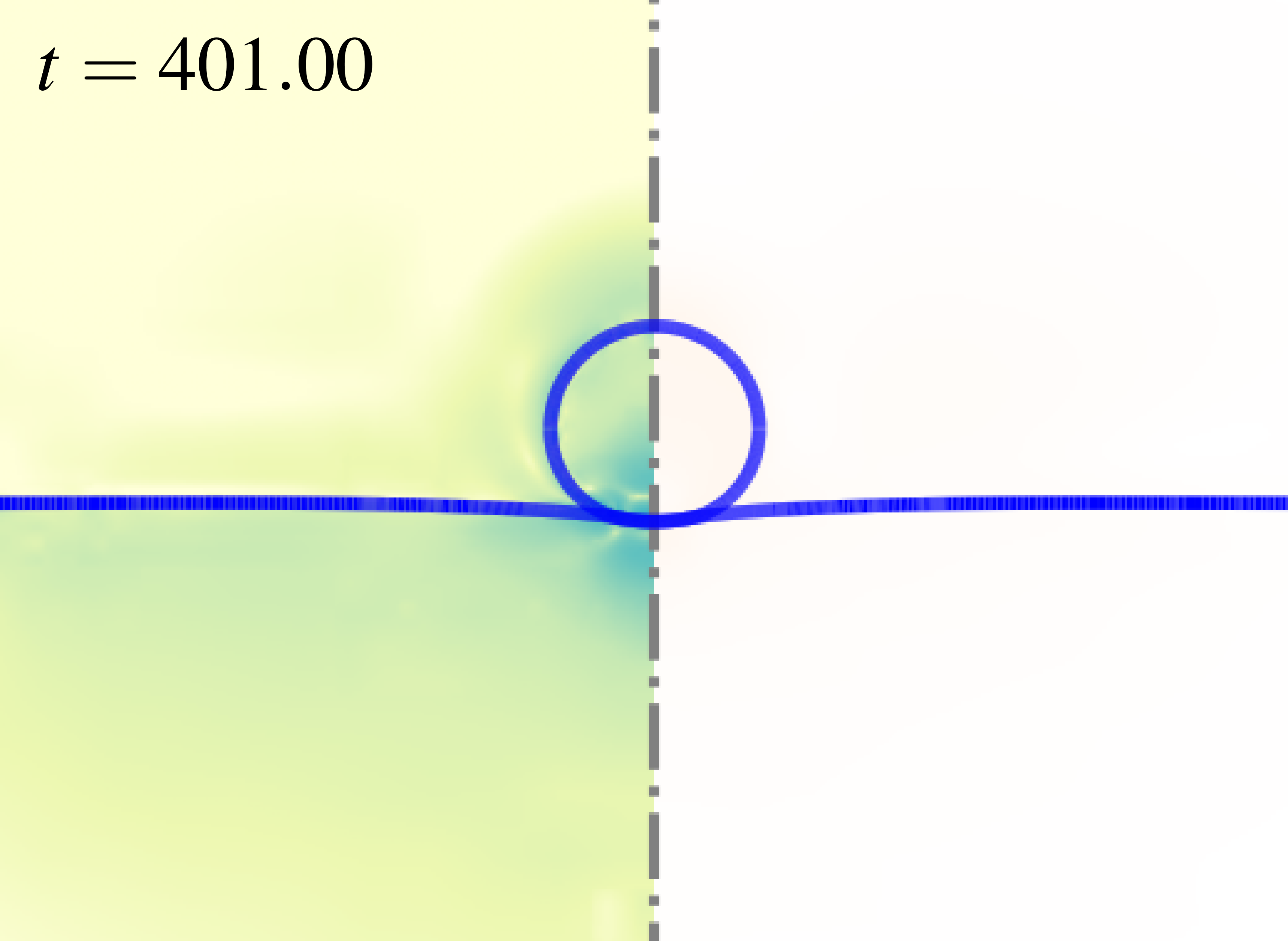}&
\includegraphics[width=0.23\linewidth]{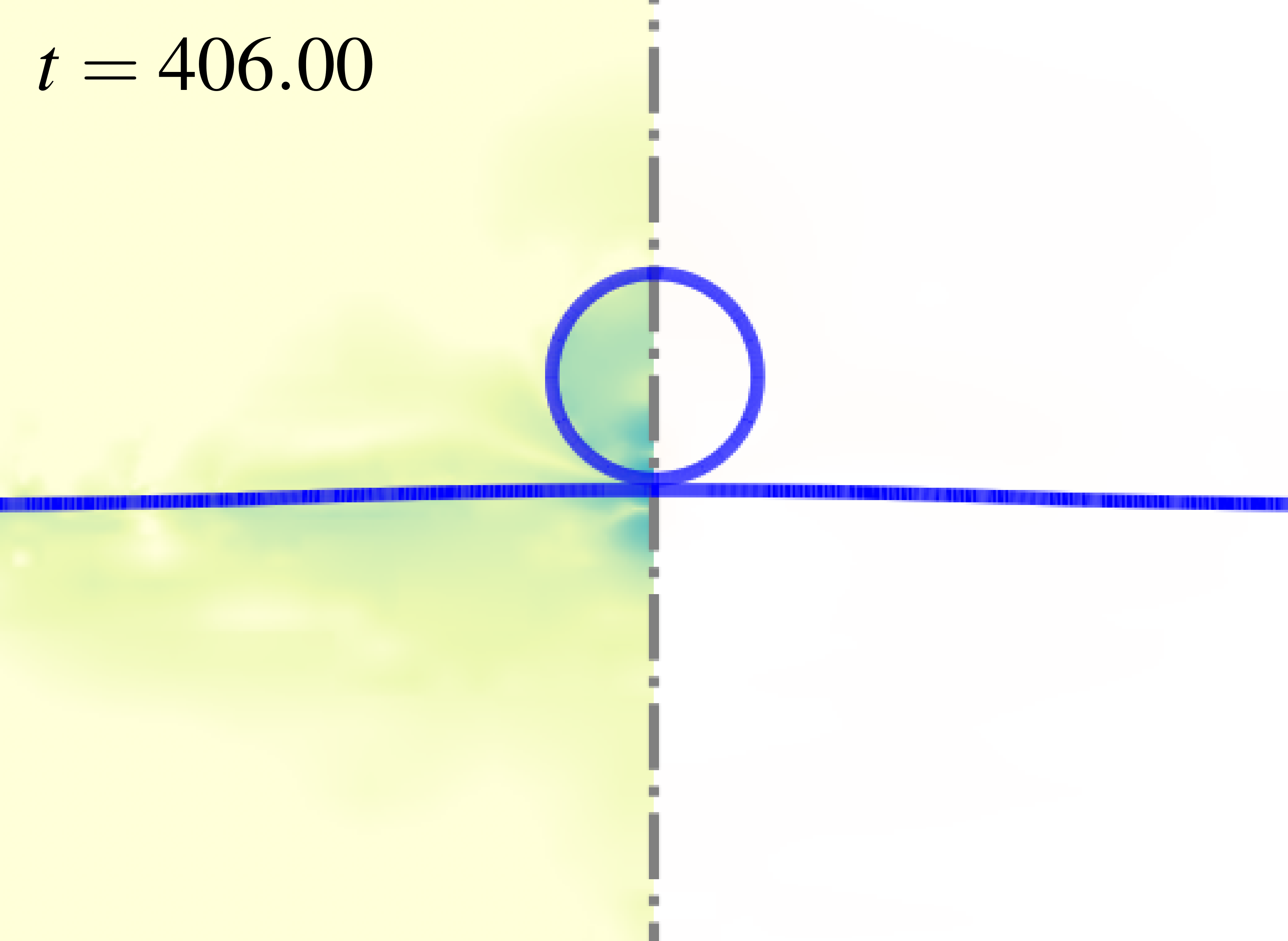}&
\includegraphics[width=0.23\linewidth]{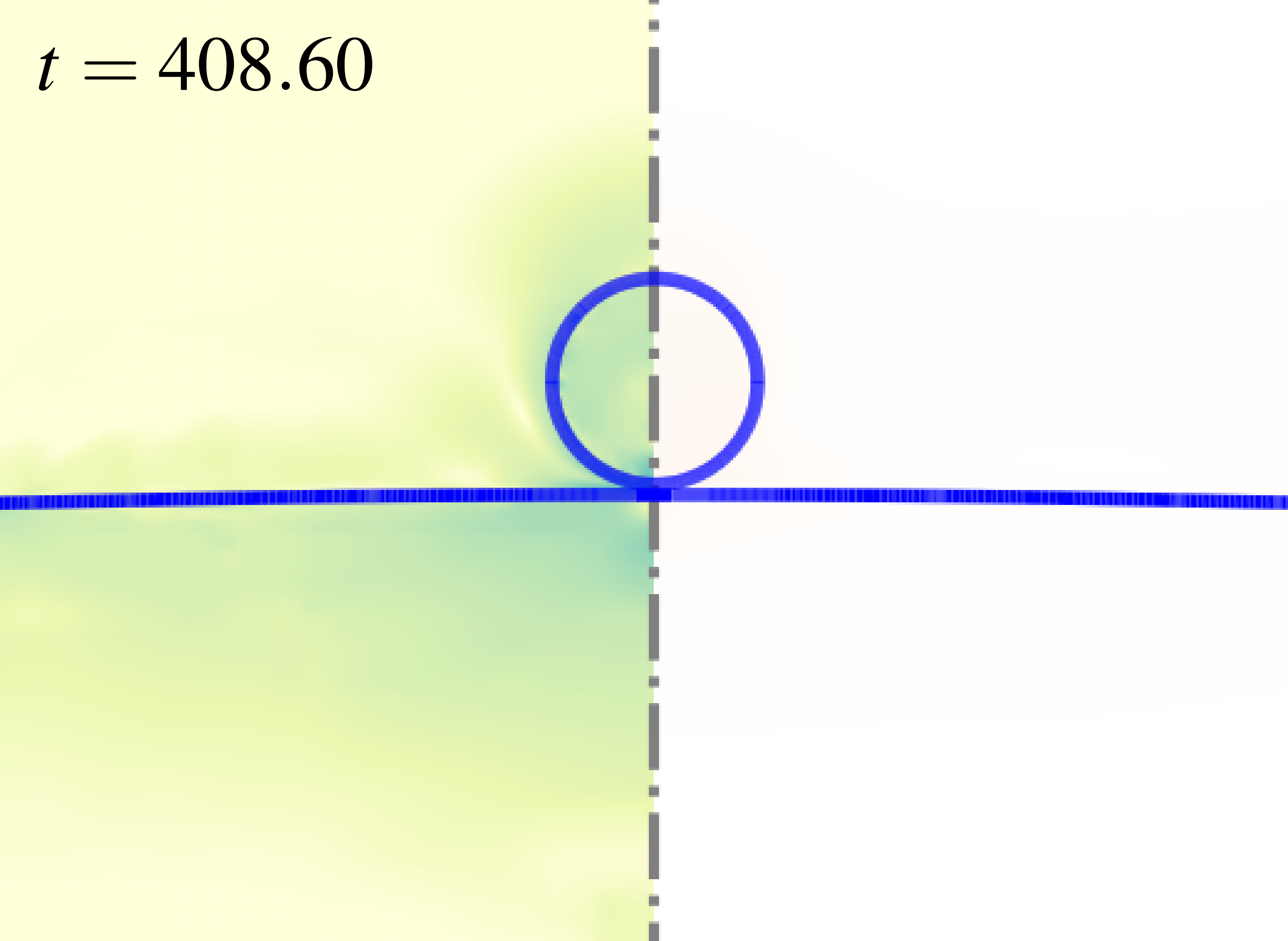}\\
\end{tabular}
\begin{tabular}{c}
(b) 
\\ 
\includegraphics[width=0.5\linewidth]{Figures/zoomed_images/colorbar.png}\\
\end{tabular}

\caption{Snapshots of the droplet--bath interface in the decaying  bath regime with  $We=1.21$ and $Re=175$. Four stages of the bounce cycle
are shown: impact, maximum spread, lift-off, and descent, corresponding to one-to-four column, respectively. The first bounce (top row) and bounce number 11 at $t=400$ (bottom row) are shown. The left and right color bars represent the
viscous-dissipation rate per unit volume, $
\Phi_\mu$, and 
the 
velocity magnitude, $|\boldsymbol{u}|$, respectively.
The blue curves denote the liquid--air interfaces and the dashed line marks the axis of symmetry.}
\label{decaying_interface}
\end{figure}

\begin{figure}
\centering
\begin{tabular}{c}
\includegraphics[width=\textwidth]{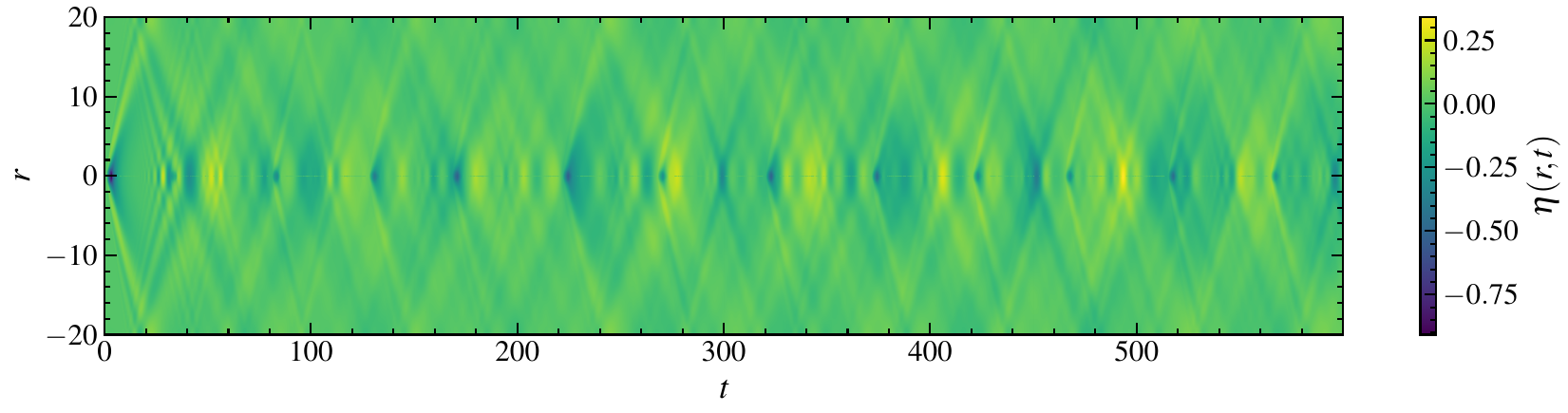} \\
(a)\\
\end{tabular}
\begin{tabular}{cc}
\includegraphics[width=0.47\linewidth]{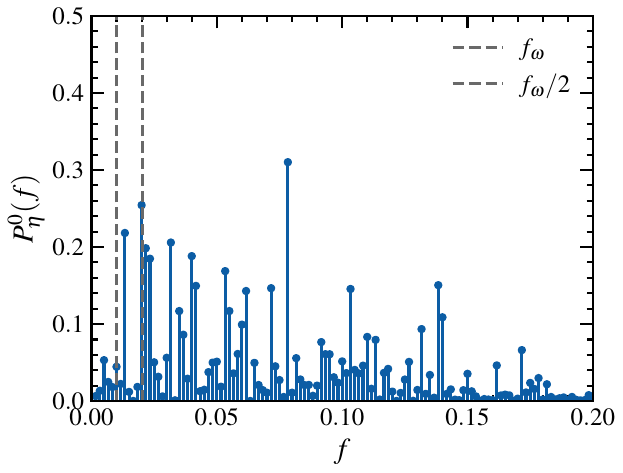} &  \includegraphics[width=0.47\linewidth]{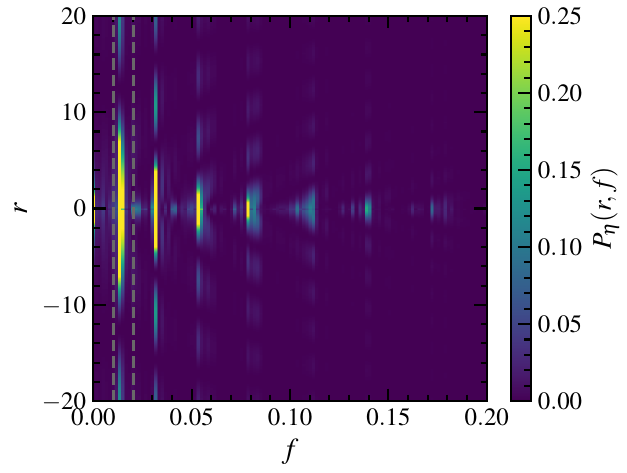}\\
(b) & (c)\\
\end{tabular}
\begin{tabular}{c}
\includegraphics[ width=\textwidth]{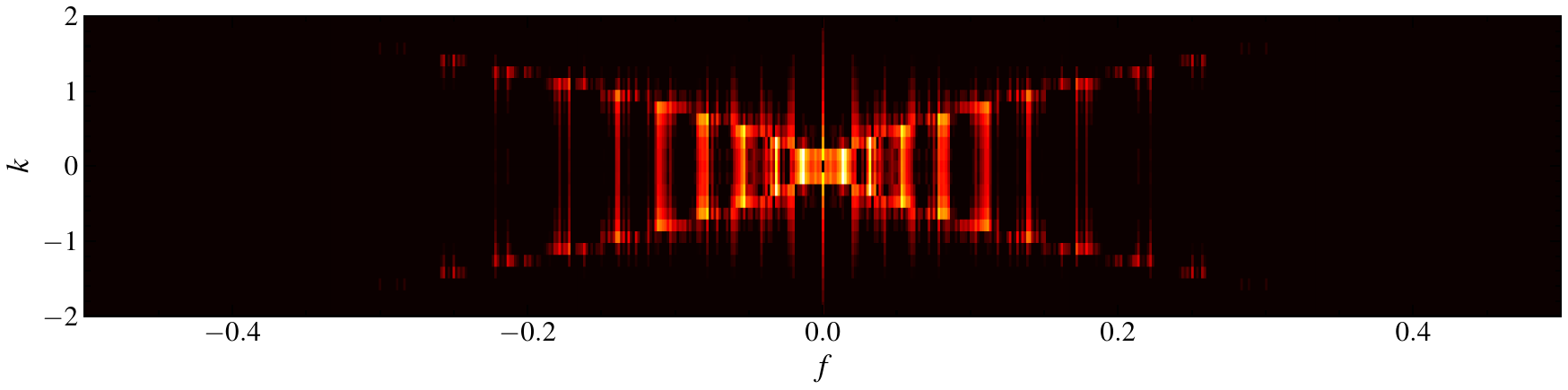}\\
(d) \\
\end{tabular}
\caption{Spectral analysis of the bath for
the decaying  bath regime with  $We=1.21$ and $Re=175$.
(a) Space--time diagram of the interface elevation
$\eta(r,t)=z_{\mathrm{int}}(r,t)-z_0$.
(b) Temporal spectrum of the centerline elevation, $P_{\eta}^{0}(f)$, with the forcing frequency $f_\omega$ and its subharmonic $f_\omega/2$ represented by the vertical dashed lines.
(c) Radial distribution of the temporal spectrum, $P_{\eta}(r, f)$.
(d) Two-dimensional power spectrum, $P_{\eta}^{2D}(k, f)$, showing combined spatial and temporal frequency content of the bath response.
}
\label{fig:decaying_spectrum}
\end{figure}

\begin{figure}
\centering
\begin{tabular}{ccc}
\includegraphics[width=0.32\linewidth]{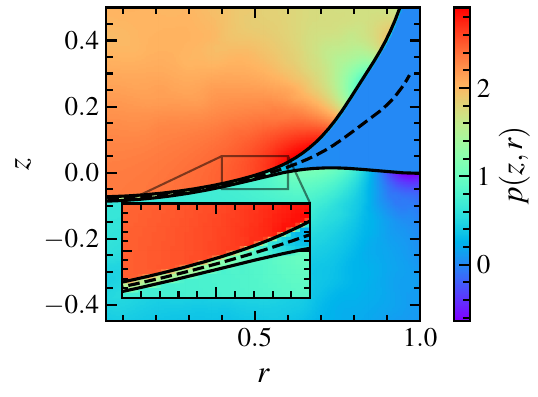} &  
\includegraphics[width=0.32\linewidth]{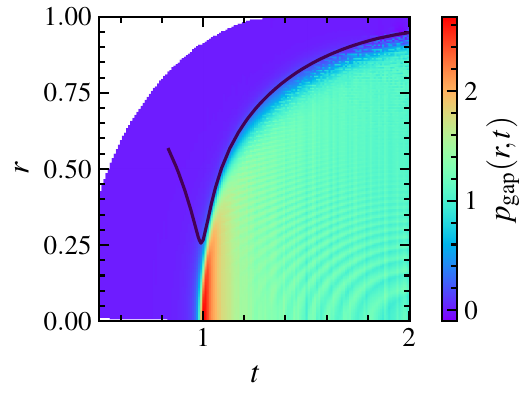} &
\includegraphics[width=0.35\linewidth]{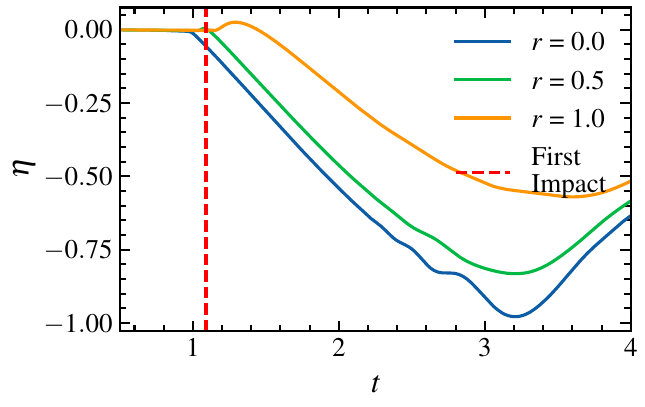} \\
  &  (a)  & \\
\includegraphics[width=0.32\linewidth]{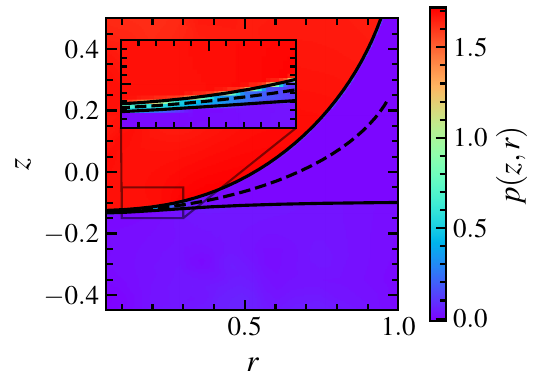} &  
\includegraphics[width=0.32\linewidth]{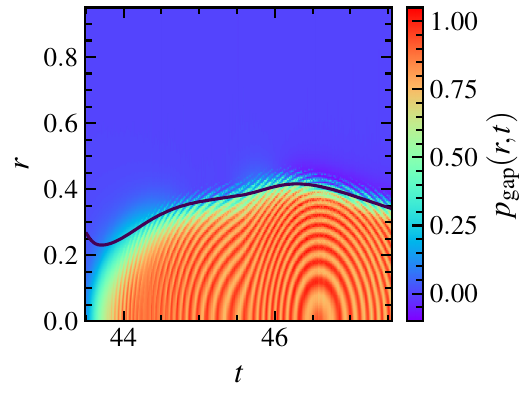} &
\includegraphics[width=0.35\linewidth]{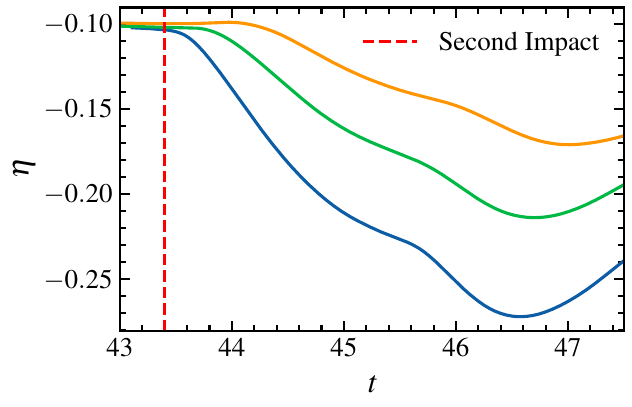} \\
  &  (b)  & \\
\end{tabular}
\caption{Impact-resolved pressure distribution, air-gap pressure evolution, and temporal bath deformation for the decaying  bath regime with  $We=1.21$ and $Re=175$. Rows (a) and (b) illustrate the first and second impact events, respectively. The left column shows the pressure field during impact, $p(z,r)$, with the inset enlarging the pressure distribution in the intervening gas layer. The middle column shows the spatio-temporal pressure evolution at the air-gap, $p_{\mathrm{gap}}(r,t)$, with the solid black line representing the radial location of the minimum droplet-bath separation, $r_{\mathrm{min}}(t)$. The right column shows the bath elevation, $\eta(r,t)$, at $r=0,~0.5,$ and 1.0, with the red dashed line indicating the corresponding impact instant.}
\label{fig: Pressure map_decaying}
\end{figure}

We next consider the decaying regime, characterised by 
$We=1.21$ and
$Re=175$.
Figure \ref{decaying_interface} shows snapshots of the droplet-bath interfaces for two different bounce instances.  The first impact produces a pronounced local deformation and an outward-propagating disturbance, with velocity and viscous dissipation concentrated near the droplet--bath interaction.  At later times, the droplet remains nearly spherical, the bath deformation is reduced, and both the velocity magnitude and viscous dissipation weaken substantially. Unlike the unstable-bath case, the impact-generated disturbances do not develop into a progressively amplified subharmonic bath response.

To assess whether the impact-generated disturbances develop into a sustained Faraday response, figure~\ref{fig:decaying_spectrum} presents the spatio-temporal evolution and corresponding spectra of the bath interface. Figure~\ref{fig:decaying_spectrum}a shows that each impact launches outward-propagating capillary waves from the axis. These disturbances reflect from the lateral walls and return towards the impact region, producing the characteristic intersecting trajectories in the space--time diagram, but their amplitude does not increase over successive bounces. The temporal spectrum at $r=0$ in figure~\ref{fig:decaying_spectrum}b contains several competing frequencies, with no dominant response at the subharmonic frequency
${f_\omega}/2$. This broadband behavior persists across the bath (see figure~\ref{fig:decaying_spectrum}c), while the two-dimensional spectrum in figure~\ref{fig:decaying_spectrum}d reflects the presence of counter-propagating waves but shows no isolated subharmonic spatial mode. Unlike the unstable-bath regime, the transient waves decay rather than developing into a sustained Faraday-wave response.

The pressure field provides a complementary measure of the weakened droplet--bath interaction. During the first bounce, a broad high-pressure region develops beneath the droplet, producing a pronounced deformation of both interfaces (see figure~\ref{fig: Pressure map_decaying}a). At the second impact,  figure~\ref{fig: Pressure map_decaying}b, the maximum gap pressure and its radial extent are reduced, while the minimum-gap $p_{\mathrm{gap}}(r,t)$ trajectory evolves more gradually than the unstable bath case. The later impacts, therefore, exert a weaker and more localized forcing on the bath, consistent with the decreasing interface deformation and the absence of cumulative wave amplification. 
Comparing the representative unstable and decaying cases
the unstable  case exhibits substantially larger bath deformation than the decaying 
case despite comparable air-gap pressure scales, with maximum depressions approximately twice and three times larger during the first and second impacts, respectively. This enhanced deformation is consistent with weaker capillary restoration, in which increasing $We$ from 1.21 to 5.0 reduces the coefficient ($We^{-1}$) by a factor of $4.1$ and shifts the capillary–gravity spectrum. Whether the resulting disturbance grows, however, is determined by the bath's parametric susceptibility rather than by deformation amplitude alone.

Finally, for completion, further details of the impact-resolved energy budgets, including the separate contributions from the droplet and bath, are provided in appendix \ref{App F}.

\section{Conclusions}
\label{sec: conclusion}

We have investigated the long-time interaction between a non-coalescing droplet and a vertically forced bath using axisymmetric two-phase direct numerical simulations. The resulting finite-time $We$–$Re$ regime map identifies quasi-periodic, decaying, faint and unstable-bath responses. Along the parameter path considered here, the transition to the unstable-bath state occurs over a comparatively narrow range of $We$ and varies only weakly with $Re$. Because $We$ is varied primarily through surface tension, this trend reflects changes in both the impact dynamics and the intrinsic capillary–gravity response of the bath and should not be interpreted as evidence for a universal Weber-number onset criterion.

In the unstable-bath regime, initially localized droplet impact-generated disturbances evolve into a spatially extended coherent response at the subharmonic forcing frequency $f_\omega/2$. By contrast, in the decaying regime the disturbances remain broadband and weaken over successive rebounds. Bath-only simulations provide a direct means of separating the role of impact from that of the external forcing. For the present geometry and fluid properties, the initially flat bath has a linear Faraday threshold of  $F_c=1.17$ . Nevertheless, at $F=1<F_c$, the droplet-present system develops a growing subharmonic response.

To test whether a finite disturbance can trigger this response, we developed a reduced estimate of the characteristic impact-induced deformation. Bath-only simulations initialized with these impact-equivalent perturbations reproduce the transition from bounded to growing subharmonic dynamics observed in the coupled simulations, despite the initially flat bath remaining linearly stable at the same forcing amplitude. Increasing $We$ increases the estimated disturbance amplitude, but also modifies the capillary--gravity spectrum and hence the susceptibility of the bath to parametric forcing. Thus, the present results should not be interpreted in terms of a universal critical perturbation amplitude. Instead, they demonstrate that droplet impact can provide a finite-amplitude seed that accesses a growing subharmonic response below the linear onset of the initially undisturbed bath.

The modal analysis identifies the radial structure associated with this response. An axisymmetric finite-domain extension of the Floquet formulation of \citet{Kumar_Tuckerman_1994} predicts that the first Neumann mode, $p=1$, possesses the lowest linear subharmonic forcing threshold. Bath-only DNS above onset independently show that the growing response is dominated by the same mode. The subharmonic component of the additional response associated with the droplet likewise projects predominantly onto the first Neumann eigenfunction. Thus, localized droplet impact couples efficiently to the radial mode to which the periodically forced bath is most susceptible.
Impact-resolved energy diagnostics further show that the first collision provides the dominant initial finite-amplitude disturbance to the bath, while later impacts occur against an already deformed and moving interface. The sustained bath response is therefore not a sequence of independent rebound events but results from the interaction between impact-generated disturbances, the residual bath motion and continuous parametric forcing.

The present axisymmetric formulation is restricted to vertical bouncing and axisymmetric bath modes, excluding walking, chaotic trajectories and non-axisymmetric Faraday patterns \citep{molavcek2013drops,perinet2009numerical}. Extending the problem to three dimensions is therefore a natural next step, although resolving the thin gas film over many forcing periods will make such simulations considerably more expensive. 
Additional interfacial physics may also modify the mechanism identified here. Viscoelasticity can alter impact-energy partitioning, interfacial deformation, and the conditions for droplet rebound, while surfactants can modify thin-film drainage, coalescence, and impact dynamics through surface-tension gradients and the associated Marangoni stresses \citep{CRCA_coalescence,zhong2026interface}. In three dimensions, non-uniform surfactant distributions may additionally break the symmetry of the rebound and promote lateral droplet motion.  Finally, the forcing frequency was fixed throughout this study, varying it would shift the discrete bath modes relative to the harmonic and subharmonic resonance conditions, thereby modifying both the instability threshold and the selected mode. This may identify regimes in which sustained bouncing can
be maintained without destabilizing the bath.\\

\noindent \textbf{Declaration of Interests}. The authors report no conflict of interest. \\

\noindent\textbf{Data availability.} 
The code and compressed datasets supporting the findings of this study will be publicly available in the CFTL-Illinois  GitHub repository.
\\

\noindent\textbf{Acknowledgments.}
We thank Andres Castillo-Castellanos for his assistance with the initial simulation setup and for fruitful discussions during the early stages of this work. This research used the Delta advanced computing and data resource, which is supported by the National Science Foundation (award OAC 2005572) and the State of Illinois.

\appendix
\section{Validation of the numerical solver}
\label{App A}
\setcounter{equation}{0}
\renewcommand{\theequation}{A.\arabic{equation}}
\setcounter{table}{0}
\renewcommand{\thetable}{A\arabic{table}}
\setcounter{figure}{0}
\renewcommand{\thefigure}{A\arabic{figure}}

To ensure the reliability of the numerical framework used throughout this work, we verify the solver against different canonical configurations involving droplet bouncing and parametrically forced surface waves. These cases test different aspects of the numerical model.

\subsection{Experimental work by \citet{alventosa2023inertio}}
\label{App A1}

To validate the numerical framework, we reproduce the experimental conditions reported by \citet{alventosa2023inertio}, in which a deionized-water droplet impacts a bath of the same liquid.
The droplet radius is $R=0.35~\mathrm{mm}$, with fluid properties corresponding to water, namely $\rho\simeq 1000~\mathrm{kg~m^{-3}}$, $\mu\simeq 1.0~\mathrm{mPa~s}$, and $\sigma\simeq 72~\mathrm{mN~m^{-1}}$. The imposed impact velocity gives $We=0.7$, while the corresponding Bond and Ohnesorge numbers are $Bo=0.017$ and $Oh=0.006$, respectively.
The gas phase is fully resolved in the DNS and is assigned properties representative of air at room temperature, with density ratio $\rho_g/\rho\simeq 1.2\times 10^{-3}$ and gas Ohnesorge number $Oh_g=O(10^{-4})\ll Oh$.
The axisymmetric computational domain and the level of refinement are similar to the DNS performed in 
\citet{alventosa2023inertio}.
The bath dynamics used in the present repeated-impact simulations are deactivated by setting $\tilde{F}=0$, while the rest of the numerical framework remains unchanged.

\begin{figure}
\centering
\begin{tabular}{cc}
\includegraphics[width=0.5\linewidth]{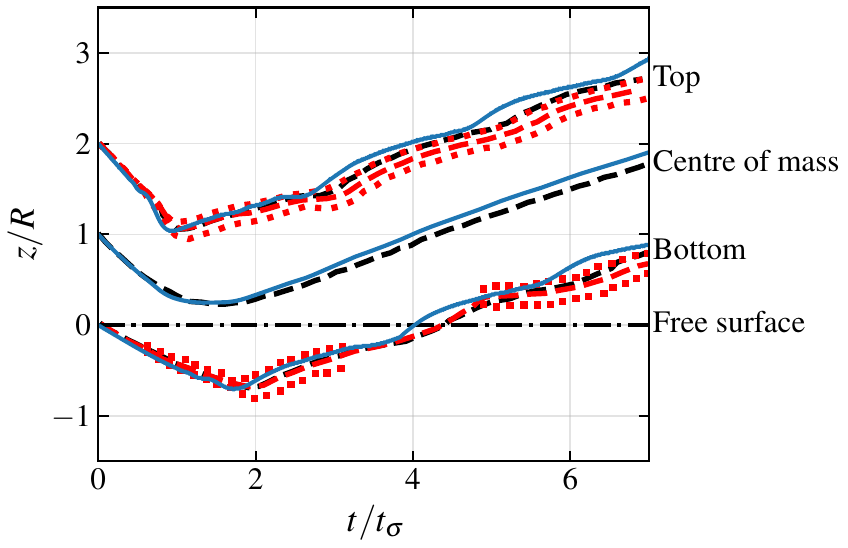} & 
\includegraphics[width=0.42\linewidth]{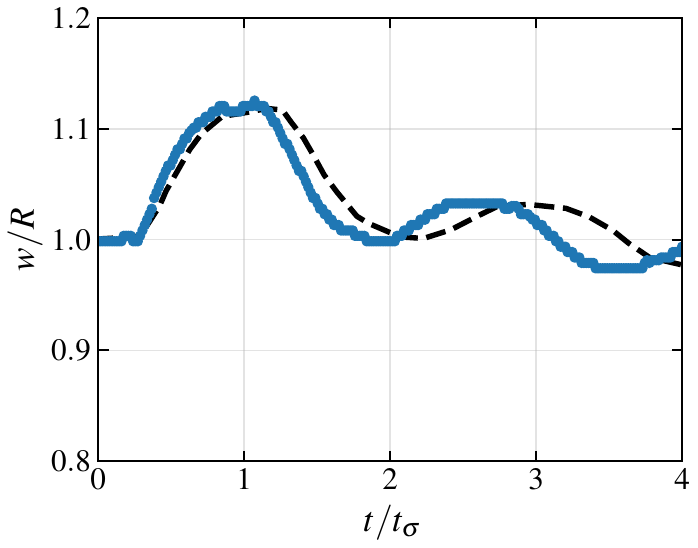}\\
(a)  & (b)
\end{tabular}
\caption{Validation of the present DNS against the experiments and DNS of \cite{alventosa2023inertio} for a deionized water droplet rebounding from a bath of the same liquid at $We=0.7$, $Bo=0.017$ and $Oh=0.006$. Time is normalized by the inertio-capillary scale $t_\sigma = \sqrt{\rho R^3/\sigma}$.  (a) Comparison of the droplet trajectory between the experiments (red dashed line, with dotted lines denoting the experimental uncertainty), the DNS of \cite{alventosa2023inertio} (black dashed line) and the present DNS (blue solid line). (b) Evolution of the maximum normalized droplet width, $w/R$, for the present DNS (blue dotted line) and the DNS of \cite{alventosa2023inertio} (black dashed line).
}
\label{fig:valid_Alventosa}
\end{figure}

Figure \ref{fig:valid_Alventosa} compares the present DNS with the experimental measurements and numerical results of \citet{alventosa2023inertio}. Figure \ref{fig:valid_Alventosa}a shows the trajectories of the droplet top, centre of mass, and bottom during impact and rebound, while figure \ref{fig:valid_Alventosa}b compares the evolution of the maximum droplet width. The present DNS reproduces both the measured rebound kinematics and the associated droplet deformation, with close agreement in the timing and amplitude of the rebound. This agreement provides validation that the numerical framework captures the dominant inertio-capillary dynamics governing droplet impact and rebound.

\subsection{Validation of the Faraday-wave response in the absence of a droplet}
\label{App A2}

\begin{figure}
\centering
\includegraphics[width=0.5\linewidth]{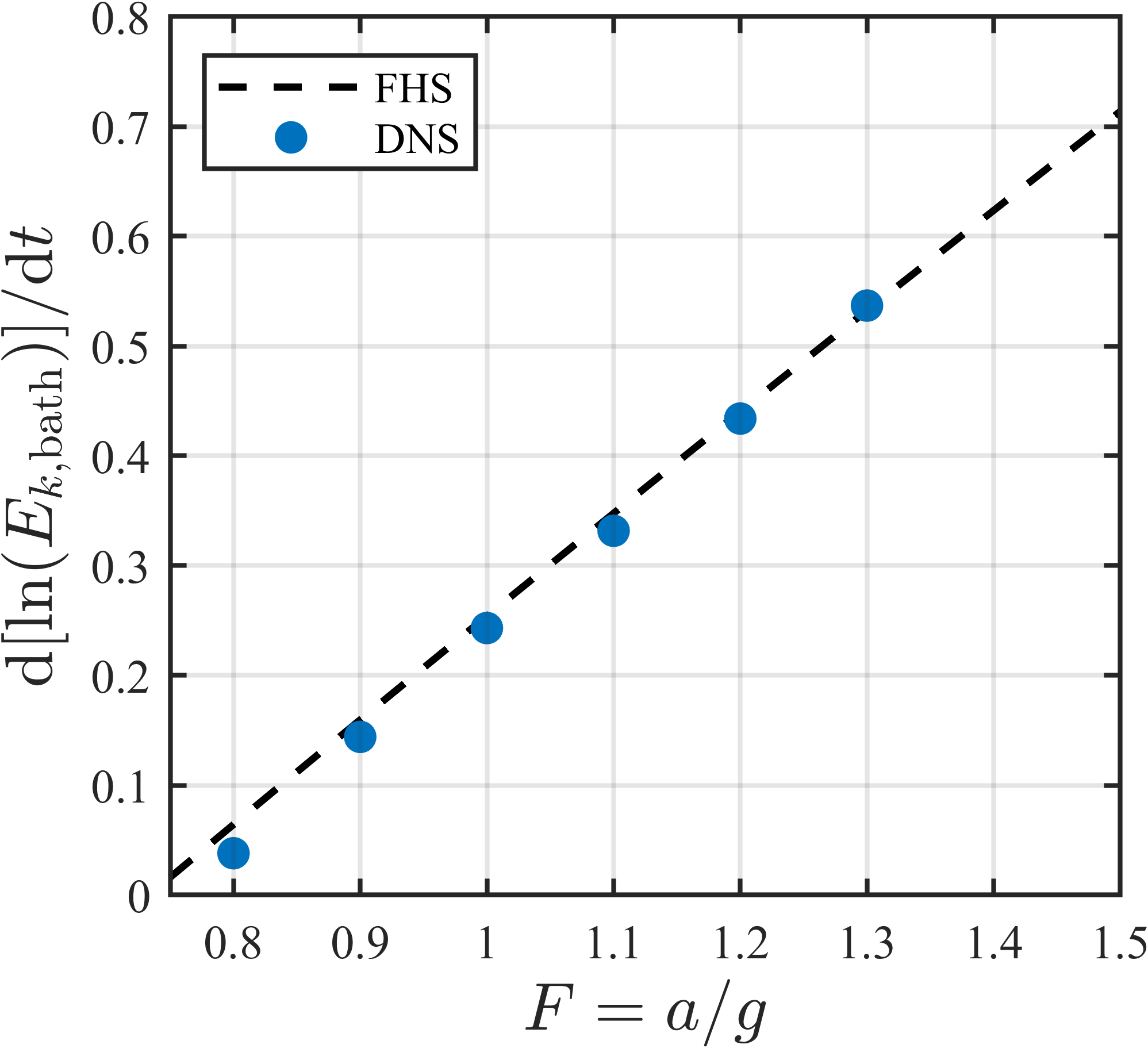}
\caption{Validation of the Faraday-wave response against the fully hydrodynamic system (FHS) formulation of \citet{Kumar_Tuckerman_1994}. The kinetic-energy growth rate, $~\mathrm{d} \left[\ln \left(E_{k,\mathrm{bath}}\right)\right] /\mathrm{d}t$, obtained from the DNS is compared with the Floquet prediction as a function of the dimensionless forcing acceleration $F=a/g$. }
\label{fig:KT94}
\end{figure}

We next validate the ability  of the numerical framework to reproduce the linear Faraday instability of a vertically vibrated liquid-gas interface. This test is important for the strongly forced droplet cases discussed in the main text, where the long-time
dynamics are no longer controlled only by the droplet--bath interaction, but also by the growth of parametrically excited waves on the bath. 
The growth rates obtained from DNS are therefore compared with the linear stability predictions of \citet{Kumar_Tuckerman_1994}.

The validation follows the two-fluid configuration used by
\citet{Kumar_Tuckerman_1994}. A silicone-oil layer of thickness
$h_1=14.5~\mathrm{mm}$, density $\rho_1=965~\mathrm{kg\,m^{-3}}$, and dynamic viscosity $\mu_1=0.02~\mathrm{kg\,m^{-1}\,s^{-1}}$ is overlaid by an air layer of thickness $h_2=29.5~\mathrm{mm}$, density $\rho_2=1.205~\mathrm{kg\,m^{-3}}$, and dynamic viscosity $\mu_2=1.82\times10^{-5}~\mathrm{kg\,m^{-1}\,s^{-1}}$. The
surface tension is $\sigma=0.02~\mathrm{kg\,s^{-2}}$. The bath is forced vertically with an acceleration amplitude $a$, and the forcing strength is reported as $F=a/g$.

For each value of $F$, a small-amplitude perturbation is imposed on the initially flat interface, and the growth rate is determined from the early-time evolution of the bath kinetic energy, while the disturbance remains within the linear regime. Specifically, the DNS kinetic-energy growth rate is obtained from $\ln(E_{k,\mathrm{bath}})$. The Floquet growth rates from the FHS are converted to the corresponding kinetic-energy growth rates before comparison with the DNS.


Figure~\ref{fig:KT94} compares the growth rates obtained from the present DNS with the fully hydrodynamic system  prediction of
\citet{Kumar_Tuckerman_1994}. Excellent agreement is obtained over the
range of forcing amplitudes considered, demonstrating that the numerical framework accurately captures the linear parametric amplification of the liquid--gas interface. This validation establishes that the bath dynamics associated with the Faraday instability are correctly represented independently of the presence of a droplet.

\subsection{Experimental work by \citet{molavcek2013drops}}
\label{App A3}

\begin{figure}
\centering
\begin{tabular}{c}
\includegraphics[width=0.45\linewidth]{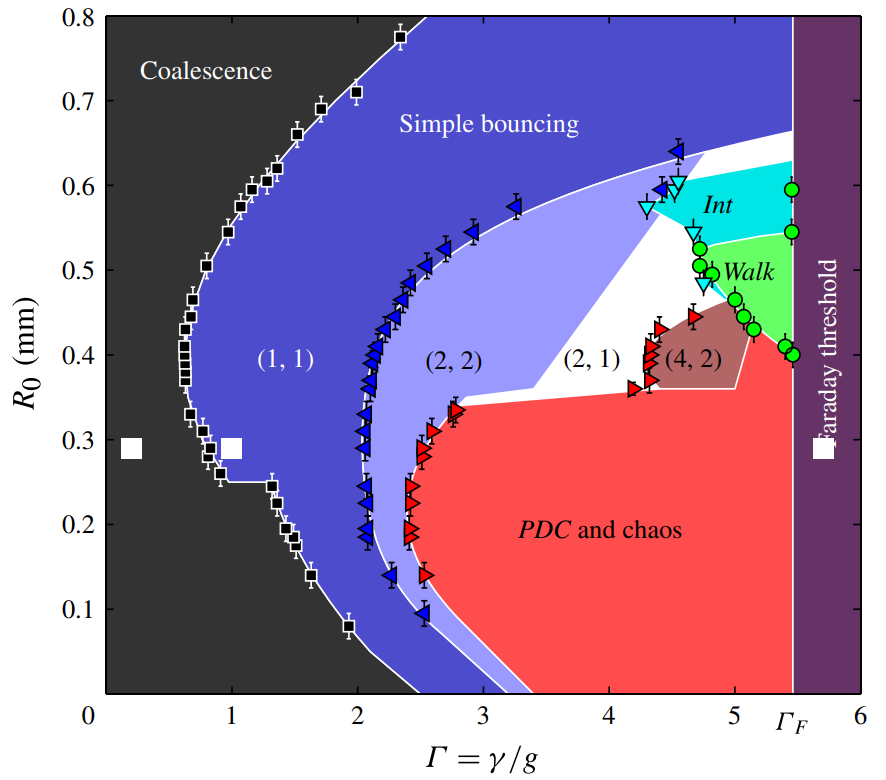}\\
(a) \vspace{8pt}\\
\end{tabular}
\begin{tabular}{cc}
\includegraphics[width=0.43\linewidth]{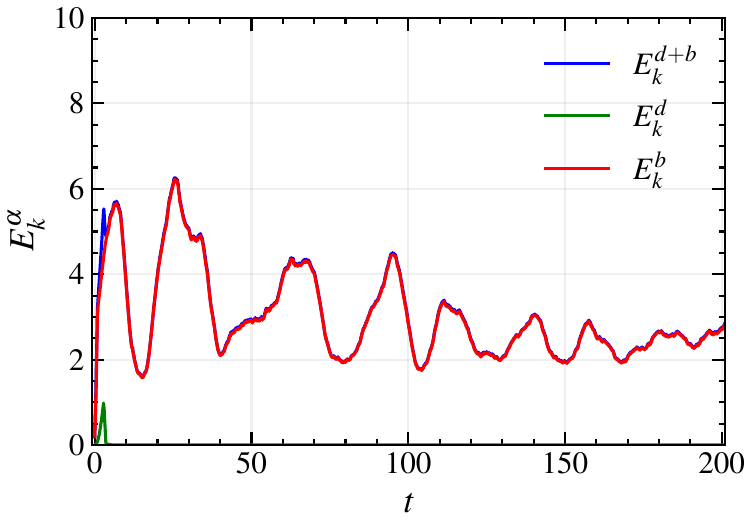}&  
\includegraphics[width=0.43\linewidth]{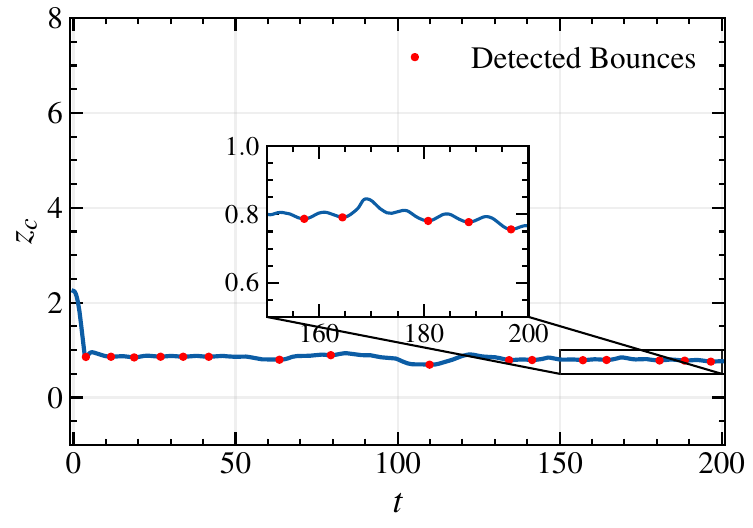}\\
(b) & (c) \vspace{8pt}\\
\includegraphics[width=0.43\linewidth]{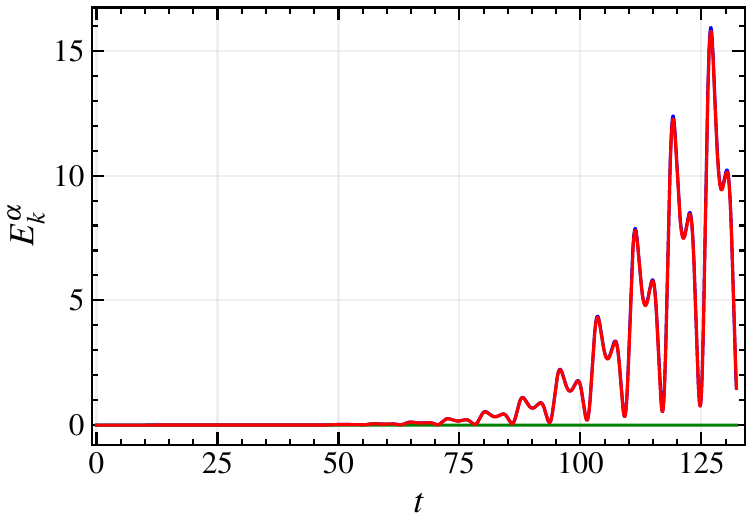}&  
\includegraphics[width=0.43\linewidth]{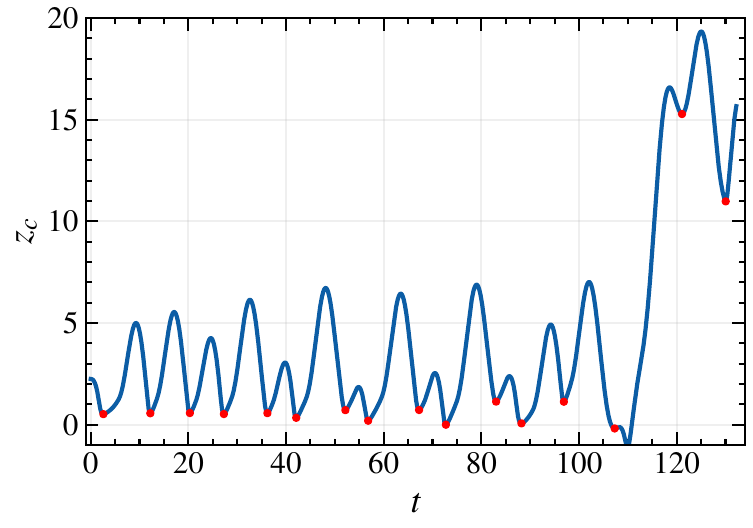}\\
(d) & (e) \\
\end{tabular}


\caption{
Validation against the bouncing-droplet regime map of \citet{molavcek2013drops}.
(a) Experimental regime map in the $(\Gamma,R_0)$ plane, where
$\Gamma=\gamma/g$ is the dimensionless bath acceleration and $R_0$ is the droplet radius. The white squares indicate the parameter values selected for the present DNS validation. 
Panels (b,c) correspond to case 1, located in the experimental coalescence region. 
Panels (d,e) correspond to case 2, selected in the strongly forced
Faraday-unstable region.
For (b-e), the left panel shows the temporal evolution of the
total, droplet, and bath kinetic energies, while the right panels  shows the vertical centroid position of the droplet and the detected rebound events.}
\label{fig:bush_validation1}
\end{figure}

\begin{figure}
\centering
\begin{tabular}{c}
\includegraphics[width=0.95\linewidth]{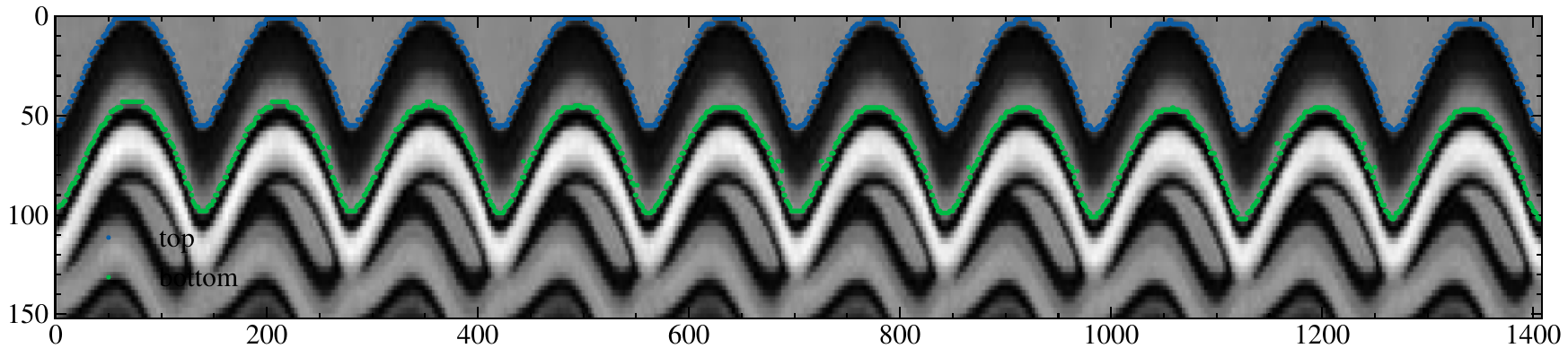}\\
(a) \\
\end{tabular}
\begin{tabular}{cc}
\includegraphics[width=0.45\linewidth]{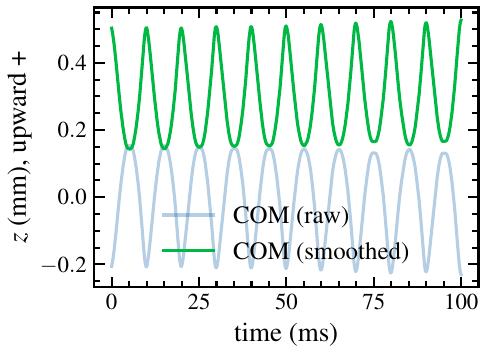} &  
\includegraphics[width=0.45\linewidth]{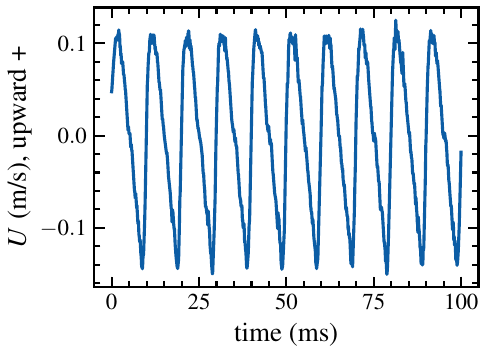}\\
(b) & (c)\\ 
\end{tabular}
\begin{tabular}{c}
\includegraphics[width=0.9\linewidth]{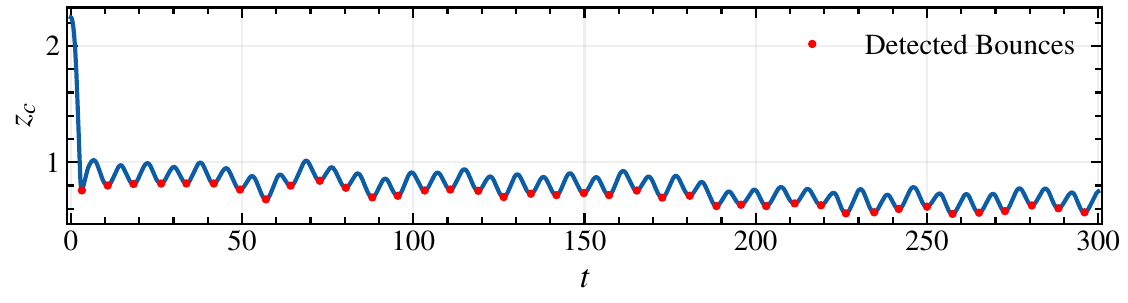}\\
(d)
\end{tabular}
\caption{
Validation of the sustained $(1,1)^1$ bouncing mode against the experiments of
\citet{molavcek2013drops}. 
(a) Processed space--time image from figure~4(a) of \citet{molavcek2013drops}, corresponding to a $50~\mathrm{cSt}$ silicone-oil droplet of radius $R_0=0.28~\mathrm{mm}$ bouncing on a bath vibrated at $f=50~\mathrm{Hz}$ and
$\Gamma=\gamma/g=1.3$. The coloured curves indicate the tracked upper droplet-edge locations used to extract the vertical motion.
(b) Raw and smoothed droplet-position signal obtained from the image processing. (c) Vertical velocity inferred from the  trajectory.  (d) DNS prediction of the vertical droplet-centroid position, with detected rebound events marked by red symbols. 
}
\label{fig:bush_validation2}
\end{figure}

To validate the ability of our numerical method  to reproduce the vertical dynamics of a bouncing droplet, we compare against the experiments of \citet{molavcek2013drops}. To the best of our knowledge, these experiments are 
among the few quantitative laboratory references reporting time-resolved vertical trajectories  of a millimetric droplet bouncing on a vertically vibrated 
liquid bath.

Figure~\ref{fig:bush_validation1}a shows the experimental regime map reported by \citet{molavcek2013drops} for droplets bouncing on a vertically vibrated bath of the same liquid. The regime map classifies the long-time droplet response in the $(\Gamma,R_0)$ plane, where $\Gamma$ is the dimensionless forcing acceleration $\Gamma=\gamma/g$. Periodic bouncing states are denoted by $(m,n)$, where the motion repeats after $m$ forcing periods and contains $n$ droplet--bath interactions. The remaining regions correspond to coalescence, period-doubling and chaos, intermittent horizontal motion, and sustained walking; $\Gamma_F$ denotes the Faraday threshold.
The white squares in figure~\ref{fig:bush_validation1}a indicate the parameter values used in our numerical validation. These points were chosen to sample the transition from coalescence to sustained bouncing and to verify that the simulations recover the experimentally observed $(1,1)$ bouncing phenomenology.

The first case, at low forcing $\Gamma=0.2$, lies within  the experimental coalescence region. We acknowledge that in the DNS, the droplet and bath are represented by two distinct color functions, so numerical coalescence is not allowed.
This case should consequently be seen as approaching the coalescence limit rather than reproducing coalescence itself. 
After the initial impact, the droplet centroid rapidly loses any clear periodic bouncing response and remains nearly stationary above  the bath interface (see figure \ref{fig:bush_validation1}c). Experimentally, such behavior would be followed by drainage of the intervening air film and eventual coalescence. Consistently, the kinetic energy evolution  shows no sustained rebound, and the
the bath kinetic energy decays toward a small baseline value, approaching the coalescence limit (see figure \ref{fig:bush_validation1}b).

The second case, at $\Gamma=5.7$, lies in the strongly forced region associated with Faraday instability of the bath. 
The bath kinetic energy initially remains small but subsequently grows rapidly, becoming virtually indistinguishable from the total kinetic energy and greatly exceeding the droplet contribution (figure~\ref{fig:bush_validation1}d). This growth indicates that the imposed forcing excites the bath itself, resulting in Faraday waves, and is consistent with the Faraday region identified in the experimental regime map.

The third case targets the simplest sustained bouncing mode.
It
corresponds to the  $(1,1)^1$ mode from  figure  4a of \citet{molavcek2013drops},
for a $50~\mathrm{cSt}$ silicone-oil droplet of radius
$R_0=0.28~\mathrm{mm}$ bouncing on a bath vibrated at $f=50~\mathrm{Hz}$ and $\Gamma=\gamma/g=1.3$. Since the published data are provided as a space--time image rather than as a tabulated centroid trajectory, we first perform image processing (see figure \ref{fig:bush_validation2}a) on the experimental strip to identify the upper location of the droplet. The extracted signal is then smoothed and differentiated to estimate the vertical velocity (see figure \ref{fig:bush_validation2}b-c) . 
The extracted  vertical velocity results in $U_0\simeq 0.1035 \mathrm{ms^{-1}}$. This value 
is consistent with a simple
ballistic estimate $U_b\simeq (2g\Delta z)^{1/2}$ based on the observed vertical excursion $\Delta z\simeq 0.35~\mathrm{mm}$, giving $U_b\simeq 0.08~\mathrm{ms^{-1}}$. This agreement provides an independent check that the image processing yields a 
consistent 
velocity scale.

Figure \ref{fig:bush_validation2}d shows the vertical droplet-centroid position, with detected rebound events marked by red symbols. The simulation recovers a sustained periodic rebound without transition to coalescence or bath-dominated Faraday motion, consistent with the experimentally observed $(1,1)^1$ response. The  dimensionless bouncing period obtained from the DNS is   $T_b\simeq 7.7$, which is in good agreement with the forcing period expressed in the present units,
$ T_f={(1/f)}/{(R_0/U_0)}\simeq 7.4 .$
Thus, the DNS reproduces the defining feature of the $(1,1)^1$ mode, which means  one complete bouncing cycle per forcing period.
We note that the successive extrema of the tracked visible droplet edge in figure \ref{fig:bush_validation2}a-b are separated by approximately 10 ms, giving a period of $T\simeq 3.8$. This shorter timescale represents a half-cycle of the visible edge motion associated with droplet deformation and should not be interpreted as the period of the complete $(1,1)^1$ bouncing state.

\section{Scaling estimate of the velocity transfer coefficient}
\label{App B}
\setcounter{equation}{0}
\renewcommand{\theequation}{B.\arabic{equation}}
\setcounter{table}{0}
\renewcommand{\thetable}{B\arabic{table}}
\setcounter{figure}{0}
\renewcommand{\thefigure}{B\arabic{figure}}

We derive a scaling estimate for the velocity-transfer coefficient $\chi$ used in the impact-equivalent perturbation model. This coefficient   relates the kinetic energy lost by the rebounding droplet to the energy imparted to a localized region of the bath.

Consider a spherical droplet of radius $R$, density $\rho_d$, and mass
$m_d=4/3\pi \rho_d R^3$ 
impacting the bath with velocity
$U_0$. 
Then we obtain the pre-impact kinetic energy as
\begin{equation}
    E_{in}=\frac{1}{2}m_d {U_0}^2
\end{equation}
We characterize the rebound using a coefficient of restitution, 
\begin{equation}
    e=\frac{U_r}{U_0}
\end{equation}
where $U_r$ denotes the rebound velocity of the droplet \citep{alventosa2023inertio, wu2020small}. The post-impact droplet kinetic energy is
\begin{equation}
    E_{reb}=\frac{1}{2}m_d {U_r}^2=\frac{1}{2}m_d e^2 {U_0}^2.
\end{equation}
However, not all of the pre-impact energy remains with the droplet after rebound. Part of it is transferred to the bath, which produces localized bath deformation.
The decrease in droplet translational kinetic energy,
\begin{equation}
    E_{loss}=E_{in}-E_{reb}=\frac{1}{2}m_d {U_0}^2 (1-e^2)
\end{equation}
is distributed among bath motion and deformation, droplet deformation, viscous dissipation and gas-film dynamics.
We therefore introduce an effective localization fraction $\eta$, representing the fraction of this energy associated with the localized bath response
\begin{equation}
    \eta=\frac{E_{bath}}{E_{loss}}
\end{equation}
which allows us to obtain the energy transferred to the bath as,
\begin{equation}
    E_{bath}= \eta \frac{1}{2}m_d {U_0}^2 (1-e^2)
\label{eq A.6}
\end{equation}
We assume the localized disturbance occupies $L=BR$, where $B$ denotes the region of contact of droplet with the bath.
Consequently, we approximate the effective mass of the bath with $m_b \sim \rho_b L^3$. If this bath region receives an initial velocity $u_i$, then the energy associated with this localized disturbance is estimated as,
\begin{equation}
    E_{bath} \sim \frac{1}{2}m_b {u_i}^2
\label{eq A.7}
\end{equation}
Since the energies are equivalent, we equate Eqs. \ref{eq A.6} and \ref{eq A.7},
\begin{equation*}
    \frac{1}{2}m_b {u_i}^2 = \eta \frac{1}{2}m_d {U_0}^2 (1-e^2)
\end{equation*}
where we determine the initial velocity $u_i$ as,
\begin{equation}
    u_i=U_0 \sqrt{\eta\frac{m_d}{m_b}(1-e^2)}
\label{eq A.8}
\end{equation}
Now we relate the incoming droplet velocity $U_0$ to the initial velocity $u_i$ as,
\begin{equation}
    u_i=\chi U_0
\end{equation}
where $\chi$ denotes the amount of velocity transferred to the bath. Substituting this relation in Eq. \ref{eq A.8} and expressing the droplet and bath masses using their densities $\rho_b$ and $\rho_d$, respectively, we deduce the velocity transfer coefficient as, 
\begin{equation}
    \chi =\sqrt{\eta \frac{4}{3}\pi\frac{\rho_d}{\rho_b}\frac{R^3}{L^3}(1-e^2)}
\end{equation}
For a system with both drop and bath composed of the same fluid ($\rho_b=\rho_d $) and assuming the footprint relation ($L=BR$) holds true, the aforementioned equation reduces to,
\begin{equation}
    \chi =\sqrt{\eta \frac{4}{3}\pi\frac{1}{B^3}(1-e^2)}
\label{eq A.11}
\end{equation}

The transfer efficiency is estimated to be $\eta=\mathcal{O}(10^{-2})$, motivated by measurements of millimetric droplets impacting liquid baths, for which approximately $1\%$ of the incident droplet kinetic energy is converted into gravity--capillary surface waves  \citep{craeye1999ring}.
The remaining impact energy is partitioned among droplet and bath deformation, bulk fluid motion, and viscous dissipation \citep{alventosa2023inertio}.

\begin{figure}
\centering
\begin{tabular}{ccc}
\includegraphics[width=0.32\linewidth]{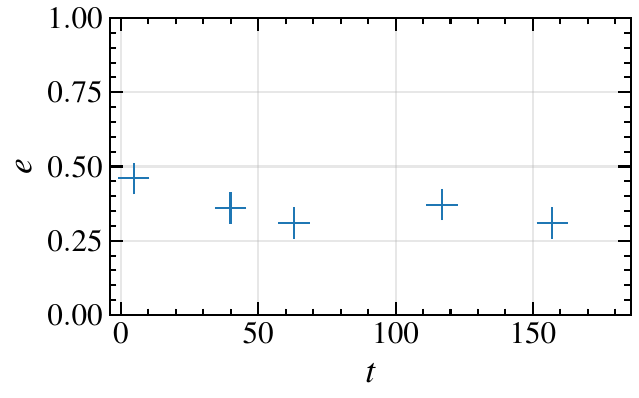} &
\includegraphics[width=0.32\linewidth]{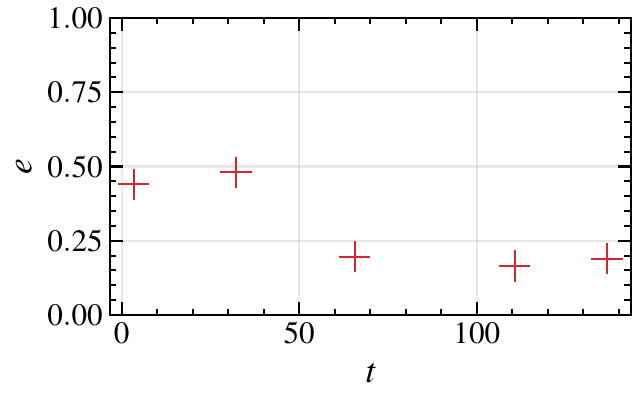} &  
\includegraphics[width=0.32\linewidth]{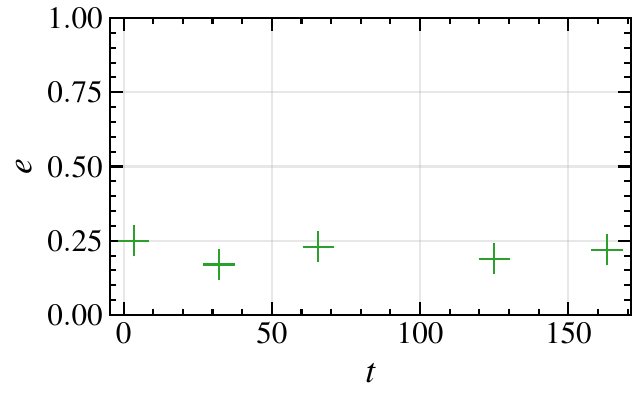} \\
(a) & (b) & (c)\\
\end{tabular}
\caption{Spatio-temporal evolution of coefficient of restitution for (a) $We=5.0 \; \mathrm{and} \; Re=200$, (b) $We=2.5 \; \mathrm{and} \; Re=200$, and (c) $We=2.0 \; \mathrm{and} \; Re=200$ 
}
\label{fig:restitution}
\end{figure}

The velocity-transfer coefficient $\chi$ is evaluated using Eq. \ref{eq A.11}, where the contact-region parameter $B$ is measured from the spatio-temporal pressure maps (see figures  \ref{fig: Pressure map_unstable} and \ref{fig: Pressure map_decaying}) and the coefficient of restitution $e$ is obtained from the post-impact droplet dynamics (see figure \ref{fig:restitution}). The resulting values of $B$, $e$, and $\chi$ corresponding to three representative cases are summarized in Table \ref{tab:transfer coefficient}.

The restitution coefficients, $e\simeq0.20-0.35$, are consistent with values reported for non-coalescing droplet–bath impacts, which approach $e\simeq 0.3$ in the inertio-capillary regime \citep{alventosa2023inertio, aguero2026droplet}. Since an equivalent velocity-transfer coefficient is not commonly defined, the $\mathcal{O}(1)$ values of $\chi$ should be interpreted as the characteristic velocity imparted to the localized bath region rather than as an energy-transfer fraction.

\begin{table}
\centering
\setlength{\tabcolsep}{15pt}
\begin{tabular}{ccccc}
$We$ & $Re$ & $B$ & $e$ & $\chi$\\
\hline
5.0 & 200  & 0.52 & 0.35 & 0.51\\
2.5 & 200  & 0.48 & 0.20 & 0.61\\
2.0 & 175 & 0.57 & 0.21 & 0.47\\
\end{tabular}
\caption{Contact-region parameter $B$, coefficient of restitution $e$, and velocity-transfer coefficient $\chi$ for the representative cases used in the  proposed perturbation-scaling model.
}
\label{tab:transfer coefficient}
\end{table}

\section{Derivation of energy budget and droplet diagnostics }
\setcounter{equation}{0}
\renewcommand{\theequation}{C.\arabic{equation}}

Here we summarize the diagnostic quantities used to characterize the droplet–bath dynamics. Section \ref{App C1} defines the phase-resolved and event-averaged energy contributions used in the impact analysis, while section \ref{App_droplet_measures} defines the droplet kinematics, trajectories, and shape measures to distinguish the different bouncing regimes.

\subsection{Phase-decomposed energy budget}
\label{App C1}

The bath and droplet can be represented by the VoF tracers $f_b$ and $f_d$, respectively. To distinguish the two liquid regions, we introduce a generic phase index $\alpha \in \{b,d\}$, where $b$ and $d$ denote the bath and droplet. The integrations are performed over the full axisymmetric computational domain $\Omega$, for which $dV=2\pi rdrdz$.
The instantaneous kinetic, gravitational potential, surface and viscous-dissipation contributions of phase $\alpha$  are 
\begin{equation}
    E_k^\alpha({t})=\int_\Omega f_\alpha \frac{\rho_\alpha}{2} \left(u_r^2+u_z^2 \right) 2 \pi r ~ dr~ dz,
\label{Eq: Ke}
\end{equation}
\begin{equation}
    E_g^\alpha({t})=\int_\Omega f_\alpha {\rho_\alpha} gz ~2 \pi r~ dr~ dz,
\end{equation}
\begin{equation}
    E_\sigma^\alpha({t})=\sigma_\alpha \int_{\Gamma_\alpha ({t})} 2 \pi r~ dl,
\end{equation}
and 
\begin{equation}
    \varepsilon_\mu^\alpha({t})=\int_\Omega f_\alpha 2\mu_\alpha {\mathcal{{D}}:\mathcal{{D}}} ~2 \pi r~ dr~ dz,
    \label{eq: dissipation}
\end{equation}
where $u_r$ and $u_z$ denote the radial and axial velocity components, respectively, and $dl$ is the differential arc length of the axisymmetric interface. $\mathcal{{D}}$ represents the rate-of-strain tensor, $\mathcal{{D}} = 1/2 \left(\nabla \boldsymbol{u} + \nabla \boldsymbol{u}^T \right)$. The dissipation rate  $\varepsilon_\mu^\alpha({t})$ is denoted with a negative sign in the energy budget to represent irreversible mechanical-energy loss. 
For a droplet impact event $m$, evaluated over $t \in \left[t_a^{(m)},t_b^{(m)} \right]$,
\begin{equation}
    \Delta t_m=t_b^{(m)}-t_a^{(m)},
\end{equation}
The event-averaged stored energy is defined as
\begin{equation}
    \overline{E}_j^{\alpha,m} = \frac{1}{\Delta {t}_m} \int_{{t}_{a}^{(m)}}^{{t}_{b}^{(m)}} E_j^{\alpha} ({t}) d{t}, \qquad j \in ({k,g,\sigma}).
\end{equation}
where $k$, $g$, and $\sigma$ indicate the kinetic, gravitational-potential and interfacial energies. For the gravitational-potential and interfacial energies, the large equilibrium contributions are removed relative to a reference time $t_r^{(m)}$ immediately preceding impact. Thus, we define the impact-induced energy variation
\begin{equation}
\Delta E_j^{\alpha,m}(t)
=
E_j^{\alpha}(t)-E_j^{\alpha}\left(t_r^{(m)}\right),
\qquad
j\in\{g,\sigma\},
\end{equation}
and the corresponding event-averaged variation is
\begin{equation}
    \Delta\overline{E}_j^{\alpha,m} = \frac{1}{\Delta {t}_m} \int_{{t}_{a}^{(m)}}^{{t}_{b}^{(m)}} \left[E_j^{\alpha} ({t}) - E_j^{\alpha} \left({t_r^{(m)}} \right) \right]  d{t}, \qquad j \in ({g,\sigma}).
\end{equation} 
while the kinetic energy is retained in absolute form. The cumulative viscous energy dissipated during the impact event $m$ is
\begin{equation}
  {E}_\mu^{\alpha,m} = \int_{{t}_{a}^{(m)}}^{{t}_{b}^{(m)}} \varepsilon_\mu^{\alpha} ({t}) d{t},
\end{equation}

\subsection{Droplet kinematic and shape diagnostics}
\label{App_droplet_measures}

The temporal droplet kinematic and shape diagnostics are assessed from the droplet VoF field $f_d$ over the axisymmetric computational domain $\Omega$, thus yielding the instantaneous droplet volume 
\begin{equation}
    V_d(t)=\int_{\Omega} f_d dV.
\end{equation}

As the droplet density is uniform, its center of mass coincides with the volume centroid. Thus, the axial centroid position is defined by
\begin{equation}
    z_c(t)=\frac{1}{V_d(t)}\int_{\Omega} f_d z dV,
\end{equation}
and the associated axial centroid velocity is
\begin{equation}
    U_c(t)=\frac{1}{V_d(t)}\int_{\Omega} f_d u_z dV.
\end{equation}
The deformation of the droplet is defined using its second axial and radial central moments,
\begin{equation}
    M_{20}(t)=\frac{1}{V_d(t)}\int_{\Omega} f_d \left(z-z_c \right)^2 dV,
\end{equation}
and
\begin{equation}
    M_{02}(t)=\frac{1}{V_d(t)}\int_{\Omega} f_d r^2 dV.
\end{equation}
For an axisymmetric ellipsoid with axial and radial semi-axes $a_z$ and $a_r$, these moments satisfy $M_{20}=a_z^2/5$ and $M_{02}=2a_r^2/5$, respectively. Thus, the equivalent axial and radial semi-axes are defined as
\begin{equation}
    a_z=\sqrt{5M_{20}}, \quad
    a_r=\sqrt{\frac{5}{2}M_{02}}.
\end{equation}
Defining $a_{\mathrm{max}}=\max \left(a_z, a_r \right)$ and $a_{\mathrm{min}}=\min \left(a_z, a_r \right)$ gives the droplet ellipticity
\begin{equation}
    \varepsilon(t)=\sqrt{1 - \left( \frac{a_{\min}}{a_{\max}} \right)^2}.
\end{equation}
where $\varepsilon=0$ indicates a spherical droplet, while increasing $\varepsilon$ implies greater deformation.

\section{Mesh study: numerical resolution of the thin air layer}
\label{App D}

\setcounter{figure}{0}
\renewcommand{\thefigure}{D\arabic{figure}}
\setcounter{table}{0}
\renewcommand{\thetable}{D\arabic{table}}

\begin{figure}
\centering
\begin{tabular}{c}
(a) $We=1.21$ and $Re=175$ \\
\end{tabular}
\begin{tabular}{cc}
\includegraphics[width=0.47\linewidth]{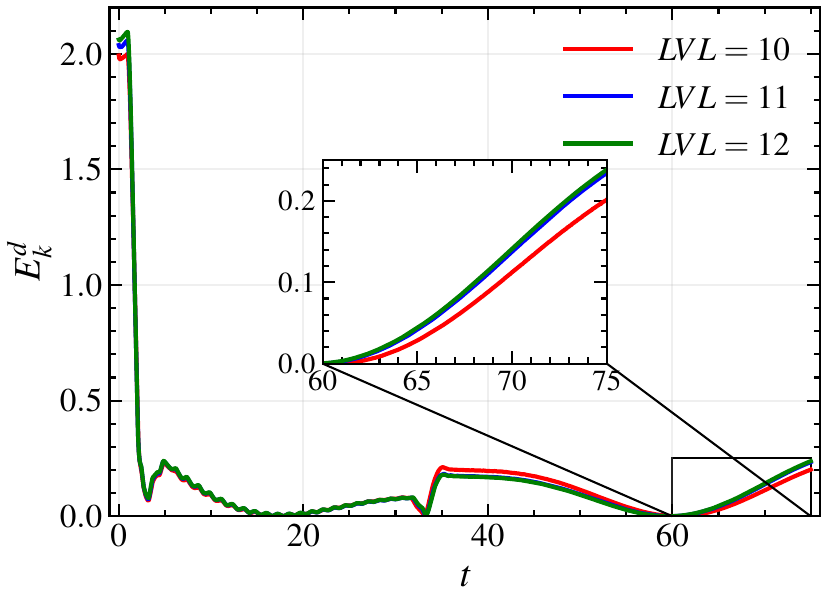}  & 
\includegraphics[width=0.47\linewidth]{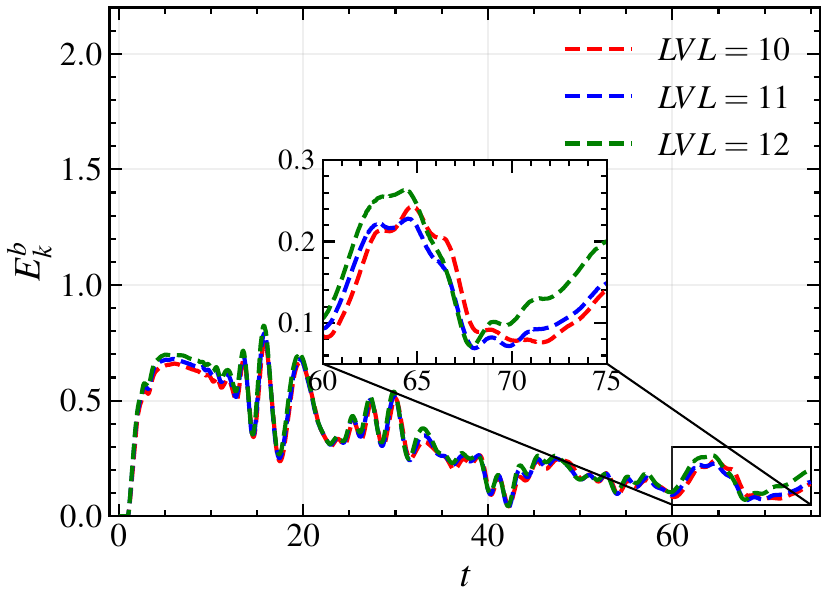} \\
\end{tabular}
\begin{tabular}{c}
(b) $We=2.5$ and $Re=200$ \\
\end{tabular}
\begin{tabular}{cc}
\includegraphics[width=0.47\linewidth]{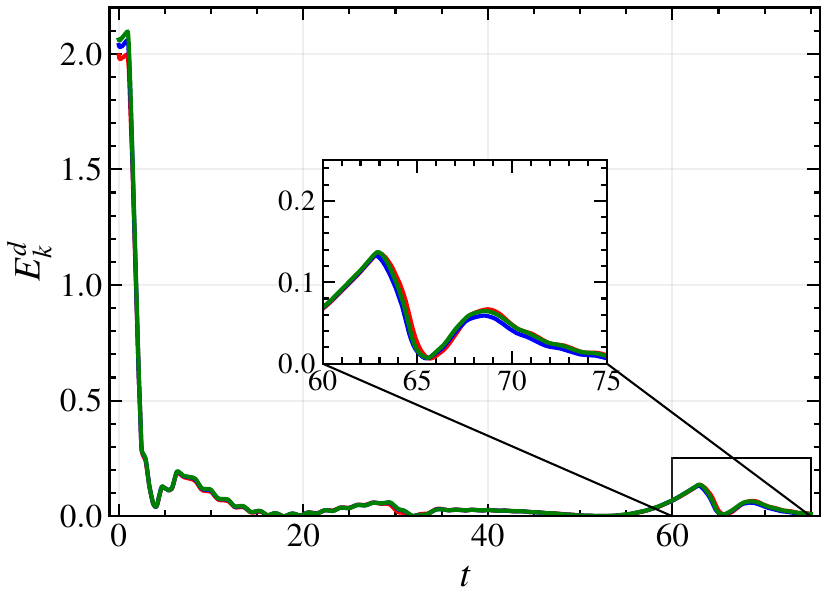}  &  
\includegraphics[width=0.47\linewidth]{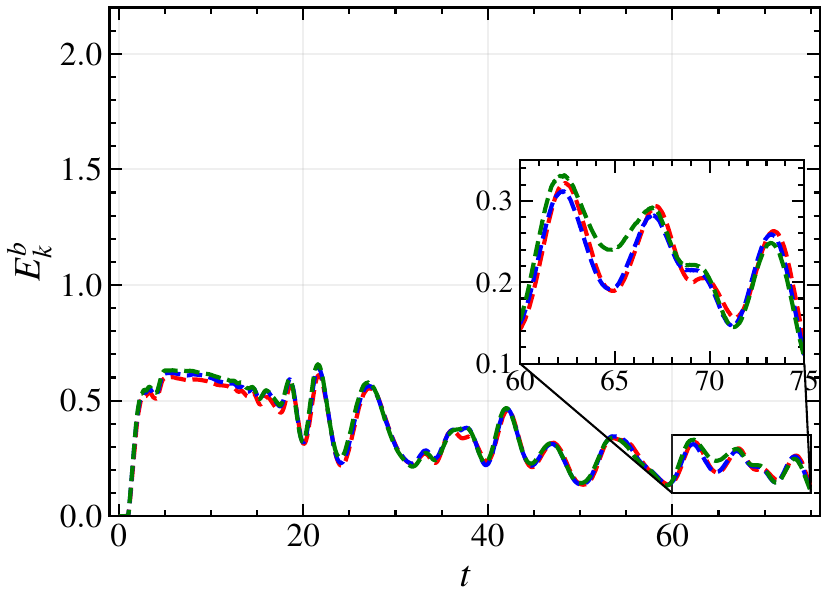} \\
\end{tabular}
\caption{Grid independence study 
for decaying  ($We=1.21$ and $Re=175$) and unstable  ($We=2.5$ and $Re=200$) regimes, corresponding to top and bottom panels, respectively.  The dynamics are shown for the first two bounces
for three different grid levels ($LVL10$, $LVL11$, and $LVL12$). The left panels show the droplet-based kinetic energies, while the right panels represent the bath-based kinetic energies, with insets highlighting the later-stage dynamics across three grid levels. }
\label{fig:Grid_study}
\end{figure}

To assess the resolution for the current simulations, we perform a grid-independent study using three maximum refinement levels, $LVL=10,~11$, and $12$, for two representative regimes, unstable and decaying. Figure \ref{fig:Grid_study} compares the droplet and bath kinetic energy histories over the first two rebound events. For both regimes, the three resolutions generate closely matching energy histories. Only minor differences appear in the post-second rebound in the bath-based kinetic energy, highlighted by the low-amplitude oscillations in the insets.

The corresponding computational cost associated with each refinement level is outlined in table \ref{tab:comp_cost}. Core-hours are computed as the elapsed wall-clock time multiplied by the eight allocated CPU cores. For the given simulation time ($t=75$), increasing the refinement from $LVL=11$ to $LVL=12$ raises the wall-clock time from 6.7 to 39.3 h, corresponding to 53.6 and 314.5 core-hours, respectively. Therefore, the finest resolution is approximately six times more expensive while showing only minor changes in the kinetic energy histories illustrated in figure \ref{fig:Grid_study}. Assuming linear scaling with simulated time, a production run up to $t=600$ would require 54 h of wall-clock time ($\sim 429$ core-hours) at $LVL=11$, compared with approximately 314 h  ($\sim 13$ days; $\sim 2516$ core-hours) at $LVL=12$. Considering both the close agreement in figure \ref{fig:Grid_study} and the significantly greater computational expense of the finest grid, $LVL=11$ has been adopted for the present long-time simulations reported in this work.

We next focus on the numerical resolution of the thin gas layer separating the bath and the droplet. In our simulations, the final drainage and rupture of the thin gas layer is not explicitly resolved. Instead, the droplet and bath are represented by two distinct color fractions, $f_d$ and $f_b$, respectively.  This treatment is motivated by the large scale of the domain in comparison to the
physical thickness of the terminal air film. In the present simulations, a  drop radius is $R = 0.035~\mathrm{cm}$, with a  computational domain of length  $L_0=20R$. The finest grid spacing at refinement level $LVL$ is therefore
$ \Delta = {L_0}/{2^{LVL}}.$ For the three levels used in the grid study, this gives
$\Delta_{10} \simeq 6.84~\mu\mathrm{m}$,
$\Delta_{11} \simeq 3.42~\mu\mathrm{m}$, and
$\Delta_{12} \simeq 1.71~\mu\mathrm{m}$.


\begin{table}
\centering
\setlength{\tabcolsep}{15pt}
\begin{tabular}{cccc}
Grid refinement level & Dimensionless time & Wall-clock time (h) & Core-hours\\
\hline
$LVL10$ & ${t}=75$  & 2.31 & 18.46\\
$LVL11$ & ${t}=75$  & 6.70 & 53.60\\
$LVL12$ & ${t}=75$  & 39.31 & 314.46\\
\end{tabular}
\caption{Computational cost for the three different grid levels at $We=2.5$ and $Re=200$, using eight MPI processes with one CPU per process. The $LVL12$ value includes two consecutive jobs required to reach $t=75$.}
\label{tab:comp_cost}
\end{table}

Previous studies of drop impact onto liquid pools have shown that the gas film reaches micrometric or submicrometric thicknesses before rupture. In the drop--pool calculations of \citet{Hendrix2016}, rupture was  imposed at a minimum air-film thickness of $0.4~\mu\mathrm{m}$, and a comparison  with $0.2~\mu\mathrm{m}$ showed little change in the entrapped bubble volume. Earlier experimental  work on same-liquid drop--pool impact by \citet{Tran2013} also identified the  trapped air film as the controlling feature during the short time preceding rupture.
In the drop–substrate experiments by \citet{Lee}, the initial film thickness was approximately $1$–$1.9 \mu\mathrm{m}$.

The corresponding number of grid cells across an air layer of thickness $h_a$ is $N_a = {h_a}/{\Delta}$. At the finest level used here, $LVL=12$, this gives $N_a(1.9~\mu\mathrm{m}) \simeq 1.11$ grid cells. Although this is consistent with the order of magnitude reported in the literature, it is insufficient to resolve the microscopic  rupture dynamics of the air layer; resolving such a film would require at least $\mathcal{O}(10)$ cells across the film thickness.
Accordingly, the focus of this work is not to provide a fully resolved model of air-film drainage or rupture but instead to investigate the macroscopic droplet-bath response over successive impacts. Since the rupture process is not explicitly resolved, no coalescence criterion is introduced. Once the air layer becomes thinner than the grid scale, the two-color formulation effectively imposes non-coalescence unless an explicit rupture or merging model is introduced. This strategy also enables us to investigate the long-time dynamics efficiently.

Finally, we note that the maximum refinement level used in the present simulations is consistent with previous DNS studies of droplet rebound from a liquid bath. In particular, \citet{alventosa2023inertio} and \citet{galeano2021capillary} performed axisymmetric Basilisk simulations in a $20R\times20R$ domain using adaptive mesh refinement up to $LVL=11$. 
There is an important distinction in scope. Both of the studies focused primarily on single-impact dynamics, for which very fine resolution can be concentrated over a relatively short physical time. In contrast, the present work follows repeated bouncing events and the cumulative droplet--bath response over many impact cycles. Long-duration simulations at $LVL=12$ would require approximately
314 wall-clock hours ($\sim 13$ days, or $\sim 2516$ core-hours) per case, making a systematic exploration of the regime map impractical. We therefore use $LVL=11$, supported by the close agreement of the energy histories in figure~\ref{fig:Grid_study}.

\section{Axisymmetric Floquet stability formulation for a  cylindrical bath}
\label{App E}
\setcounter{equation}{0}
\renewcommand{\theequation}{E.\arabic{equation}}
\setcounter{table}{0}
\renewcommand{\thetable}{E\arabic{table}}
\setcounter{figure}{0}
\renewcommand{\thefigure}{E\arabic{figure}}

In this appendix, we  extend the 
hydrodynamic stability formulation of \citet{Kumar_Tuckerman_1994} 
to a finite axisymmetric case.
In the original formulation, the horizontal dependence is represented by Fourier modes with a continuous wavenumber. Here we replace these modes by the discrete Bessel eigenfunctions admitted by our cylindrical bath forming a $90^\mathrm{o}$ contact angle at the sidewall. We consider the vertical structure, Floquet expansion, and viscous interfacial conditions directly from the full hydrodynamic system (FHS) of \citet{Kumar_Tuckerman_1994}.

\subsection{Geometry and eigenfunctions}
We consider two immiscible, incompressible fluid layers contained within a cylindrical domain of radius $R_c$. Fluid 1  occupies the bottom layer, $-h_1 \le z \le 0$, and fluid 2 resides above it, $0 \le z \le h_2$. Their densities, dynamic viscosities and kinematic viscosities are denoted by $\rho_j$, $\mu_j$, and $\nu_j=\mu_j/\rho_j$, respectively, with $j=$1, 2.  In the reference frame of the vertically oscillating cylinder, the effective gravitational acceleration is
\begin{equation}
    G(t)=g-f(t), 
\end{equation}
where $f(t)$ denotes the imposed vertical acceleration. The unperturbed interface is located at $z=0$, and its small displacement is given by the separated form $\eta(r,t)=\hat{\eta}(t) \Phi(r)$, where the radial eigenfunction fulfills
\begin{equation}
    \frac{1}{r} \frac{d}{dr} \left(r \frac{d \Phi}{dr} \right)=-\kappa^2 \Phi.
\end{equation}
Setting regularity at the axis ($r=0$) gives
\begin{equation}
    \Phi(r) = J_0(\kappa r),
\end{equation}
Therefore, the interfacial displacement assumes the form
\begin{equation}
    \eta(r,t)=\hat{\eta}(t) J_0(\kappa r).
\end{equation}
At the cylindrical wall ($r=R_c$), the radial velocity follows impermeability $u_r=0$, while the tangential stress vanishes, leading to $\tau_{rz}=0$. In addition, the interface satisfies a static contact angle of $90^\mathrm{o}$. Under the small-slope approximation, this gives
\begin{equation}
    \frac{\partial \eta}{\partial r} \bigg|_{r = R_c} = 0
\label{Eq. contact angle}
\end{equation}
Since $\partial \eta/ \partial r = -\kappa \hat{\eta}(t) J_1(\kappa r)$, the admissible radial wavenumbers fulfill
\begin{equation}
    J_1(\kappa_p R_c) = 0, \qquad \kappa_p=\frac{\beta_p}{R_c}, \qquad p = 1,2,....,
\end{equation}
where $\beta_p$ is the $p$th positive zero of $J_1$. Hence, the present cylindrical bath admits the discrete radial modes $\kappa_1, \kappa_2, ..., \kappa_p$, unlike the continuous wavenumber $k$ in the horizontally unbounded \citet{Kumar_Tuckerman_1994} formulation.
Condition \ref{Eq. contact angle} fixes the interfacial slope but not the contact-line elevation. The contact line thus may slide vertically while maintaining the given contact angle, so that, 
\begin{equation}
    \eta(R_c,t) \ne 0.
\end{equation}
The uniform mode $\kappa_0=0$ is excluded because it defines a spatially uniform displacement of the interface. Therefore each non-zero mode satisfies

\begin{equation}
    \int_0^{R_c} r J_0(\kappa_p r)dr=\frac{R_c}{\kappa_p}J_1(\kappa_p R_c)=0,
\end{equation}
so that each admissible mode preserves the volume of the two fluid layers to linear order.

\subsection{Axisymmetric full hydrodynamic system}
For a radial mode $p$, the interface displacement and vertical velocity are written as
\begin{equation}
    \eta_p(r,t)=\hat{\eta}_p(t) J_0(\kappa_p r), \qquad w_{jp}(r,z,t)=\hat{w}_{jp}(z,t) J_0(\kappa_p r).
\end{equation}
Using the identity,
$$
    \frac{1}{r} \frac{\partial}{\partial r} \left(r \frac{\partial J_0(\kappa_p r)}{\partial r} \right)=-\kappa_p^2 J_0(\kappa_p r),
$$
the linearized vertical-velocity equation becomes
\begin{equation}
    \left[\frac{\partial}{\partial t}-\nu_j \left(\frac{\partial^2}{\partial z^2} - \kappa_p^2 \right) \right] \left(\frac{\partial^2}{\partial z^2} - \kappa_p^2 \right) \hat{w}_{jp}=0.
\label{Eq:Gov_eq_floq}
\end{equation}
Since both fluid layers are finite, impermeable free-slip conditions are imposed  
at the lower and upper boundaries, yielding
\begin{equation}
    \hat{w}_{1p}=\frac{\partial \hat{w}_{1p}}{\partial z}=0 \qquad \mathrm{at} ~~z=-h_1,
\label{Eq:floquet_BC1}
\end{equation}
and
\begin{equation}
    \hat{w}_{2p}=\frac{\partial \hat{w}_{2p}}{\partial z}=0 \qquad \mathrm{at} ~~z=h_2.
\end{equation}
At the interface ($z=0$), we define the jump of any quantity $\phi$ as
$\Delta\phi=\phi_2-\phi_1$, where subscripts 1 and 2 denote the lower and upper fluid layers, respectively. Continuity of the vertical and radial velocities then gives
\begin{equation}
    \Delta\hat{w}_{p}=0, \qquad \Delta\left(\frac{\partial \hat{w}_{p}}{\partial z} \right)=0,
\label{Eq. No-slip}
\end{equation}
and continuity of tangential stress condition gives
\begin{equation}
    \Delta \left[\mu  \left(\frac{\partial^2}{\partial z^2} + \kappa_p^2 \right) \hat{w}_{p}\right] = 0.
\end{equation}
The kinematic condition is set at the interface by
\begin{equation}
    \frac{\partial \hat{\eta}_{p}}{\partial t}=\hat{w}_{1p}(0,t)=\hat{w}_{2p}(0,t).
\end{equation}
Finally, the jump in the normal component of the stress tensor is
\begin{equation}
    \Delta \left[ \left\{ \rho \frac{\partial}{\partial t}
- \mu \left( \frac{\partial^2}{\partial z^2} - \kappa_p^2 \right)
+ 2\mu \kappa_p^2 \right\}
\frac{\partial \hat{w}_p}{\partial z} \right]
= - \left[ \Delta \rho G(t) - \sigma \kappa_p^2 \right] \kappa_p^2 \hat{\eta}_p.
\label{Eq. Strees jump}
\end{equation}
where $\sigma$ denotes the interfacial tension. The governing equation \ref{Eq:Gov_eq_floq}, along with the boundary and interfacial conditions
\ref{Eq:Gov_eq_floq} -- \ref{Eq. Strees jump}, constitutes the axisymmetric fully hydrodynamic system (FHS) for radial mode $p$.

\subsection{Floquet expansion and vertical structure}
The interface displacement is expanded in Floquet form as
\begin{equation}
    \eta_{p}(r, t) = J_0(\kappa_p r)e^{(s + i\alpha)t}
\sum_{n=-\infty}^{\infty} \xi_{pn}e^{in\omega t},
\end{equation}
and the vertical velocity in fluid $j$ as
\begin{equation}
    w_{jp}(r, z, t) = J_0(\kappa_p r)e^{(s + i\alpha)t}
\sum_{n=-\infty}^{\infty} w_{jpn}(z)e^{in\omega t}.
\end{equation}
where $s$ is the instability growth rate, $\omega$ is the angular frequency of the oscillation, and $n$ is the temporal Fourier index. Harmonic and subharmonic responses are given by $\alpha=0$ and $\alpha={\omega}/{2}$, respectively. The neutral stability is obtained by setting $s=0$. Defining $s_n=s+i(\alpha+n\omega)$, the vertical velocity condition satisfies
\begin{equation}
    \left[s_n-\nu_j \left(\frac{d^2}{d z^2} - \kappa_p^2 \right) \right] \left(\frac{d^2}{d z^2} - \kappa_p^2 \right) {w}_{jpn}=0,
\end{equation}

with the general solution
\begin{equation}
    w_{jpn}(z) = a_{jpn} e^{\kappa_p z} + b_{jpn} e^{-\kappa_p z}
+ c_{jpn} e^{q_{jn} z} + d_{jpn} e^{-q_{jn} z},
\label{Eq. Gen_sol}
\end{equation}
where
\begin{equation}
    q_{jpn}^2=\kappa_p^2+\frac{s_n}{\nu_j}, \qquad \Re(q_{jpn}) > 0.
\end{equation}

\subsection{Boundary conditions for the finite fluid layers}

The indices $p$ and $n$ are dropped on the velocity coefficients for compactness, while we retain $\kappa=\kappa_p$, $q_j=q_{jpn}$, $s_n=s+i(\alpha+n \omega)$, and $\xi=\xi_{pn}$. For each $(p, n)$, the eight coefficients $a_1, b_1, c_1, d_1, a_2, b_2, c_2, d_2$ are determined in terms of $\xi$ by these eight equations. The kinematic condition gives
\begin{equation}
    a_1+b_1+c_1+d_1=s_n\xi.
\label{Eq 8_1}
\end{equation}

Continuity of vertical velocity, radial velocity, and the tangential stress at the interface gives
\begin{equation}
    a_1+b_1+c_1+d_1=a_2+b_2+c_2+d_2.
\end{equation}
\begin{equation}
    \kappa(a_1-b_1)+q_1(c_1-d_1)=\kappa(a_2-b_2)+q_2(c_2-d_2).
\end{equation}
\begin{equation}
    \mu_1[2\kappa^2(a_1+b_1)+(q_1^2+\kappa^2)(c_1+d_1)]=\mu_2[2\kappa^2(a_2+b_2)+(q_2^2+\kappa^2)(c_2+d_2)].
\end{equation}
At the lower finite boundary, $z=-h_1$,
\begin{equation}
 a_{1} e^{-\kappa h_1} + b_{1} e^{\kappa h_1}
+ c_{1} e^{-q_{1} h_1} + d_{1} e^{q_{1} h_1}=0,
\end{equation}
\begin{equation}
 \kappa \left(a_{1} e^{-\kappa h_1} - b_{1} e^{\kappa h_1} \right)
+ q_1 \left(c_{1} e^{-q_{1} h_1} - d_{1} e^{q_{1} h_1}\right)=0.
\end{equation}
At the upper finite boundary, $z=h_2$,

\begin{equation}
 a_{2} e^{\kappa h_2} + b_{2} e^{-\kappa h_2}
+ c_{2} e^{q_{2} h_2} + d_{2} e^{-q_{2} h_2}=0,
\end{equation}
\begin{equation}
 \kappa \left(a_{2} e^{\kappa h_2} - b_{2} e^{-\kappa h_2} \right)
+ q_2 \left(c_{2} e^{q_{2} h_2} - d_{2} e^{-q_{2} h_2}\right)=0.
\label{Eq 8_8}
\end{equation}
The set of equations \ref{Eq 8_1} -- \ref{Eq 8_8} form an $8 \times 8$ complex linear system that expresses all velocity coefficients as linear multiples of $\xi_{pn}$, which are the finite layer counterparts of the boundary conditions presented in Appendix B of \citet{Kumar_Tuckerman_1994}.

\subsection{Stress jump condition and marginal stability}
Using the general solution \ref{Eq. Gen_sol} at the interface ($z=0$),
\begin{equation}
    w'_{jpn}(0) = \kappa_p (a_{jpn} - b_{jpn}) + q_{jpn}(c_{jpn} - d_{jpn}),
\end{equation}
and 
\begin{equation}
    w'''_{jpn}(0) = \kappa_p^3 (a_{jpn} - b_{jpn}) + q_{jpn}^3(c_{jpn} - d_{jpn}),
\end{equation}
Therefore, the $n$th Fourier component of the normal-stress condition is 
\begin{equation}
    \Delta \left[ (\rho_j s_n + 3\mu_j \kappa_p^2) w'_{jpn}(0) - \mu_j w'''_{jpn}(0) \right]
+ (\Delta \rho g - \sigma\kappa_p^2) \kappa_p^2 \xi_{pn}
= \Delta \rho \kappa_p^2 [f(t) \eta_p(t)]_n.
\label{Eq. stress_jump}
\end{equation}
For the single-frequency forcing $f(t)=a ~\mathrm{cos}(\omega t)$, we get
\begin{equation}
    [f(t)\eta_p(t)]_n = \frac{a}{2}(\xi_{p,n-1} + \xi_{p,n+1}).
\end{equation}

After solving the finite-layer velocity problem, the left-hand side of
(\ref{Eq. stress_jump}) may be written as
$D_{pn}\xi_{pn}$, giving
\begin{equation}
D_{pn}\xi_{pn}=\frac{a}{2}\Delta\rho\,\kappa_p^2 \left(\xi_{p,n-1}+\xi_{p,n+1}\right).
\label{Eq. jump_xi}
\end{equation}

Since the interface displacement is real, the Fourier coefficients satisfy, for the harmonic response $(\alpha=0)$,
\begin{equation}
    \xi_{p,-n}=\xi_{pn}^*,
\end{equation}
while for the subharmonic response $(\alpha=\omega/2)$,
\begin{equation}
\xi_{p,-n-1}=\xi_{pn}^*.
\end{equation}

At marginal stability, $s=0$, such that $s_n=i(\alpha+n \omega)$. After truncating the Floquet expansion and separating real and imaginary
parts, equation~(\ref{Eq. jump_xi}) takes the generalized eigenvalue form
\begin{equation}
\mathbf A_p\mathbf x_p
=
a\,\mathbf B_p\mathbf x_p.
\end{equation}

For each discrete radial wavenumber $\kappa_p$, the smallest positive real eigenvalue gives the marginal forcing amplitude $a_p$. When scaled by the gravitational acceleration $g$, the dimensionless acceleration reads
\begin{equation}
    F_p=\frac{a_p}{g},
\end{equation}
and the critical acceleration is obtained from
\begin{equation}
    F_c=\min_p F_p
\end{equation}
where the minimizing $\kappa_p$ identifies the radial Bessel mode selected at the onset of the axisymmetric instability.

\section{Energy budget for the decaying case}
\label{App F}
\setcounter{equation}{0}
\renewcommand{\theequation}{F.\arabic{equation}}
\setcounter{table}{0}
\renewcommand{\thetable}{F\arabic{table}}
\setcounter{figure}{0}
\renewcommand{\thefigure}{F\arabic{figure}}

\begin{figure}
\centering
\begin{tabular}{c}
\includegraphics[width=\linewidth]{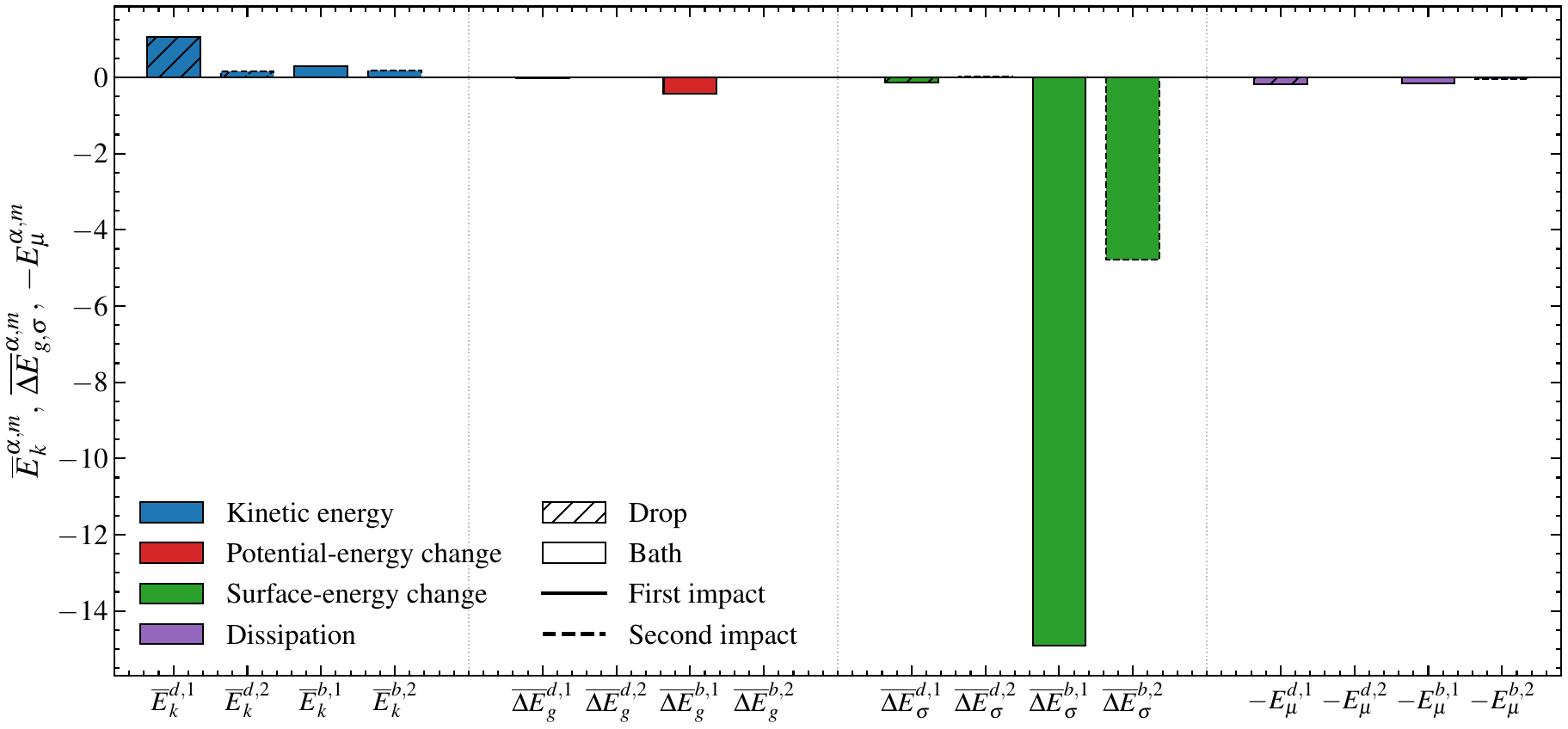}\\
(a)\\
\end{tabular}
\begin{tabular}{cc}
\includegraphics[width=0.47\linewidth]{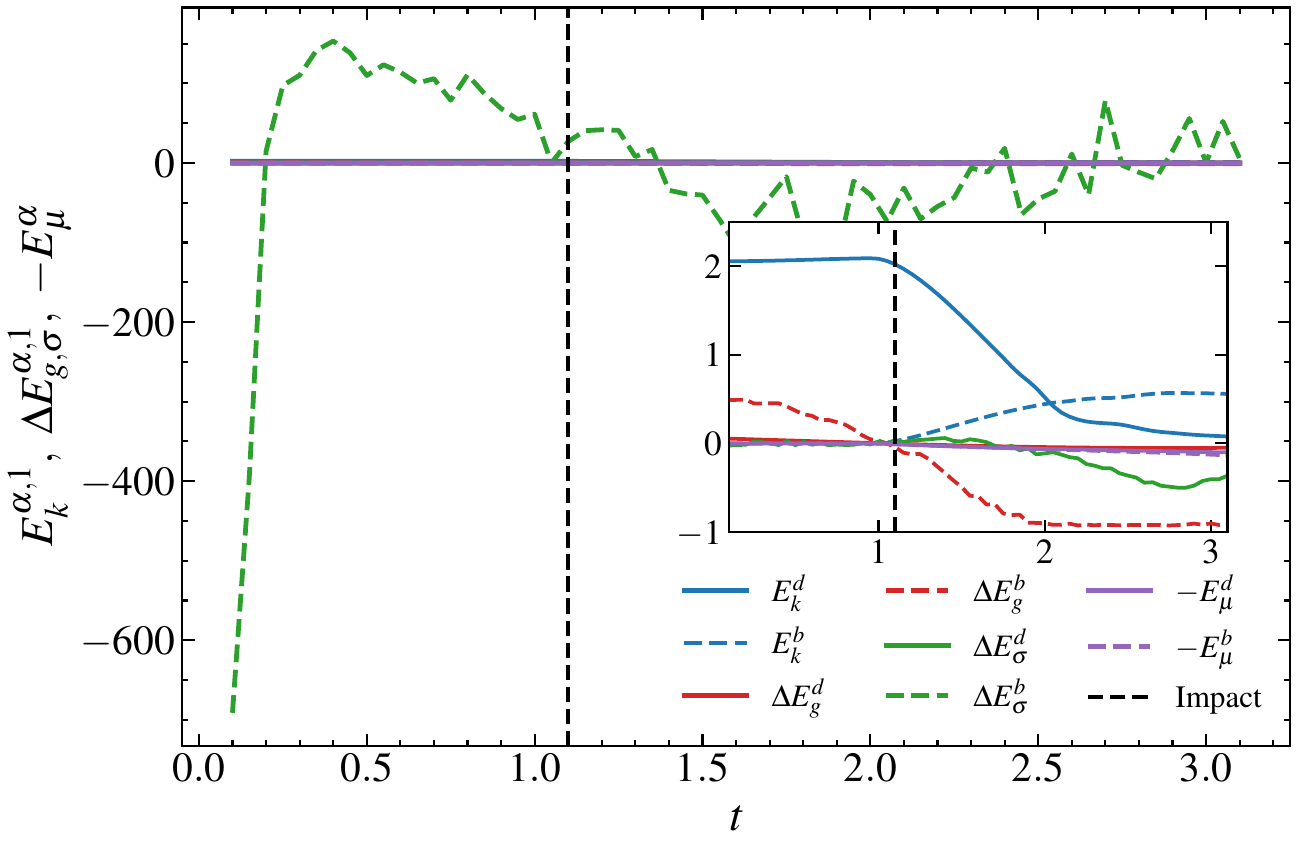}&
\includegraphics[width=0.47\linewidth]{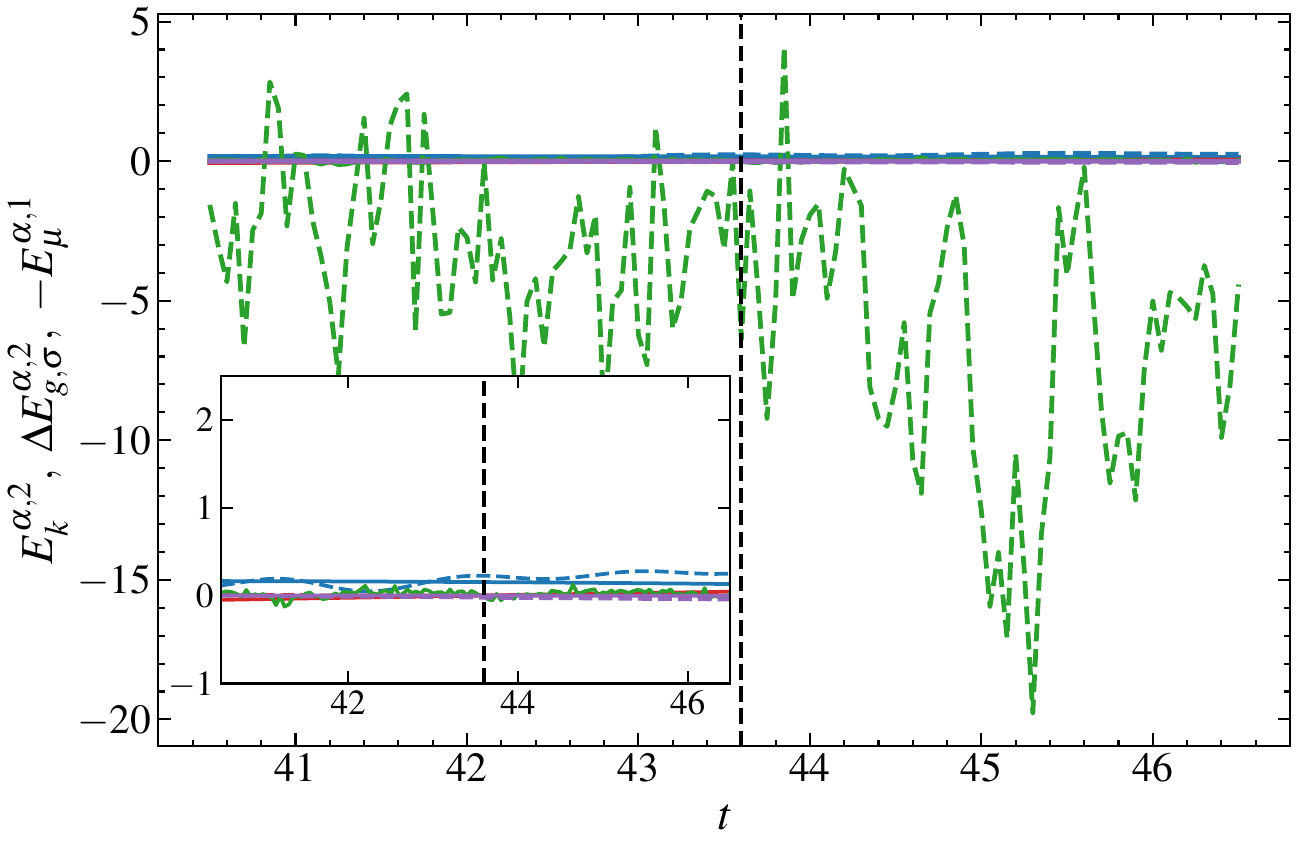}\\
(b)&(c)\\
\end{tabular}
\caption{Energy redistribution during successive droplet impacts for $We=1.21$ and $Re=175$. (a) Phase-decomposed  contributions for the first $(m=1)$ and second $(m=2)$ impacts: event-averaged kinetic-energy $\overline{E}_k^{\alpha,m}$, gravitational-potential energy change $\Delta\overline{E}_g^{\alpha,m}$, surface-energy change $\Delta\overline{E}_\sigma^{\alpha,m}$, and cumulative viscous dissipation $-E_\mu^{\alpha,m}$, where $\alpha=d,b$ represents the droplet and bath, respectively. Hatched and unhatched bars distinguish the droplet and bath, while solid and dashed outlines show the first and second impacts. Panels (b) and (c) illustrate the corresponding temporal energy contributions during first and second impacts, respectively, with the dashed line marking the reference time $t_r^{(m)}$ immediately preceding impact. 
}
\label{fig: energy_decaying}
\end{figure}

For completeness, the impact-resolved energy budget for the representative decaying case, $We=1.21$ and $Re=175$, is presented in this appendix. The first impact is dominated by the droplet kinetic energy and a significant change in the bath surface energy, as shown by figure \ref{fig: energy_decaying}a. Following impact, the droplet's kinetic energy decreases while the bath acquires a smaller but non-zero contribution to kinetic energy, indicating redistribution of the impact energy into motion of the bath and deformation of the interface. Changes in gravitational potential energy are small by comparison. The temporal evolution in figure \ref{fig: energy_decaying}b also shows that the bath surface energy experiences a large excursion after the first collision, followed by an oscillatory relaxation. The droplet kinetic energy and the kinetic-energy response of the bath are both much smaller at the second impact, as well as the magnitude of the surface-energy variation of the bath (see figure \ref{fig: energy_decaying}c). These trends are in agreement with the progressive attenuation of the decaying regime. For both impact events, the cumulative viscous dissipation is negligible compared to the kinetic and surface energy contributions.

The decaying regime has a much larger capillary-energy excursion during the first impact than the unstable regime shown in figure \ref{fig: energy budget}; however, it contains less kinetic energy in the bath during the following collision. At low $We$, the stronger effect of surface tension results in a large temporary storage of the impact energy in interfacial deformation, followed by capillary relaxation of the bath surface. Therefore, there is relatively little bath motion left before the next collision. Conversely, for the high-$We$ unstable case, the excursion in surface energy is weaker, and a larger fraction of the bath kinetic activity remains between successive impacts. This residual bath motion can interact with the subsequent droplet impact in the presence of continuous vertical forcing and contribute to the progressive amplification of the bath response. This comparison indicates that the difference between decaying and unstable regimes arises mainly from the partition of the impact energy between transient capillary deformation and persistent bath motion, rather than from a significantly larger viscous-energy loss in the decaying case.

\bibliographystyle{jfm}
\bibliography{jfm}

\end{document}